\documentclass[%
 preprint,
 amsmath,amssymb,
 aps,
prb,
]{revtex4-2}

\usepackage{graphicx}
\usepackage{dcolumn}
\usepackage{bm}
\usepackage{dsfont}
\usepackage{amsthm}
\usepackage{amsmath}
\usepackage{amsfonts}
\usepackage{tensor}
\usepackage{mathtools}
\usepackage{amssymb}
\usepackage{bm}
\usepackage{setspace}
\usepackage{enumitem}
\usepackage{mdwlist}
\usepackage{parskip}
\usepackage{listings}
\usepackage{color}
\usepackage{hyperref}
\usepackage{subcaption}
\usepackage{mathrsfs}
\usepackage[normalem]{ulem}
\usepackage{bbold}

\usepackage[mathlines]{lineno}
\newcommand{\ads}{\text{AdS}}

\newcommand{\dif}{\mathrm{d}}
\newcommand{\avg}[1]{\left\langle #1 \right\rangle}

\DeclarePairedDelimiter\abs{\lvert}{\rvert}%

\newcommand{\WZI}{Van der Waals - Zeeman Institute, Institute of Physics, University of Amsterdam, Sciencepark 904, 1098 XH Amsterdam, The Netherlands}
\newcommand{\UUE}{Institute for Theoretical Physics and Center for Extreme Matter and Emergent Phenomena, Utrecht University, Princentonplein 5, 3584 CC Utrecht, The Netherlands}
\newcommand{\FBK}{Fondazione Bruno Kessler, Via Sommarive 18, 38123, Povo (TN), Italy}
\newcommand{\UUH}{Institute for Theoretical Physics and Center for Extreme Matter and Emergent Phenomena, Utrecht University, Princentonplein 5, 3584 CC Utrecht, The Netherlands}
\newcommand{\DIEP}{Dutch Institute for Emergent Phenomena (DIEP), Sciencepark 904, 1098 XH Amsterdam, The Netherlands}

\begin{document}

\preprint{APS/123-QED}

\title{Gauge-gravity duality comes to the lab: evidence of momentum-dependent scaling exponents in the nodal electron self-energy of cuprate strange metals}

\author{E. Mauri}
 \email{emauri.research@proton.me}
 \affiliation{\UUE}\affiliation{\FBK}
 \author{S. Smit}%
 \affiliation{\WZI}
\author{M.S. Golden}
\affiliation{\WZI}
\affiliation{\DIEP}
\author{H.T.C. Stoof}%
 \email{H.T.C.Stoof@uu.nl}
\affiliation{\UUH}

\begin{abstract}
We show that the momentum-dependent scaling exponents of the holographic fermion self-energy of the conformal-to-AdS$_2$ Gubser-Rocha model can describe new findings from angle-resolved photoemission spectroscopy experiments on a single layer (Pb,Bi)$_{2}$Sr$_{2-x}$La$_x$CuO$_{6+\delta}$ copper-oxide. In particular, it was recently observed in high-precision measurements on constant energy cuts along the nodal direction that the spectral function departs from the Lorentzian line shape that is expected from the power-law-liquid model of a nodal self-energy, with an imaginary part featureless in momentum as $\Sigma''_{\text{PLL}}(\omega) \propto (\omega^2)^\alpha$. By direct comparison with experimental results, we provide evidence that this departure from either a Fermi liquid or the power-law liquid, resulting in an asymmetry of the spectral function as a function of momentum around the central peak, is captured at low temperature and all dopings by a semi-holographic model that predicts a momentum-dependent scaling exponent in the electron self-energy as $\Sigma(\omega,k) \propto \omega (-\omega^2)^{\alpha (1 - (k - k_F)/k_F) - 1/2}$, with $\hbar k_F$ the Fermi momentum.
\end{abstract}

\maketitle


 \section{\label{sec:message} Introduction}

Back in 1986, in one of the most exciting experimental discoveries in condensed-matter physics, the 
phenomenon of high-temperature superconductivity was observed for the first time in a layered
copper-oxide perovskite  by Bednorz and M\"uller \cite{Bednorz1986}. Since then, other copper-oxide compounds, or cuprates---a class of materials whose  common trait is a layered structure of $\text{CuO}_2$ planes---have been found with increasingly higher critical temperatures. These sit well above the expected limit from the BCS theory of superconductivity \cite{Bardeen1957} that successfully describes the underlying physics of ``conventional'' superconductors. The desire to understand this phenomenon sparked a huge effort from both the experimental and the theoretical community to unveil the mystery behind the anomalous behavior of copper-oxide materials \cite{Keimer2015, Greene2020}, that eludes an explanation within the standard Fermi-liquid framework. This effort is still ongoing, underlining the challenges that these materials present due to the strongly interacting physics at play \cite{Keimer2015, Lee2006, Scalapino2012}.

Peculiarities do not lie within the superconducting phase only, but also in the normal phase of the cuprates just above the maximum critical temperature in the phase diagram, known as the strange-metal regime. As the name suggests, this phase is characterized by a non-Fermi-liquid behavior as highlighted, for example, by an anomalous temperature behavior of the Hall angle \cite{Chen1991} and by a linear-in-$T$ resistivity, that does not saturate at high temperatures \cite{Custers2003, Cooper603, bruin_similarity_2013, Analytis2014, Legros2018, Licciardello2019} and persists at low temperatures even if superconductivity is suppressed by a magnetic field \cite{Hussey_2018}.
There have been a variety of attempts and techniques to model the properties of high-$T_c$ cuprates, such as $t$-$J$ models, starting from the physics of the Mott insulator in the underdoped region of the cuprate phase diagram \cite{Lee2006, tJ_review,AFM_multilayer}, the marginal Fermi liquid for describing the optimally doped strange metal \cite{marginal_FL, marginal_FM_2}, and stripe phases in high-temperature superconductors \cite{Zaanen1996, Emery1999, Berg_2009, Matthias_2009, Zhang68}, to mention a few. One technique, with roots in high-energy and particle physics, that has been used to describe a class of non-Fermi liquids that at low energies shares some of the properties of the strange-metal phase, is based on the gauge/gravity (holographic) duality \cite{Maldacena1998, Witten1998, Gubser1998, Faulkner2011fixed}. It relates the response of a strongly interacting system to a higher-dimensional gravitational theory and it has proven to be a powerful tool when applied to strongly interacting condensed-matter systems to model their qualitative behavior \cite{hartnoll2016holographic, Zaanen2015, ammon_erdmenger_2015}, as it is able to describe some of the anomalous properties observed in transport experiments on cuprates \cite{hartnoll2016holographic}. Moreover, angle-resolved photoemission spectroscopy (ARPES) measurements pointed to a possible explanation of the phenomenology of the strange metal in the presence of a particular quantum critical phase that is local in space, and hence featureless in momentum \cite{Keimer2015, Reber2019}, in accordance with the marginal Fermi-liquid model \cite{PhysRevLett.63.1996}. This is also well captured by the holographic realization of a strongly interacting fermion system \cite{Faulkner2011fixed, Iqbal2009, Iqbal2011, Liu2011, Faulkner2011, Cubrovic2009}, that reproduces the marginal Fermi-liquid results near the Fermi surface. However, moving away from the Fermi surface these holographic models start to deviate from the completely featureless in momentum scaling of the marginal Fermi liquid, as they predict momentum-dependent scaling exponents \cite{Faulkner2011fixed, Iqbal2009, Iqbal2011}. 

Our main objective in this paper is to bring holography to the test of experimental ARPES measurements along the nodal line of a single-layer cuprate. In particular, we aim to verify if its prediction of momentum-dependent scaling exponents in the electron self-energy can explain the recently observed peak asymmetry in experimental data \cite{Smit2021}, that show deviations from the previously proposed power-law liquid (PLL) model \cite{Reber2019}. Such a model is characterized by a momentum-independent self-energy, with an imaginary part obeying $\Sigma''_{PLL}(\omega; T = 0) \propto (\omega^2)^\alpha$. Here $\alpha$ is a scaling exponent increasing (approximately linearly) with doping, from $\alpha = 1/2$ at optimal doping towards, but never reaching, the Fermi-liquid value of $1$ \cite{Smit2021, Negele1998Quantum} at higher dopings. The analysis of the experimental data from ARPES measurements is performed on each momentum distribution curve (MDC), which measures the spectral function as a function of momentum at a fixed (negative) energy $\hbar \omega$. For a range of energies close to the Fermi surface, the PLL model predicts a Lorentzian lineshape for the distribution peaks as
\begin{align}\label{eq:fit_function_pll}
    \mathcal{A}(k; \omega) = \frac{W(\omega)}{\pi} \frac{\Gamma(\omega)/2}{(k - k_*(\omega))^2 + (\Gamma(\omega)/2)^2} \text{ ,}
\end{align}
where $\Gamma(\omega) = 2 \Sigma''_\text{PLL}(\omega)/v_F + G_0(\omega)$ is the full-width at half maximum (FWHM), with $G_0(\omega)$ describing contributions other than the electron self-energy to the width in the data, e.g., due to phonons, impurities, and instrument sensitivity. We define $\Sigma \equiv \Sigma' - i \Sigma''$, thus a negative imaginary part of the self-energy requires $\Sigma'' > 0$. The real part instead modifies the dispersion relation determining $k_*(\omega) \simeq k_F + \omega/v_F$ with $k_F$ the Fermi wave number and $v_F$ the renormalized Fermi velocity. 

\begin{figure}
    \centering
    \includegraphics[width=1.1\linewidth]{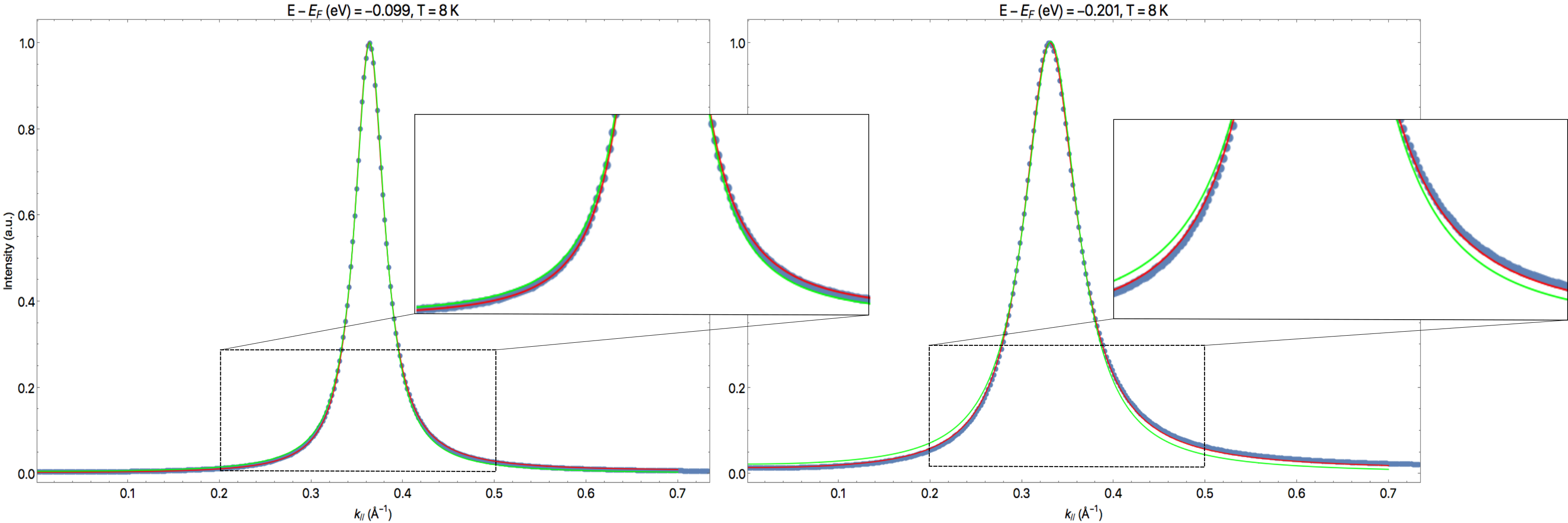}
    \caption{Comparison of the smoothed experimental data (\textcolor{blue}{blue dots}) for an overdoped sample with $\alpha = 0.65$ (see Eq.\ \eqref{eq:self_energy_hol}) at $T = 8 K$, with a Lorentzian fit based on the PLL (\textcolor{green}{green line}), and a fit based on the holographic prediction in Eq.\ \eqref{eq:self_energy_hol} (\textcolor{red}{red line}). While there is little difference up to energies $E-E_F = -0.1 \text{ev}$, (left panel), it is evident that the holographic model accurately captures the peak asymmetry as we move further away from the Fermi surface (right panel). The details of the fit are presented in section \ref{sec:comparison_with_experiment}.}
    \label{fig:experimental_comparison}
\end{figure}

In holography, a prediction, common to a large class of models proposed for the theoretical description of non-Fermi liquid, is that the electron self-energy is dominated by its frequency dependence $\Sigma \propto \omega(-\omega^2)^{\nu_k - 1/2}$, with the momentum dependence confined to its scaling exponent $\nu_k$, and $\omega = \omega + i 0$. Notice that in the literature this result is often quoted as $\Sigma \propto \omega^{2\nu_k}$, here however, in the range of interest to us $1/2 < \nu_k < 1$, we want to make the analytic structure of the self-energy explicit, with a branch cut everywhere on the real axis. In particular, we show that in the model analyzed in this paper, we have
\begin{align}\label{eq:self_energy_hol}
    \Sigma(\omega, k; T = 0) \propto \omega(-\omega^2)^{\alpha (1 - (k - k_F)/k_F)} \text{ ,}
\end{align}
and we explain how this peculiar momentum dependence, which reduces to the PLL form for the sharp distribution peaks near the Fermi surface, provides a much better description of the experimental data away from the Fermi surface. Indeed, this is found to well describe deviations from the typical symmetric Lorentzian shape of the peaks, as observed in very recent high-quality angle-resolved photoemission measurements \cite{Smit2021}, that are reproduced in Fig.\ \ref{fig:experimental_comparison}, convincingly breaking the long-standing assumption of a self-energy that is completely independent of momentum. Note that, while our analysis arises from a holographic calculation of the self-energy, momentum-dependent exponents have also been theorized in a one-dimensional nonlinear-Luttinger liquid model \cite{Jin2019}. The successful description of nodal MDCs by a momentum-dependent scaling exponent could, then, also hint at the emergence of one-dimensional physics along the nodal line.

In the hope of making this paper more accessible to a wider audience, we start first with a brief summary of holographic fermions in Section \ref{sec:fermions}, where we introduce the gravitational background used and its main properties, and explain how to compute fermionic spectral functions in such a background. 
The reader already familiar with holography might want to skip straight to Section \ref{sec:comparison_with_experiment} where the main new results are presented, and refer back to the first sections to check the notation, the conventions adopted as well as details of the derivations that led to the model given there. In Section \ref{sec:IR_geometry} we then show how the low-energy behavior of the spectral function is related to the solution of the Dirac equation in the geometry of the deep interior of the spacetime, and in particular how this leads to an imaginary part of the self-energy described by momentum-dependent scaling exponents. We also explain how this gives rise to an asymmetry in the spectral function peaks. In Section \ref{sec:semi-holography} we introduce a semi-holographic construction as proposed in Ref.\ \cite{Gursoy2012} and explain the limitations of this approach in describing quantitatively experimental data due to the non-universal real part of the self-energy. We then proceed to decouple the non-universal physics from the low-energy emergent quantum critical behavior to accurately model ARPES measurements on a single $\text{CuO}_2$-layer cuprate. The details of the modeling of the experimental data are the topic of Section \ref{sec:comparison_with_experiment}, where we, in particular, explain how we need to interpret the holographic fermion response as the one for the holes with an emergent particle-hole symmetry. We also show that changing parameters in the Dirac equation allows for a proper fit of the cuprate fermionic self-energy at different dopings, and we finally introduce also the self-energy corrections due to the coupling to phonon degrees of freedom, to correctly capture the experimentally observed behavior of the MDC's at both low temperatures and higher frequencies.

\section{\label{sec:fermions}Introduction to fermions in Einstein-Maxwell-Dilaton theories}

In this section, we summarize how to use the tools provided by the so-called AdS/CFT correspondence to compute fermionic spectral functions for a strongly interacting system. In particular, we introduce the gravitational background used for the computation and explain how it captures some key characteristics of the cuprates, as well as remind the reader how to add fermions in this background to compute the Green's function of probe fermions in the holography framework. Although the foundations on which this section is based can be found in Refs. \cite{Liu2011, Iqbal2011, Gubser2012}, we use it here to introduce the notation adopted throughout the paper,
 with special emphasis on deriving the solution in terms of the boundary energy scales, $k_B T$, and the chemical potential $\mu$, and in keeping track of dimensionful factors. 

\subsection{Gravitational background}
For our background we use a 3D Einstein-Maxwell-dilaton theory \cite{Gubser_2010} (where with the notation $n$D, we use the condensed-matter convention to denote with $n$ the number of spatial dimensions only). This has been proposed as a holographic dual for the description of the low-energy physics of the strange-metal phase of the 2D and strongly interacting cuprates, as it captures the linear in temperature behavior of the resistivity, it does not become unstable in the zero-temperature limit and, as we will show in this paper, it describes extremely well their fermionic response obtained by ARPES measurements. This is part of a large class of models characterized by a dynamical scaling exponent $z = \infty$---of which the most studied one is the Reissner-Nordstr\"om background \cite{Liu2011}---that describes an emergent low-energy quantum critical phase where, under a scaling transformation, time scales while space does not. This implies that momentum becomes dimensionless under scaling and the electron self-energy is, hence, dominated by the frequency dependence. The Gubser-Rocha model considered here is further characterized by a hyperscaling-violating exponent $\theta = -\infty$, with $-\theta/z = 1$, that allows, once broken translational symmetry is accounted for in the model, for the linear in temperature resistivity $\rho$ and entropy $S$ as $\rho \propto S \propto T^{(d-\theta)/z}$ in $d$ spatial dimensions of the boundary \cite{Zaanen2015}. To clarify the terminology, we want to point out that the model presented in this paper is that of a non-Fermi liquid with translation invariance, so it has infinite conductivity at all temperatures. In this paper we are interested in the fermionic self-energy alone and not on transport properties so we do not consider disorder in the system \footnote{Without disorder, our model is translationally invariant and it, hence, has an infinite conductivity at all $T$. Given that a strange metal is defined by its linear-in-$T$ resistivity, it would be more appropriate to refer to the model as that of a \textit{non-Fermi liquid}. However, since we do not consider transport here but only the fermionic self-energy that we compare to the spectral function measured by ARPES of the \textit{strange metal} phase of a cuprate, we still sometimes refer to the model as one for a strange metal with a little abuse of nomenclature.}, for transport properties in such a model with broken translational symmetry, see for example Ref.\ \cite{Balm2023}.

The gravitational action for the Gubser-Rocha model \cite{Gubser_2010} is 
\begin{align}\label{eq:background_action}
    S_{\text{EMD}} = \frac{c^3}{16\pi G} \int \dif r \dif (c t) \dif^2 x \sqrt{-g}\left[R - \frac{(\partial_\mu\phi)^2}{2} + \frac{6}{L^2} \cosh\left(\frac{\phi}{\sqrt{3}}\right) - \frac{e^{\phi/\sqrt{3}}}{4 g_F^2} F_{\mu\nu}^2 \right] \text{ ,}
\end{align}
with $r$ denoting the additional spacial direction of the curved bulk spacetime, $R$ the Ricci scalar, $\phi$ a dimensionless scalar field known as the dilaton, and $F_{\mu\nu}$ the electromagnetic tensor with coupling constant
\begin{align}
    g_F^2 = \frac{c^4 \mu_0}{16 \pi G} \text{ ,}
\end{align}
where $[\mu_0] = \text{m kg}/\text{C}^2$ is a constant with the dimension of a magnetic permittivity. Finally, $L$ is the anti-de Sitter (AdS) radius. 
The equations of motion following from the action in Eq.\ \eqref{eq:background_action} define a metric of the form 
\begin{align}
  \dif s^2 = e^{2\alpha(r)}(-f(r) \dif t^2 + \dif x^2) + \dif r^2 e^{-2\alpha(r)}/f(r)
\end{align}
and are solved by
\begin{align}\label{eq:background_eom}
   \begin{split}
      A_0(r) =& g_F \frac{\sqrt{3 Q (r_0 + Q)}}{L}\left(1 - \frac{r_0 + Q}{r + Q}\right) \text{ ,}\\
      \phi(r) =& \frac{\sqrt{3}}{2} \log\left(1 + \frac{Q}{r}\right) \text{ ,}\\
      \alpha(r) =& \log\left(\frac{r}{L}\right) + \frac{3}{4}\log\left(1 + \frac{Q}{r}\right) \text{ ,}\\
      f(r) =& 1 - \left(\frac{r_0 + Q}{r + Q}\right)^3 \text{ ,}
   \end{split}
\end{align}
with $Q$ a positive integration constant with the dimension of a length and $r_0 \in [0, \infty)$ the horizon radius. The chemical potential $\mu$ of the boundary field theory is defined through the asymptotic behavior of the bulk gauge field
\begin{align}
    \lim_{r \to \infty} A_0(r) = g_F \frac{\sqrt{3 Q (r_0 + Q)}}{L} \equiv \frac{\mu}{c q} \text{ ,}
\end{align}
with $q$ the electric charge so that $\mu$ has dimensions of an energy, and $c$ the speed of light. At high energies, it describes a system with a relativistic linear dispersion $\hbar \omega = \pm \hbar c k - \mu$ in the spectral function of the density operator dual to the Maxwell field. However, we are ultimately interested in the effective description of the electronic response in the cuprates, where near the Fermi energy the dispersion can also be linearized, but with a velocity $v_F$ much smaller than the speed of light. Hence, we interpret the speed $c$ in the holographic model as the (bare) Fermi velocity of the electron system near the Fermi surface. We come back to this point later on in Sec.\ref{sec:comparison_with_experiment}.
The temperature of the boundary field theory can be computed from the black-hole horizon in the bulk spacetime
\begin{align}
    k_B T = \frac{c \hbar}{L}\frac{3 \sqrt{r_0 (r_0 + Q)}}{4 \pi L}\text{ ,}
\end{align}
and we can see that, contrary to the Reissner-Nordstr\"om solution that is unstable at low temperatures, in the Gubser-Rocha model $T \to 0$ for $r_0 \to 0$, hence the entropy, $S \propto A_{\text{bh}} = \int \dif x \dif y\, e^{2 \alpha(r_0)} \sim \sqrt{r_0} \text{ for } r_0 \to 0$, vanishes in the zero-temperature limit. 

For better clarity as well as for numerical computations, we want to work with dimensionless quantities, by expressing everything in terms of physical constants and the dimensionful scale $L$ of the theory, that is by measuring distances in units of $L$ and energies in terms of $\hbar c/L$. We, therefore, define the dimensionless coordinates $(\tilde r, \tilde t, \tilde{\bm x}) \equiv (r, ct, \bm x)/L$, and we adsorb the gauge coupling into the field $\tilde A_{\mu} \equiv A_\mu/g_F$. The action in terms of these dimensionless coordinates and fields then becomes
\begin{align}\label{eq:dimless_background_action}
    \tilde S_{\text{EMD}} = \frac{c^3 L^2}{16\pi \hbar G} \int \dif \tilde r\, \dif \tilde t\, \dif^2 \tilde{x}\sqrt{-g} \left[R - \frac{(\partial_\mu\phi)^2}{2} + 6 \cosh\left(\frac{\phi}{\sqrt{3}}\right) - \frac{e^{\phi/\sqrt{3}}}{4} \tilde{F}_{\mu\nu}^2 \right] \text{ .}
\end{align}
We further define a dimensionless electric charge $\tilde q \equiv q L g_F/\hbar$ so that $\lim_{r \to \infty} \tilde A_0 = L \mu/\tilde q \hbar c \equiv \tilde \mu/\tilde q$, and temperature ${\tilde T} \equiv L k_B T/\hbar c$. From now on, we will use dimensionless quantities only, unless explicitly stated otherwise, so we will drop the tilde for notational convenience. 

The solutions in Eq.\ \eqref{eq:background_eom} can then be expressed in terms of the boundary field theory (dimensionless) chemical potential and temperature, to take the form:
\begin{align}\label{eq:background_eom_v2_tilde}
   \begin{split}
      A_0(r) =& \frac{\mu}{q}\left(1 - \frac{1 + q^2 \frac{(4 \pi T)^2}{3 \mu^2}}{1 + \sqrt{3} q \frac{r}{\mu} \sqrt{1 + q^2 \frac{(4 \pi T)^2}{3 \mu^2} }}\right)\\
      \phi(r) =& \frac{\sqrt{3}}{2} \log\left(1 + \frac{1}{\sqrt{3} q \frac{r}{\mu} \sqrt{1 +  q^2 \frac{(4 \pi T)^2}{3\mu^2}}}\right)\\
      \alpha(r) =& \log\left(r\right) + \frac{3}{4}\log\left(1 + \frac{1}{\sqrt{3}q \frac{r}{\mu} \sqrt{1 +  q^2 \frac{(4 \pi T)^2}{3 \mu^2}}}\right)\\
      f(r) =& 1 - \left(\frac{1 +  q^2 \frac{(4 \pi T)^2}{3 \mu^2}}{1 +  \sqrt{3} q \frac{r}{\mu}\sqrt{1 +  q^2 \frac{(4 \pi T)^2}{3 \mu^2} }}\right)^3 \text{ ,}
   \end{split}
\end{align}
making it explicit that the solution depends only on the energy scale $\mu/q$ and on the dimensionless ratio $q T/\mu$, as expected from the scaling symmetry of a deformed conformal field theory (CFT).

\subsection{Holographic fermions}

In order to use the tools of holography to compute the spectral function of a fermionic boundary operator $\mathcal{O}$, we need to add a Dirac action to the higher-dimensional gravitational background action $S_{\text{EMD}}$ \cite{Liu2011, Iqbal2009, Iqbal2011, Zaanen2015}, that, reverting to dimensionful units for a moment, takes the form:
\begin{align}\label{eq:dirac_action}
    S = S_{\text{EMD}} - i g_f \int \dif r \dif (ct) \dif^{2} x\, \sqrt{-g} \bar\psi \left[ e^\mu_a \Gamma^a \left(\hbar c \left[\partial_\mu + \frac{1}{4}\omega_{\mu b c} \Gamma^{bc}\right] - i q c A_\mu \right) - m c^2 \right] \psi~,
\end{align}
with $\Gamma^a$ the Dirac gamma matrices, $\bar\psi \equiv \psi^\dagger \Gamma^0$, $\Gamma^{bc} \equiv 2 [\Gamma^b, \Gamma^c]$, the vielbein $e^\mu_a$ is defined by $e^\mu_a e^\nu_b g_{\mu\nu} = \eta_{\mu\nu}$ and the spin connection is $\omega_{\mu b c} $, which ensures that local Lorentz symmetries are preserved and is defined as 
\begin{align}
    \tensor{\omega}{_\mu_c_b} = \eta_{ca} \tensor{\omega}{_\mu^a_b} = \eta_{ca}(e^a_\lambda e^\nu_b \Gamma^\lambda_{\mu\nu} - e^\nu_b \partial_\mu e^a_\nu) \text{ ,}
\end{align}
where $\Gamma^\lambda_{\mu\nu} \equiv 1/2 g^{\lambda\sigma}(\partial_\mu g_{\sigma\nu} + \partial_\nu g_{\sigma\mu} - \partial_\sigma g_{\mu\nu})$ are the Christoffel symbols (the index structure should avoid the confusion with the $\Gamma$ matrices).
Finally, $g_f$ is a coupling constant with the dimension of an inverse velocity. Introducing the dimensionless variables of the previous section, we see that there are only two dimensionless parameters characterizing the fermions, namely  $\tilde q$ and the dimensionless mass $\tilde m = m c L/\hbar$, as the action takes the form 
\begin{align}\label{eq:dirac_action_dimless}
    \tilde S = \tilde S_{\text{EMD}} - i \tilde g_f \int \dif \tilde r \dif \tilde t \dif^{2} \tilde{x} \sqrt{-g} \bar\psi \left[ e^\mu_a \Gamma^a \left(\left[\partial_\mu + \frac{1}{4}\omega_{\mu b c} \Gamma^{bc}\right] - i \tilde q \tilde A_\mu \right) - \tilde m \right] \psi \text{ ,}
\end{align}
where again, we are going to drop the tilde from now on, as we always use rescaled quantities unless explicitly mentioned otherwise. 

In particular, since we are interested in a 2D boundary theory, we specified above already the case $d = 2$. In agreement with this, we now choose a representation of the $\Gamma$ matrices of the form
\begin{align}
  \begin{array}{cc}
    \Gamma^{r} = \left(\begin{array}{cc}
        \mathbb{1} & \mathbb{0} \\
        \mathbb{0} & -\mathbb{1}
    \end{array}\right) \text{ , and} & \Gamma^{\mu} = \left(\begin{array}{cc}
        \mathbb{0} & \gamma^\mu \\
        \gamma^\mu & \mathbb{0}
    \end{array}\right) \text{ ,} 
   \end{array}
\end{align}
with $\gamma^\mu$ the 2D gamma matrices. For definiteness, we choose a basis as in Ref.\ \cite{Liu2011}: $\gamma^0 = i \sigma_2$, $\gamma^1 = \sigma_1$, $\gamma^2 = \sigma_3$, with $\sigma$ the Pauli matrices.
We can then decompose $\psi$ in terms of the chirality eigenvectors of $\Gamma^r$, namely $\psi = \psi_R + \psi_L$ with $\Gamma^r \psi_{R} = \psi_R$ and $\Gamma^r \psi_{L} = -\psi_L$. 
Upon variation of the Dirac action in Eq.\ \eqref{eq:dirac_action_dimless}, we straightforwardly obtain the Dirac equation in curved spacetime for the field $\psi$ as
\begin{align}\label{eq:dirac_equation}
  \left[ e^\mu_a \Gamma^a \left(\left[\partial_\mu + \frac{1}{4}\omega_{\mu b c} \Gamma^{bc}\right] - i q A_\mu \right) - m \right] \psi = 0\text{ ,}
\end{align}
and similarly, we can obtain the equation of motion for $\bar\psi$. 
The boundary term arising from the variation of the action is given by
\begin{align}
    \delta S_{\partial} =& -i g_f \int_{\partial} \dif t \dif^2 {x} \sqrt{-h} \bar\psi \Gamma^r \delta\psi =\\
    =& -i g_f \int_\partial \dif t \dif^2{x} \sqrt{-h} (\bar\psi_L\delta\psi_{R} - \bar\psi_R\delta\psi_{L}) \text{ ,}
\end{align}
since terms of the form $\bar\psi_{C}\delta\psi_{C} = \psi_{C}^\dagger \Gamma^0 \delta\psi_{C} = 0$, with $C = R, L$. Here $h$ is the
determinant of the induced metric on the boundary $h_{\mu\nu} = g_{\mu\nu} - n_\mu n_\nu$, $n_\mu$ being the normal vector orthogonal to the boundary.

If we now introduce a boundary term of the form
\begin{align}
   S_\partial = - i g_f \int \dif t \dif^2 { x} \sqrt{-h} \bar\psi_{R}\psi_{L} \text{ ,}
\end{align}
the on-shell action is stationary upon imposing the boundary condition $\delta \psi_R = 0$. We then interpret the boundary value of the field $\psi_R$ as the source of a boundary fermionic operator, with $\psi_L$ determining its one-point function. In order to make this split explicit, we define $\psi_{\pm}$ such that $\psi_{R} = \left(\begin{array}{c}
     \psi_{+}  \\
     0
\end{array}\right)$ and $\psi_{L} = \left(\begin{array}{c}
     0 \\
     \psi_{-}
\end{array}\right)$, where $\psi_\pm$ are two-components Dirac spinors. These are not independent, but related by the Dirac equation by 
\begin{align}\label{eq:xi_matrix}
    \psi_{-}(r, \omega, \bm k) = -i\xi(r, \omega, \bm k) \psi_{+}(r, \omega, \bm k) \text{ ,}
\end{align}
where we adopted momentum-space representation for later convenience.
Inserting Eq.\ \eqref{eq:xi_matrix} into the Fourier transform of the  boundary action we obtain
\begin{align}
    S_\partial = -i g_f \lim_{\Lambda \to \infty} \int_{r = \Lambda} \frac{\dif \omega \dif^2 k}{(2\pi)^3}\,\sqrt{-h} \psi_+^\dagger(r, \omega, \bm k) \gamma^0 (-i \xi(r, \omega, \bm k)) \psi_{+}(r, \omega, \bm k)\text{ ,}
\end{align}
where $\Lambda$ is an ultra-violet (UV) cutoff scale.

Near the boundary (as $r \to \infty$) the mass term becomes the dominant one in the Dirac equation and $\psi$ behaves as 
\begin{align}\label{eq:psi_near_boundary}
    \psi = \left(\begin{array}{c}
         \psi_+^{(0)}  \\
         0
    \end{array}\right) r^{-3/2 + m}(1 + \dots) + \left(\begin{array}{c}
         0 \\
         \psi_-^{(0)}  
    \end{array}\right) r^{-3/2 - m}(1 + \dots) \text{ ,}
\end{align}
where the ``$\dots$'' stand for lower-order terms in the large-$r$ limit and we restricted ourselves to $m \in (-1/2, 1/2)$. We can then interpret $\psi_+^{(0)} = \lim_{r \to \infty} r^{3/2 - m} \psi_+$ as the source of the two-dimensional boundary Dirac operator $\cal \bar O_-$, with conformal dimension $\Delta = 3/2 + m$, whose one-point function is $\avg{\bar{\cal O}_-} = \bar{\psi}_-^{(0)}$. 
We then define the Green's function from the effective boundary action that can be written as
\begin{align}\label{eq:effective_boundary_action}
 \begin{split}
    S_\partial =& -i\int \frac{\dif \omega \dif^2 k}{(2\pi)^3}\, \psi_{+}^{(0)\dagger}(\omega, \bm k)\gamma^0\left(-i g_f \lim_{\Lambda \to \infty} \Lambda^{2 m} \xi(r = \Lambda, \omega, \bm k) \right)\psi_{+}^{(0)}(\omega, \bm k) \\
    =& -i \int \frac{\dif \omega \dif^2 k}{(2\pi)^3}\, \psi_{+}^{(0)\dagger}(\omega, \bm k)\gamma^0 \left(-i G_H(\omega, \bm k) \gamma^0\right)\psi_{+}^{(0)}(\omega, \bm k) \text{ ,}
  \end{split}
\end{align}
where we used $\avg{\mathcal{O}_{-}} = -i G_H \gamma^0 \psi_{+}^{(0)}$ \cite{Liu2011}, so that the holographic Green's function is 
\begin{align}\label{eq:green_probe_fermions}
    G_H 
     =\avg{\cal{O}_{-} \cal{O}_{-}^\dagger} 
    \equiv \left(-g_f \lim_{\Lambda \to \infty} \Lambda^{2m} \xi(r = \Lambda) \gamma^0 \right) \text{ .}
\end{align}

Notice that in the mass range $-1/2 < m < 1/2$, both terms in the expansion in Eq.\ \eqref{eq:psi_near_boundary} are normalizable and there are, therefore, two possible quantizations. The one implicitly used above is known as \textit{standard quantization}, where $\psi_{+}^{(0)}$ is the fixed source. However, in what is known as \textit{alternative quantization} we exchange the role of coefficients, i.e., $\psi_{+}^{(0)}$ is now considered a dynamical field we can integrate over and we then have that $\xi$ is proportional to the inverse of the Green's function, i.e., $G_{H}^{\text{alt}} \equiv -G_H^{-1} = \avg{\cal O_{+} \cal{O}_{+}^\dagger}$, where the last equality underlies that this alternative Green's function is the two-point function associated to a fermionic operator $\cal O_{+}$ with conformal dimension $\Delta = 3/2 - m$ \footnote{As you can see, the dimension of the operator in standard and alternative quantization is related by $m \rightarrow -m$, if we indeed always define the source as the leading order term in the expansion in Eq.\ \eqref{eq:psi_near_boundary}, then changing the sign of $m$ exchanges the roles of the coefficients as the source and operator response, effectively going from standard to alternative quantization.}. It is also important to point out the fact that we are considering the Green's function for \textit{probe fermions}, as we are interested in the behavior of the electronic self-energy near the Fermi surface due to interactions, and not on transport. To study the latter, one needs to go beyond the probe limit and consider the background charged matter as, e.g., in Ref.\ \cite{Balm2023, Jacobs2014, Jacobs2015}. The relationship between probe fermions and background spectrum is a matter of ongoing research, see also the discussion at the end of section 4.4 in Ref.\ \cite{hartnoll2016holographic}.

Given that we are ultimately interested in studying the response function, it is typical to derive an equation for the components of the matrix $\xi$ directly. 
Introducing the rescaled fields $\psi_{\pm} = (-g g^{rr})^{-1/4} e^{-i \omega t + i \bm{k}\cdot\bm{x}} \left(\begin{array}{c}
      -i y_\pm\\
       z_\pm
\end{array}\right)$ and using the rotational symmetry to set without loss of generality the momentum along the x-axis, i.e., $\bm k = (k, 0, 0)$, the bulk Dirac equation derived from the gravitational action in Eq.\ \eqref{eq:dirac_action} reads
\begin{align}
    \left(\partial_r \mp m \sqrt{g_{rr}} \right) y_{\pm} =& \pm \sqrt{\frac{g_{rr}}{g_{xx}}} \left(k - \sqrt{\frac{g_{rr}}{-g_{tt}}}(\omega + q A_t)\right) z_\mp \text{ ,}\\
    \left(\partial_r \pm m \sqrt{g_{rr}} \right) z_{\mp} =& \pm \sqrt{\frac{g_{rr}}{g_{xx}}} \left( k + \sqrt{\frac{g_{rr}}{-g_{tt}}}( \omega + q  A_t)\right) y_\pm \text{ .}
\end{align}
Finally, defining the field ratios $\xi_{+} = y_{-}/z_{+}$, $\xi_{-} = z_{-}/y_{+}$, we can now show that the Green's function is
\begin{align}\label{eq:green_matrix}
    G_H = g_f \lim_{\Lambda \to \infty} \Lambda^{2 m} \left(\begin{array}{cc}
        \xi_{+}(r = \Lambda, \omega, k) & 0 \\
        0 & \xi_{-}(r = \Lambda, \omega, k) 
    \end{array}\right) \text{ ,}
\end{align}
where $\xi_{\pm}$ are solutions to 
\begin{align}\label{eq:xi_equations}
    \partial_r \xi_\pm = -2 m \sqrt{g_{rr}}\xi_\pm \mp \sqrt{\frac{g_{rr}}{g_{xx}}} \left[\left( k \mp \sqrt{\frac{g_{rr}}{-g_{tt}}}( \omega + q A_t) \right) - \left(k \pm \sqrt{\frac{g_{rr}}{-g_{tt}}}( \omega + q  A_t) \right)\xi^2_\pm\right] \text{ ,}
\end{align}
with the infalling boundary conditions at the black-hole horizon for the fermionic field $\psi$ corresponding to $\xi_{\pm}(r = r_0) = i$ at any non-zero frequency (the boundary conditions at zero frequency become $\xi = \pm \sqrt{g_{xx}(r_0)}m/k - \sqrt{g_{xx}(r_0) m^2/k^2 + 1}$) \cite{Son_2002, Iqbal2009}.
Furthermore, it follows from Eq.\ \eqref{eq:xi_equations}, that the solutions for $\xi_\pm$ satisfy $\xi_{\pm}^{(m)}(\omega, k) = -1/\xi_{\mp}^{(-m)}(\omega, k)$. Hence, the Green's function for $-m$, when $m \in (-1/2, 1/2)$, is simply equivalent to the Green's function in alternative quantization. We can also see that $\xi_+(\omega, k) = \xi_-(\omega, -k)$, allowing us to focus only on one component of the Green's function matrix. We then solve the equation numerically by integrating from the horizon to a UV cutoff $\Lambda = 10^{-8}$ to compute $G_H$ according to Eq.\ \eqref{eq:green_matrix} Notice, from the large-$r$ expansion of Eq.\ \eqref{eq:xi_equations} that extra care is needed to treat the limiting cases $m \to \pm 1/2$, see e.g.\ Ref.\ \cite{Iqbal2009}. This is not a problem, however, in the case analyzed in this paper, where we fix $m = -0.49$.

\begin{figure}
    \centering
    \includegraphics[width=0.95\linewidth]{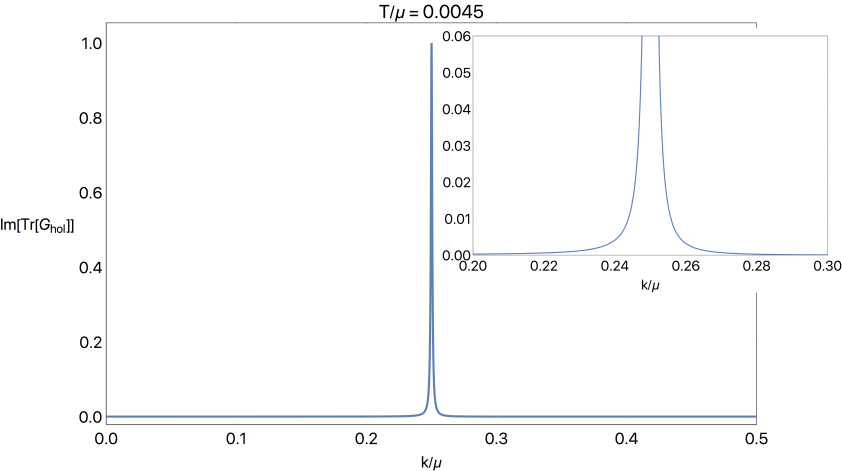}
    \caption{Peak in the fermionic spectral function at the Fermi energy, normalized to peak height, and a zoomed-in version (top right), showing the presence of a Fermi surface.  Here we used $m = -0.49$, $q = 0.27$ at $T/\mu = 0.0045$.}
    \label{fig:fermi_surface}
\end{figure}

\begin{figure}
    \centering
    \includegraphics[width=\linewidth]{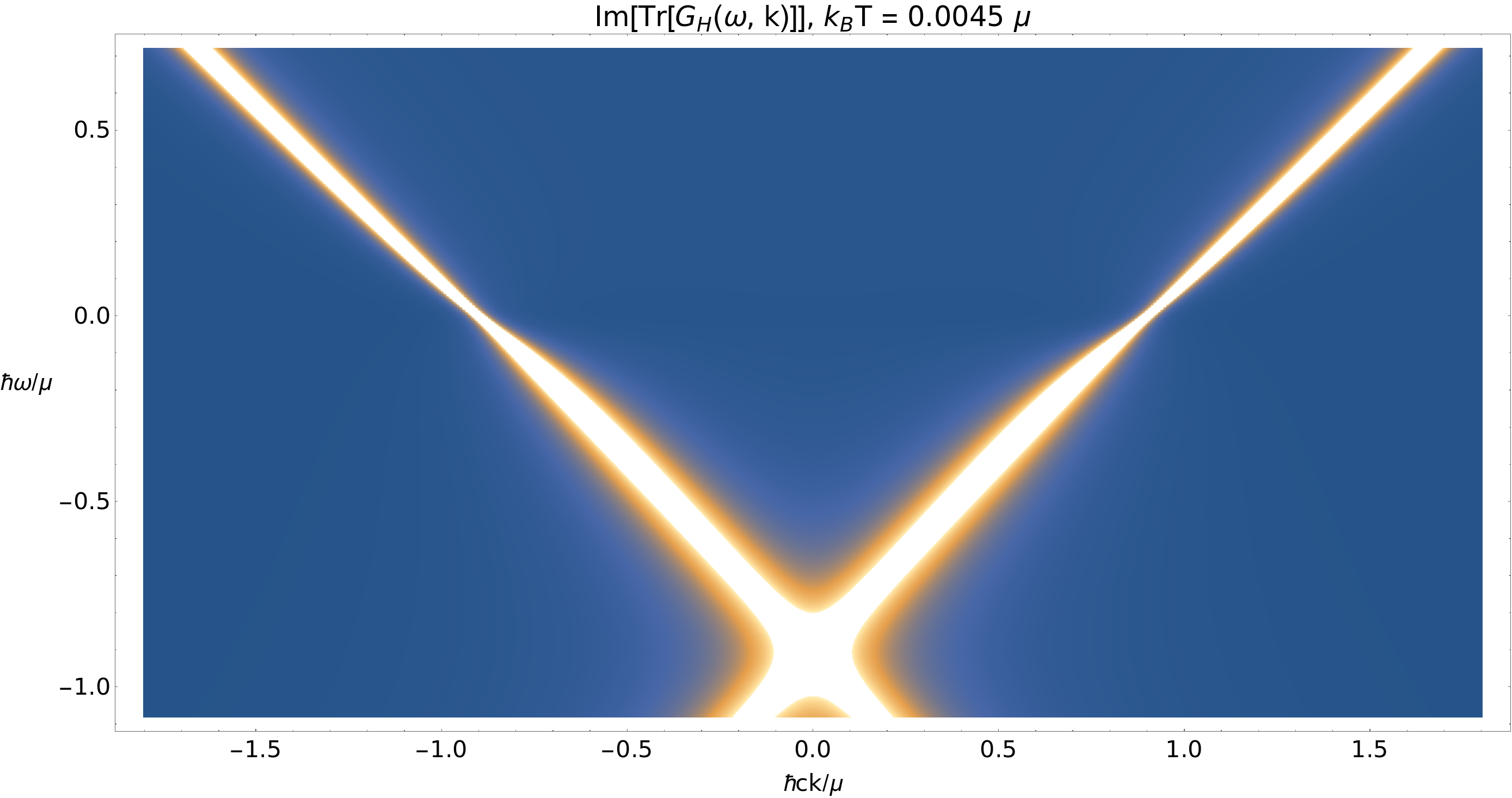}
    \caption{Spectral function for a fermionic operator computed from holography. It is symmetric in momentum and it shows a linear dispersion expected for a massless fermion, with the cone shifted down by the chemical potential. Here we used $m = -0.49$ and $q = 0.27$.}
    \label{fig:holographic_spec}
\end{figure}

The spectral function for the holographic fermion highlights the presence of a Fermi  surface at a non-zero $k \equiv k_F$, as shown in Fig.\ \ref{fig:fermi_surface} by a sharp peak at $\omega = 0$, corresponding to the Fermi energy. Depending on the $m$ and $q$ parameters in the Dirac equation, there can be zero, one, or multiple Fermi surfaces \cite{Liu2011, Gubser2012}. However, for the values of the mass and the charge that we use in the rest of the paper, we always deal with a single well-defined Fermi surface, as in the example shown in Fig.\ \ref{fig:holographic_spec}.

\begin{figure}[h]\centering
  \begin{subfigure}{.5\textwidth}
    \centering
    \includegraphics[width=\linewidth]{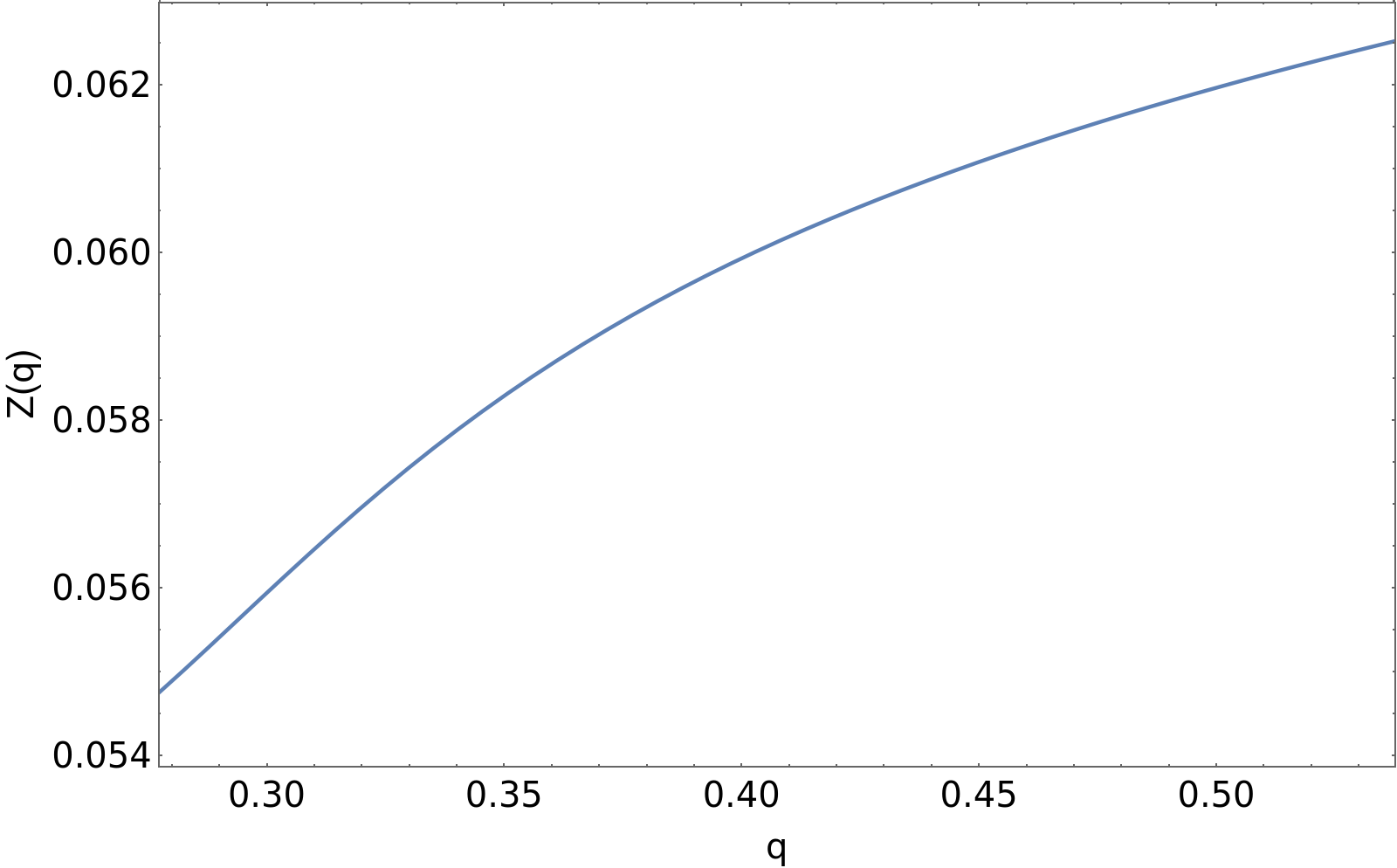}
  \end{subfigure}%
\begin{subfigure}{.5\textwidth}
  \centering
  \includegraphics[width=\linewidth]{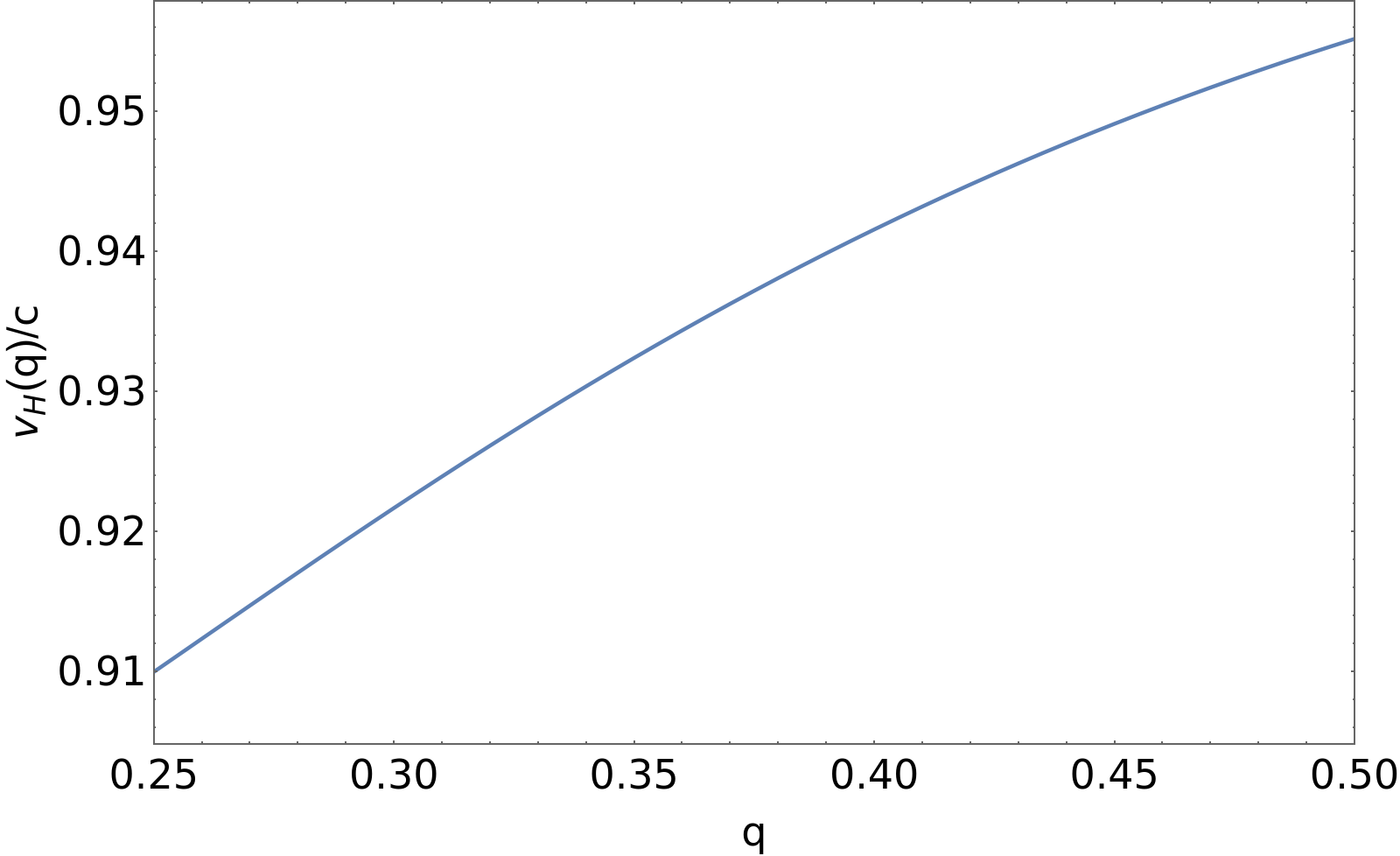}
\end{subfigure}
\vskip\baselineskip
 \begin{subfigure}{.5\textwidth}
    \centering
    \includegraphics[width=\linewidth]{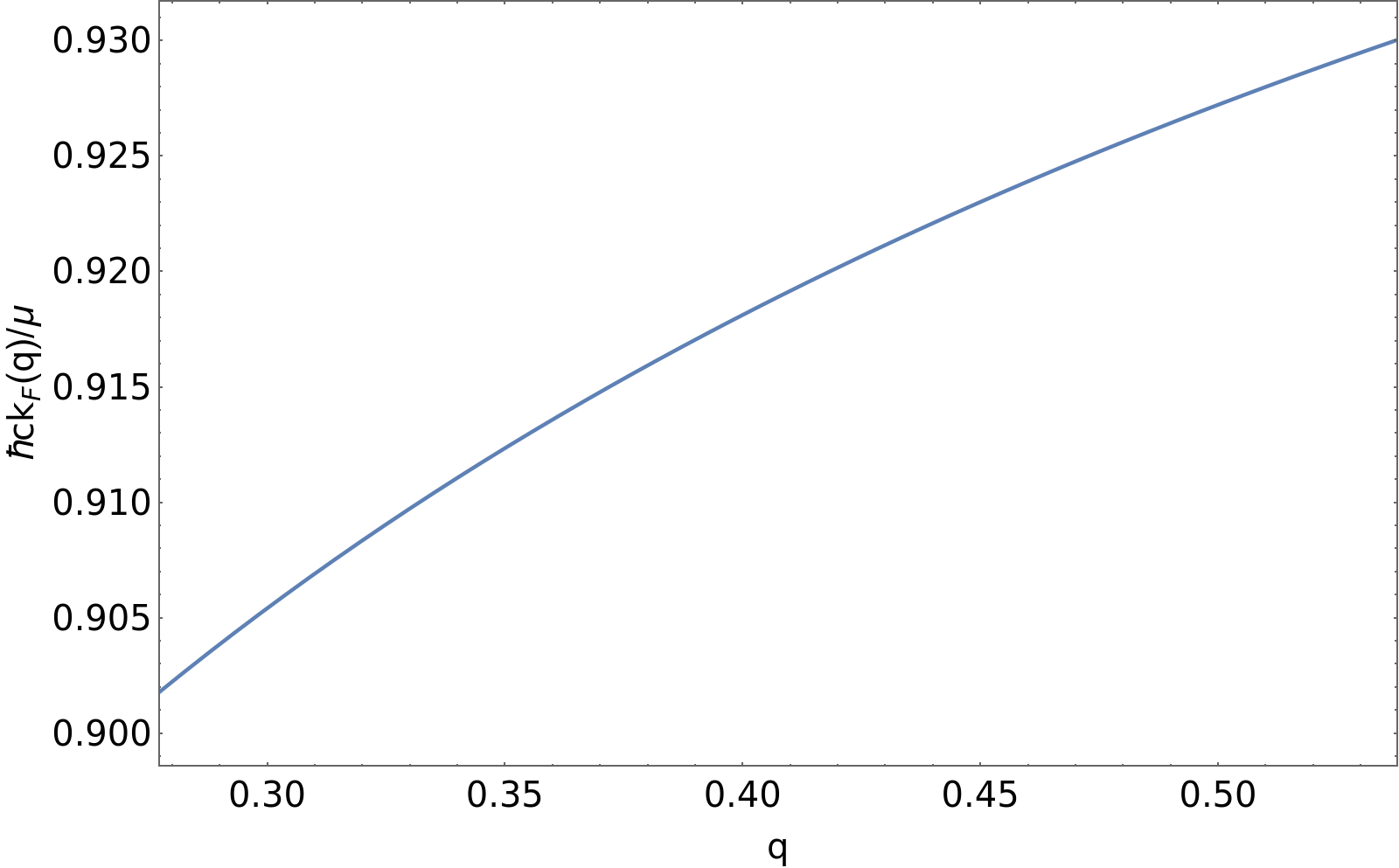}
  \end{subfigure}%
\caption{\label{fig:fit_green_parameters} Numerical results for the dependence on the charge $q$ of the quantities determining the Green's function near the Fermi surface. We fixed the mass $m = -0.49$ and we used a value $k_B T/\mu = 4.5 \times 10^{-3}$}
\end{figure}

The physics of the Green's function near the Fermi surface is well captured by a familiar form for the two-point function of a fermionic particle
\begin{align}\label{eq:holographic_green_fermi}
    G_H \simeq \frac{Z}{-\omega + v_H (k - k_F) - i \Sigma''(\omega, k)} \text{ ,}
\end{align}
with $\Sigma''(\omega, k) > 0$ governing the decay rate of the excitations and, hence, determining the shape of the peaks observed in ARPES experiments. For this reason, $\Sigma''(\omega, k)$ is the main focus of this paper and we analyze it in detail in the next section. In many holographic theories, as in the Gubser-Rocha model used here, $k_F$, $v_H$, and $Z$ can only be determined numerically, for general $m$ and $q$, from the full solution of Eq.\ \eqref{eq:xi_equations}. In particular, we keep the value of $m = -0.49$ fixed throughout the paper (the reason for this is mentioned in section \ref{subsec:non_zero_T}) while we vary the value of $q$ to fit ARPES data at different doping levels. The values for the above quantities as a function of $q$ are shown in Fig.\ \ref{fig:fit_green_parameters}. 

\section{\label{sec:IR_geometry} Near-horizon geometry and IR emergent semi-local quantum liquid}
Here we show that the near-horizon geometry implies a low-energy scaling of the zero-temperature Green's function of the form $\omega(-\omega^2)^{\nu_k - 1/2}$, where the leading momentum dependence only enters through the $k-$dependent exponent. In particular, we show that in the solvable EMD background solution proposed by Gubser and Rocha \cite{Gubser_2010} the exponent depends only on the rescaled momentum $q k/\mu$. 

\subsection{Near-horizon Dirac equation}

The holographic duality has found application in condensed-matter physics as it allows for a qualitative description of the emergent infrared (IR) physics in strongly interacting systems governed by a universal quantum critical phase. The extra ``radial" dimension in the dual gravitational theory geometrizes the energy scale of the field theory, and the near-horizon geometry controls all the low-energy dissipative processes. As such, in application to condensed-matter problems we are interested in this inner region, and holography can be thought of as an effective field theory that describes the low-energy physics up to ultraviolet (UV) coefficients that depend on the particular UV completion, i.e., on the asymptotic (large-$r$ in our conventions) form of the dual spacetime. That is to say, in the regime where $T, \omega \ll \mu$, we have $\text{Im}[G_H(\omega,k;T)] \propto \text{Im}[\mathcal{G}_k(\omega;T)]$, with $\mathcal{G}_k$ the IR Green's function that can be obtained by 
solving the Dirac equation in the geometry of the deep interior of the spacetime only. This is exactly the focus of this section. In the case at hand, we show that in the emergent geometry---that is conformal to $AdS_2 \times \mathbb{R}^2$ at $T = 0$, with the $AdS_2$ replaced by a Schwarzschild black-hole geometry at non-zero $T$---the Dirac equation can be solved analytically \cite{Gubser2012}. This gives rise to a self-energy scaling $\Sigma(\omega, k) = C_k \omega(- \omega^2)^{\nu_k -1/2}$, with $C_k \in \mathbb{R}$ and the momentum-dependent scaling exponent $\nu_k \propto q \abs{k}/\mu$. In a narrow momentum range near $k_F$ the scaling exponent can then be considered constant $\nu_k \simeq \nu_{k_F}$ and we immediately see that we recover the physics of the power-law-liquid self-energy proposed to describe the results of ARPES measurements on cuprates \cite{Reber2019}. In later sections, we however argue that away from the Fermi surface, the experimentally observed peaks span a momentum range large enough to be able to observe an effect of the momentum dependence in the scaling exponent. In particular, we see that the momentum dependence implies that the peaks in the MDCs, described by a symmetric Lorentzian in the power-law liquid, become asymmetric as more spectral weight is shifted towards $\abs{k} < \abs{k_F}$, as also noticeable by looking closely at the tails of the distribution in Fig.\ \ref{fig:fermi_surface}. Given the relevant role of the geometry of the deep interior of the spacetime, below we explicitly compute its form. 

In the following we are going to focus on the $\xi_{-}$ component of the Green's function, hence the relevant equations from Eq.\ \eqref{eq:dirac_equation} are 
\begin{align}\label{eq:dirac_xi_m}
    \partial_r\left(\begin{array}{c}
         y_{+}  \\
         z_{-}
    \end{array}\right) =
    \left(\begin{array}{cc}
         m \sqrt{g_{rr}} & \sqrt{\frac{g_{rr}}{g_{xx}}} k - \sqrt{\frac{g_{rr}}{-g_{tt}}} (\omega + q A_t)  \\
        \sqrt{\frac{g_{rr}}{g_{xx}}} k + \sqrt{\frac{g_{rr}}{-g_{tt}}} (\omega + q A_t)  & -m \sqrt{g_{rr}}
    \end{array}\right) \left(\begin{array}{c}
         y_{+}  \\
         z_{-}
    \end{array}\right) \text{ .}
\end{align}
Expanding the EMD metric from Eq.\ \eqref{eq:background_eom_v2_tilde}, in the low-energy, low-temperature limit, to leading order in $r/\mu$ we find
\begin{align}\label{eq:ir_background}
  \begin{split}
    \sqrt{g_{rr}} &\simeq \frac{1}{3^{3/8} \frac{\mu}{  q^{1/4}} \left(\frac{r}{\mu}\right)^{3/4} \sqrt{1 - q\frac{(4\pi T/\mu)^2}{3 \sqrt{3}}\frac{\mu}{r}}} \text{ ,}\\
    \sqrt{\frac{g_{rr}}{g_{xx}}} &\simeq \frac{1}{\frac{\mu^2}{q} \frac{r}{\mu} \sqrt{1 - q\frac{(4\pi T/\mu)^2}{3 \sqrt{3}}\frac{\mu}{r}}} \text{ ,}\\
    \sqrt{\frac{g_{rr}}{-g_{tt}}} &\simeq \frac{1}{
    3^{3/4} \frac{\mu^2}{\sqrt{q}}\left(\frac{r}{\mu}\right)^{3/2} (1 - q\frac{(4\pi T/\mu)^2}{3 \sqrt{3}}\frac{\mu}{r})} \text{ ,}\\
     A_0 &\simeq \sqrt{3} \mu \frac{r}{\mu} \text{ .}
  \end{split}
\end{align}
In order to compute the IR Green's function $\mathcal{G}_k(\omega,k;T)$ we now need to solve Eq.\ \eqref{eq:dirac_xi_m} in such a geometry.
The computation for fermions in a 3D boundary theory, in the $(4 + 1)$-dimensional generalization of the background used here was first done by Gubser and Ren \cite{Gubser2012}. Here, however, we are interested in describing a single-layer cuprate strange metal and hence we consider a boundary theory in two spatial dimensions dual to a $(3 + 1)$-dimensional bulk spacetime. In the following, we go through the details of the computation for fermions in two dimensions to show that it can be recast in the same form as the one for the three-dimensional theory in Ref.\ \cite{Gubser2012}, to obtain the same temperature and frequency dependence, with the only difference being in the prefactor. 
The Dirac equation in the background of Eqs.\ \eqref{eq:ir_background} becomes 
\begin{align}
 \begin{split}
  &r \sqrt{1 - q\frac{(4\pi T/\mu)^2}{3 \sqrt{3}} \frac{\mu}{r}} ~\partial_r\left(\begin{array}{c}
         y_{+}  \\
         z_{-}
    \end{array}\right) =\\
    &\left(\begin{array}{cc}
         \frac{m}{3^{3/8}} q^{1/4} \left(\frac{r}{\mu}\right)^{1/4} &  q\frac{k}{\mu} -  \sqrt{q} \frac{\omega/\mu}{3^{3/4} \sqrt{r/\mu}} \frac{1}{\sqrt{1 - q\frac{(4\pi T/\mu)^2}{3 \sqrt{3}}\frac{\mu}{r}}}  + \frac{q^{3/2}}{ 3^{1/4}} \sqrt{\frac{r}{\mu}}  \\
       q\frac{k}{\mu} +  \sqrt{q} \frac{\omega/\mu}{3^{3/4}  \sqrt{r/\mu}} \frac{1}{\sqrt{1 - q\frac{(4\pi T/\mu)^2}{3 \sqrt{3}}\frac{\mu}{r}}}  + \frac{q^{3/2}}{3^{1/4}} \sqrt{\frac{r}{\mu}}  & -\frac{m}{3^{3/8}} q^{1/4} \left(\frac{r}{\mu}\right)^{1/4}
    \end{array}\right)\\ &\left(\begin{array}{c}
         y_{+}  \\
         z_{-}
    \end{array}\right) \text{ .}
    \end{split} 
\end{align}

We can immediately see an important distinction compared to the Reissner-Nordstr\"om background (see for example Ref.\ \cite{Iqbal2011}). In the Reissner-Nordstr\"om near-horizon regime, the momentum term scales with the same power of $r/\mu$ as the terms involving the mass and charge of the fermion. On the contrary, here the mass and gauge terms are subleading in the small $r/\mu$ expansion and the IR solution then does not depend explicitly on these two quantities. 
In this limit, we are then left to solve the differential equation 
\begin{align}
   r \partial_r \left(\begin{array}{c}
         y_{+}  \\
         z_{-}
    \end{array}\right) =
    \left(\begin{array}{cc}
         0 & \mathcal{F}(-\omega, k, r) \\
       \mathcal{F}(\omega, k, r) &  0
    \end{array}\right) \left(\begin{array}{c}
         y_{+}  \\
         z_{-}
    \end{array}\right) \text{ ,}
\end{align}
where we defined 
\begin{align}
\mathcal{F}_{2D}(\omega, k, r) \equiv  q\frac{k}{\mu} +  \sqrt{q} \frac{\omega}{3^{3/4} \sqrt{r \mu}} \frac{1}{\sqrt{1 - \frac{q}{3 \sqrt{3}} \frac{(4\pi T)^2}{\mu r}}} .
\end{align}
Similarly, for a three-dimensional boundary theory we have \cite{Gubser2012}
\begin{align}
  \mathcal{F}_{3D}(\omega, k, r) \equiv q \frac{k}{\mu} +  \frac{\omega}{2 r} \frac{1}{\sqrt{1 - \frac{(\pi T)^2}{r^2}}} \text{ .}
\end{align}
By the variable redefinitions shown in table \ref{tab:2d_vs_3d}, the Dirac equation takes the same form in both boundary dimensions, namely
\begin{table}
    \centering
    \begin{tabular}{c | c}
        2D boundary theory & 3D boundary theory \\
        \hline
        \\
        $\zeta = 2 \sqrt{q} \frac{\omega}{3^{3/4} \sqrt{r \mu}}$ & $\zeta = \frac{\omega}{2 r}$\\
        $\nu_k = 2 q \frac{k}{\mu}$  & $\nu_k =  q \frac{k}{\mu}$\\
         $\delta_0 = \frac{2\pi T}{\omega}$ & $\delta_0 = \frac{2\pi T}{\omega}$
    \end{tabular}
    \caption{Change of variables for the low-energy solution in a 2D and 3D boundary theory.}
    \label{tab:2d_vs_3d}
\end{table}
\begin{align}\label{eq:dirac_ir_nonzero_T}
  -\zeta \sqrt{1 - \zeta^2 \delta_0^2} ~\partial_\zeta\left(\begin{array}{c}
         y_{+}  \\
         z_{-}
    \end{array}\right) =
    \left(\begin{array}{cc}
         0 & -\zeta/\sqrt{1 - \zeta^2 \delta_0^2} + \nu_k \\
       \zeta \sqrt{1 - \zeta^2 \delta_0^2} + \nu_k &  0
    \end{array}\right) \left(\begin{array}{c}
         y_{+}  \\
         z_{-}
    \end{array}\right) \text{ .}
\end{align}

We start by carrying out the computation at zero temperature, i.e., we first set $\delta_0 = 0$. 
The boundary of the conformal-to-$\ads_2$ spacetime is at $\zeta \rightarrow 0$, and we can already see that the asymptotic behavior of the solution takes the form of $C_1 \zeta^{-|\nu_k|} + C_2 \zeta^{|\nu_k|}$.  
In order to put the equation in a more familiar form, it is convenient to perform a change of variables \cite{Gubser2012} as
\begin{align}
    \left(\begin{array}{c}
         u_{+}  \\
         u_{-}
    \end{array}\right) = 
    \frac{1}{\sqrt{2}}  \left(\begin{array}{cc}
         1 & i \\
        1 &  -i
    \end{array}\right) \left(\begin{array}{c}
         y_{+}  \\
         z_{-}
    \end{array}\right) 
\end{align} 
so that we get 
\begin{align}\label{eq:ir_diff_equation}
    & \partial_\zeta^2 u_{+} + \frac{\partial_\zeta u_{+}}{\zeta} + u_{+} \left(\frac{i}{\zeta} - \frac{\nu_k^2}{\zeta^2} + 1\right) = 0 \text{ ,}\\
    & u_{-} = -\frac{\zeta}{i \nu_k} (\partial_\zeta u_{+} + i u_{+}) \text{ ,}
\end{align}
where we recognize in Eq.\ \eqref{eq:ir_diff_equation} the Whittaker equation with two possible solutions of the form
\begin{align}
  u_{+}(\zeta) =& C \frac{i \nu_k}{\sqrt{\zeta}} W_{\pm1/2, \nu_k}(\pm 2 i \zeta) \text{ .}
\end{align}
Near the horizon $\zeta \to \infty$, the solution behaves as $u_{+} \propto e^{\mp i \zeta}$, and by imposing the infalling-wave condition at the horizon (corresponding to $e^{i \zeta}$) the solution becomes  
\begin{align}
    u_{+}(\zeta) =& C \frac{i \nu_k}{\sqrt{\zeta}} W_{-1/2, \nu_k}(- 2 i \zeta) \text{ ,}\\
    u_{-}(\zeta) =& C \frac{1}{\sqrt{\zeta}} W_{1/2, \nu_k}(-2 i \zeta)\text{ ,}
\end{align}
that in terms of the original variables is 
\begin{align}
\begin{split}\label{eq:ir_original_var}
  y_{+}(\zeta) =& \frac{C}{2\sqrt{\zeta}} \left(i \nu_k W_{-1/2, \nu_k}(-2 i \zeta) + W_{1/2, \nu_k}(-2 i \zeta)\right)\text{ ,}\\
  z_{-}(\zeta) =& \frac{C}{2\sqrt{\zeta}} \left(\nu_k W_{-1/2, \nu_k}(-2 i \zeta) + i W_{1/2, \nu_k}(-2 i \zeta)\right)\text{ .}
  \end{split}
\end{align}
The IR Green's function can then be extracted by expanding this exact IR solution for $\zeta \to 0$
\begin{align}\label{eq:ir_near_boundary_expansion}
 \begin{split}
    \left(\begin{array}{c}
         y_{+}  \\
         z_{-}
    \end{array}\right)  = \frac{C}{2} \bigg[&\left(\begin{array}{c}
         -1  \\
          1
    \end{array}\right) \left(\frac{(-2 i)^{1 + \nu_k} \Gamma[-2\nu_k]}{-\Gamma[-\nu_k]} + \mathcal{O}(\zeta)\right) \zeta^{\nu_k} + \\
    &\left(\begin{array}{c}
          1  \\
          1
    \end{array}\right) \left(\frac{(2 i)^{\nu_k} \Gamma[1/2 + \nu_k]}{\sqrt{\pi}} - \frac{(2 i)^{\nu_k} \Gamma[-1/2 + \nu_k]}{2 \sqrt{\pi}} \zeta\right) \zeta^{-\nu_k} \bigg] \text{ ,}
 \end{split}
\end{align} 
where we used the notation $\Gamma[x]$ for the gamma function, and we get
\begin{align}
    \mathcal{G}_k = i \frac{(-i)^{2 \nu_k} 4^{-\nu_k} \Gamma[1/2 - \nu_k]}{\Gamma[1/2 + \nu_k]} \zeta^{2\nu_k} r^{\abs{2 q k/\mu}} \text{ .}
\end{align}
Explicitly we thus obtain 
\begin{align}\label{eq:ir_green_2d}
 \begin{split}
    q^{2\abs{q k}/\mu} \mathcal{G}_k^{2D}/\mu^{2\abs{q k}/\mu} =& i (-i)^{4\abs{qk}/\mu} \frac{ \Gamma[1/2 - 2 \abs{q k}/\mu]}{\Gamma[1/2 + 2 \abs{q k}/\mu]} \left(\frac{q\omega}{3^{3/4} \mu}\right)^{4 \frac{\abs{q k}}{\mu}}\\ 
    =& \omega (-\omega^2)^{2 \frac{\abs{q k}}{\mu} - 1/2}\frac{ \Gamma[1/2 - 2 \abs{q k}/\mu]}{\Gamma[1/2 + 2 \abs{q k}/\mu]} \left(\frac{q^2}{3^{3/2} \mu^2}\right)^{2 \frac{\abs{q k}}{\mu}} \text{ ,}
    \end{split}\\
    \label{eq:ir_green_3d}
    \begin{split}
     q^{2\abs{q k}/\mu}\mathcal{G}_k^{3D}/\mu^{2\abs{q k}/\mu} =& i (-i)^{2 \abs{q k}/\mu} \frac{ \Gamma[1/2 - \abs{q k}/\mu]}{\Gamma[1/2 + \abs{q k}/\mu]} \left(\frac{q \omega}{4 \mu}\right)^{2\frac{\abs{q k}}{\mu}}\\
     =& \omega(-\omega^2)^{\frac{2 \abs{q k}}{\mu} - 1/2}\frac{ \Gamma[1/2 - \abs{q k}/\mu]}{\Gamma[1/2 + \abs{q k}/\mu]} \left(\frac{q^2 }{16 \mu^2}\right)^{\frac{\abs{q k}}{\mu}}
    \text{ ,}
    \end{split}
\end{align}
where in the second line of Eqs. \eqref{eq:ir_green_2d} and \eqref{eq:ir_green_3d}, $\omega$ has to be thought as usual as the limit $\omega + i0$.
At non-zero temperature, but with $T/\mu \ll 1$, 
generalizing the result obtained by Gubser \cite{Gubser2012} according to the definitions in table \ref{tab:2d_vs_3d}, we find ultimately 
\begin{align}\label{eq:ir_green_2d_nonzero_T}
     q^{\nu_k}\mathcal{G}_k^{2D}/\mu^{\nu_k} = i \left(q\frac{2 \pi T}{3^{3/4} \mu}\right)^{2 \nu_k} \frac{\Gamma[1/2 - \nu_k]}{\Gamma[1/2 + \nu_k]}\frac{\Gamma[1/2 + \nu_k - i \frac{\omega}{2\pi T}]}{\Gamma[1/2 - \nu_k - i \frac{\omega}{2\pi T}]}  \text{ .}
\end{align}
Upon expanding the gamma functions for $\omega/2\pi T \to \infty$ we recover the zero-temperature solution in Eq.\ \eqref{eq:ir_green_2d}.

First, a few observations about the result just obtained. The first peculiarity of this background is that $\nu_k \propto \abs{q k/\mu}$ with no other terms coming from the mass and charge of the fermion. In comparison, in the Reissner-Nordstr\"om (see e.g.\ Ch.\ 9 of Ref.\ \cite{Zaanen2015}) the scaling exponent takes the form $\nu_k = \sqrt{2 q^2/\mu^2 k^2 + m^2/6 - q^2/3}$. In such models, there is an instability at small momenta and large values of the ratio $q^2/m^2$, where we find a pathological log-oscillatory behavior (as the exponent becomes imaginary) \cite{Faulkner2011,Liu2011}. In the opposite limit of small $q^2/m^2$ the scaling exponent assumes the general form $\nu_k \propto \sqrt{(k/\mu)^2 + 1/\xi^2}$, with $\xi = \xi(m^2, q^2)$ setting a correlation length with the Green's function decaying (at least) exponentially as $G \propto e^{-x/\xi}$ at large $x \gg \xi$ \cite{Iqbal2011}. These features are both absent in the Gubser-Rocha model, as shown in Eq.\ \eqref{eq:ir_green_2d}, allowing for more freedom in setting the values of $m$ and $q$.

The imaginary part of the analytical result in Eqs. \eqref{eq:ir_green_2d} and \eqref{eq:ir_green_2d_nonzero_T} is finite for every value of $\nu_k$. However, this is not the case for the real part that contains a pole for every value of momenta such that $\nu_k = 1/2 + n$, $n \in \mathbb{N}$. These divergences, though, are not present in the full Green's function, as it can be seen in Fig.\ \ref{fig:holographic_spec}, where, in the full solution obtained numerically, we observe a linear dispersion for every value of $k/\mu$ in the range of interest, according to Eq.\ \eqref{eq:holographic_green_fermi}. This is because, when the IR solution is matched with the outer solution to compute the full Green's function, as explained below, these simple poles are in fact canceled by divergences in the matching coefficients \cite{Faulkner2011}. 

\subsection{Matching procedure and comparison with the ansatz}\label{subsec:matching}
\begin{figure}[h]\centering
  \begin{subfigure}{.5\textwidth}
    \centering
    \includegraphics[width=\linewidth]{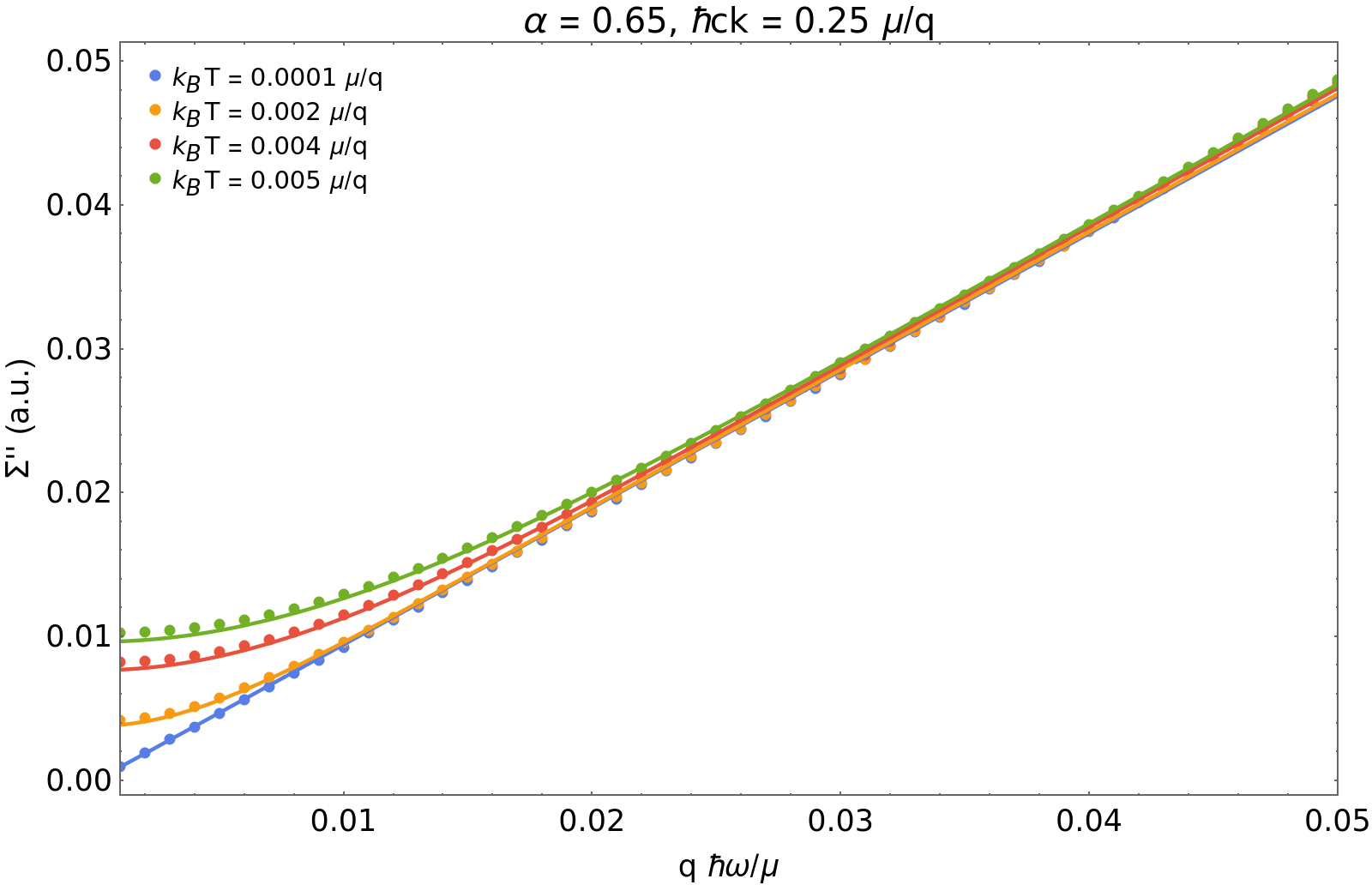}
  \end{subfigure}%
\begin{subfigure}{.5\textwidth}
  \centering
  \includegraphics[width=\linewidth]{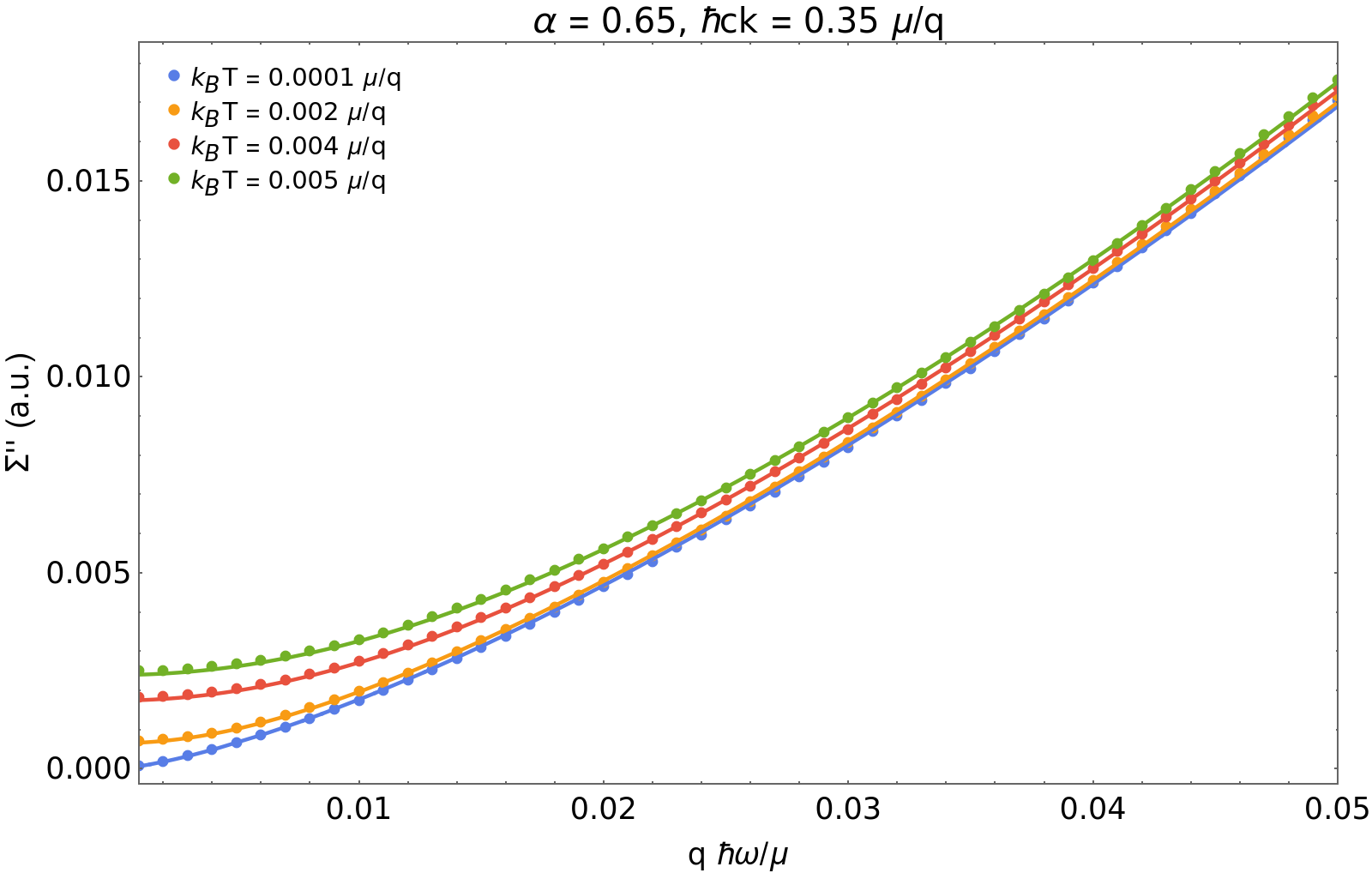}
\end{subfigure}
\vskip\baselineskip
 \begin{subfigure}{.5\textwidth}
    \centering
    \includegraphics[width=\linewidth]{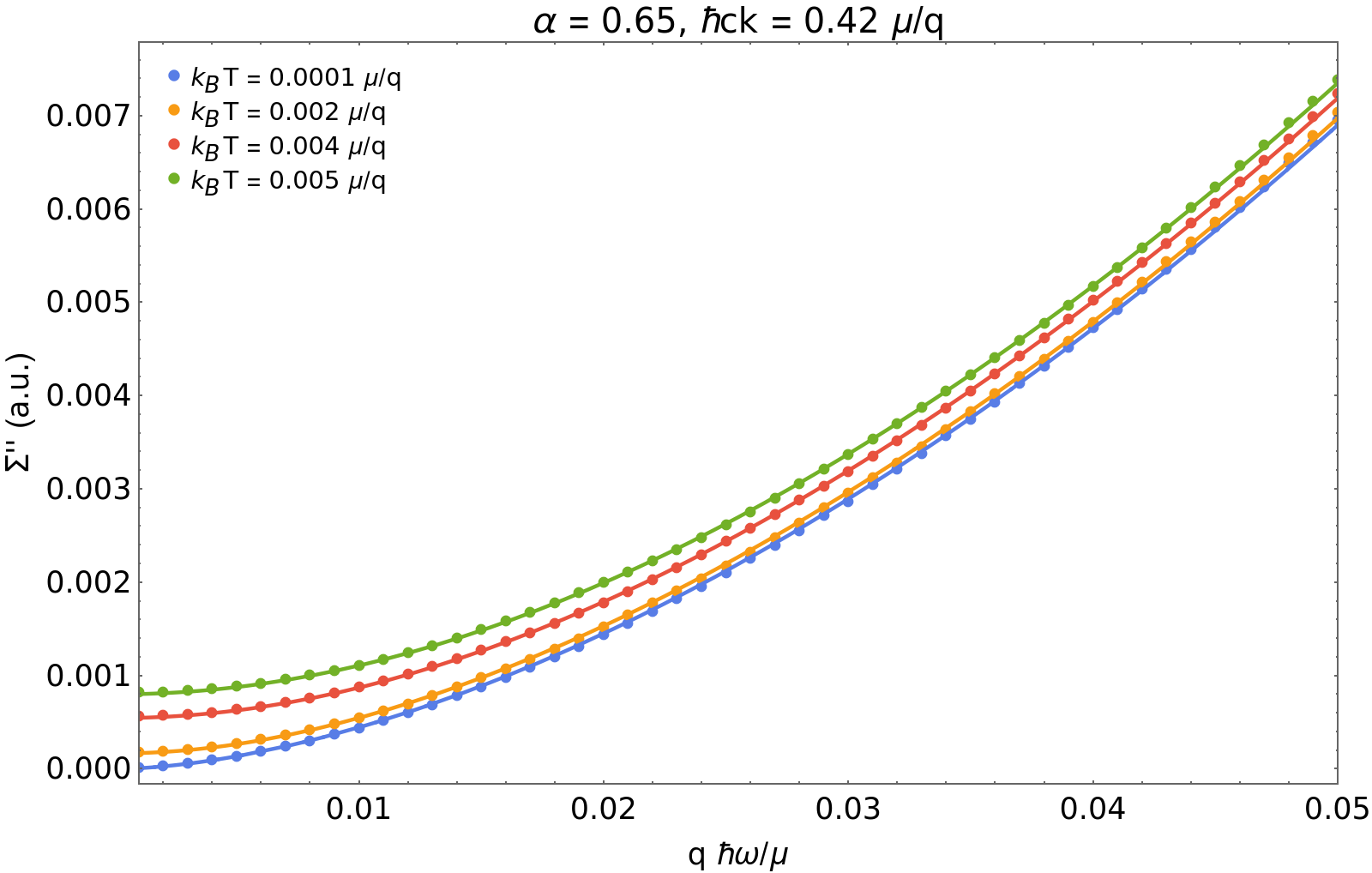}
  \end{subfigure}%
\begin{subfigure}{.5\textwidth}
  \centering
  \includegraphics[width=\linewidth]{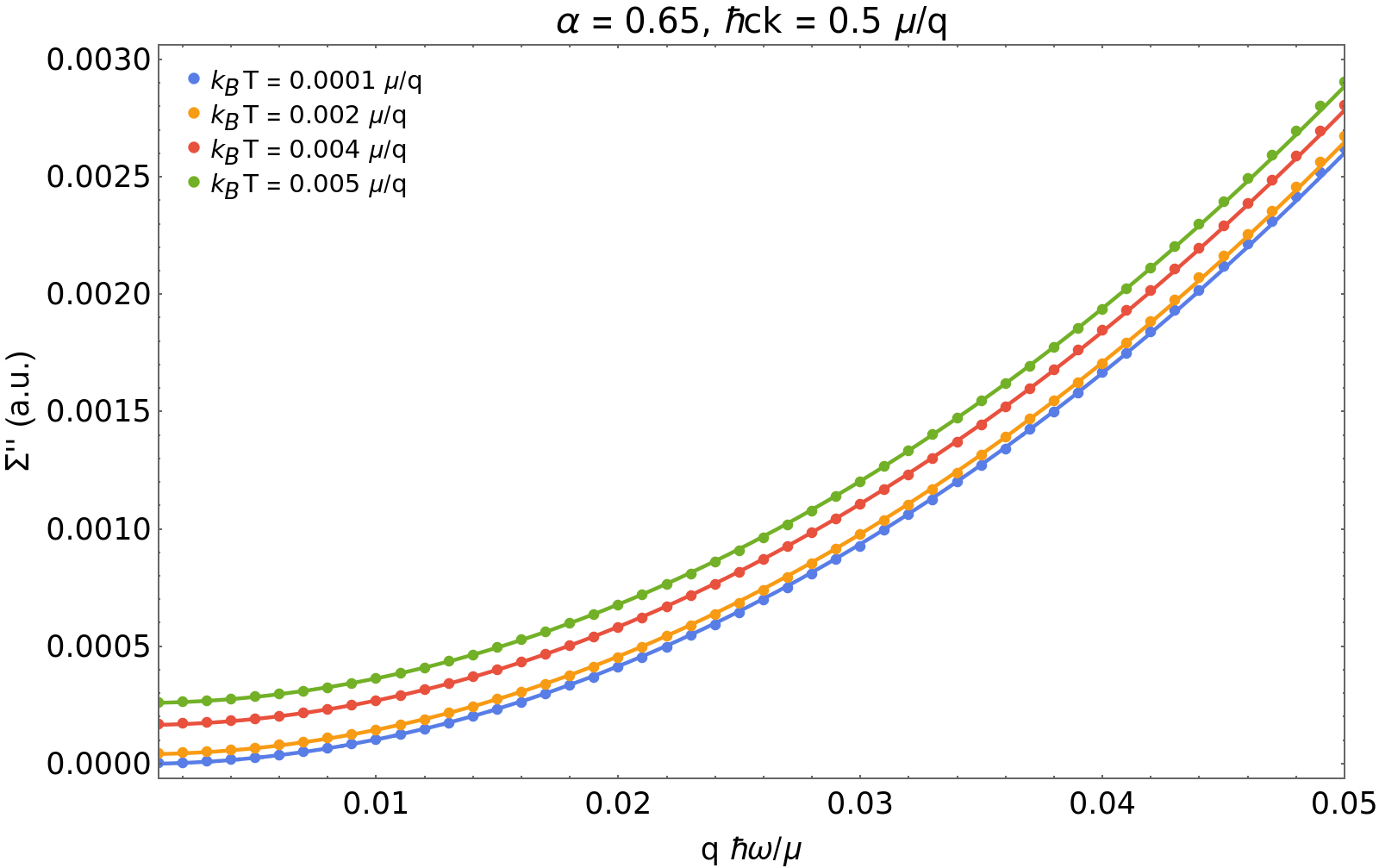}
\end{subfigure}
\caption{\label{fig:fit_exact} Imaginary part of the numerical self-energy (circles) for different values of fixed momentum and temperatures in the range of interest, together with a fit to the analytical low-energy result (lines). We can clearly see that the low-energy behavior of the imaginary part of the self-energy is indeed well described by the $\ads_2 \times \mathbb{R}^2$ solution.}
\end{figure}

\begin{figure}
\centering
\begin{subfigure}{.5\textwidth}
  \centering
  \includegraphics[width=.95\linewidth]{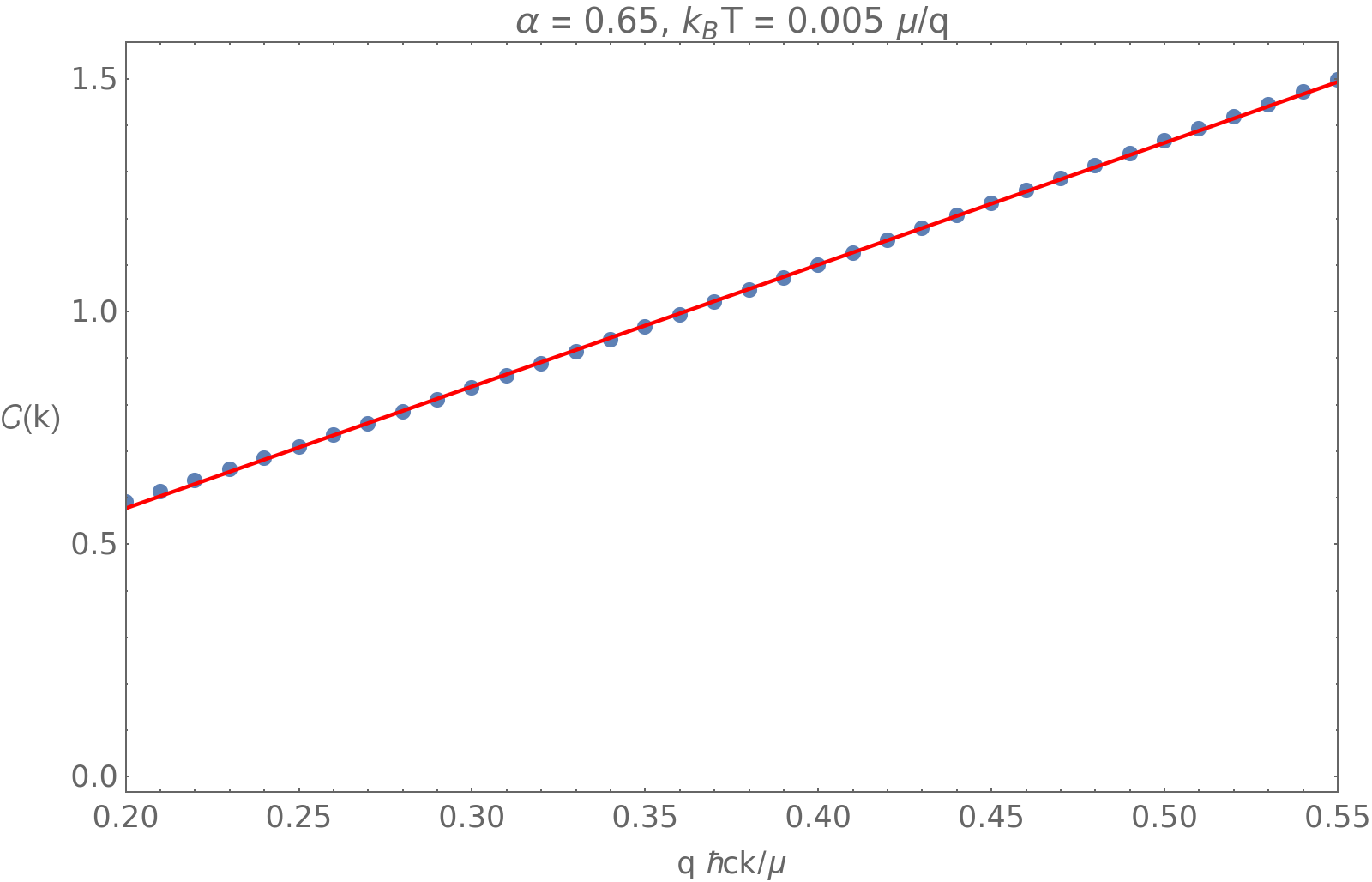}
  \caption{Momentum dependence of the coefficient $\mathcal{C}$ with a linear fit (red line)}
  \label{fig:c_k}
\end{subfigure}%
\begin{subfigure}{.5\textwidth}
  \centering
  \includegraphics[width=.95\linewidth]{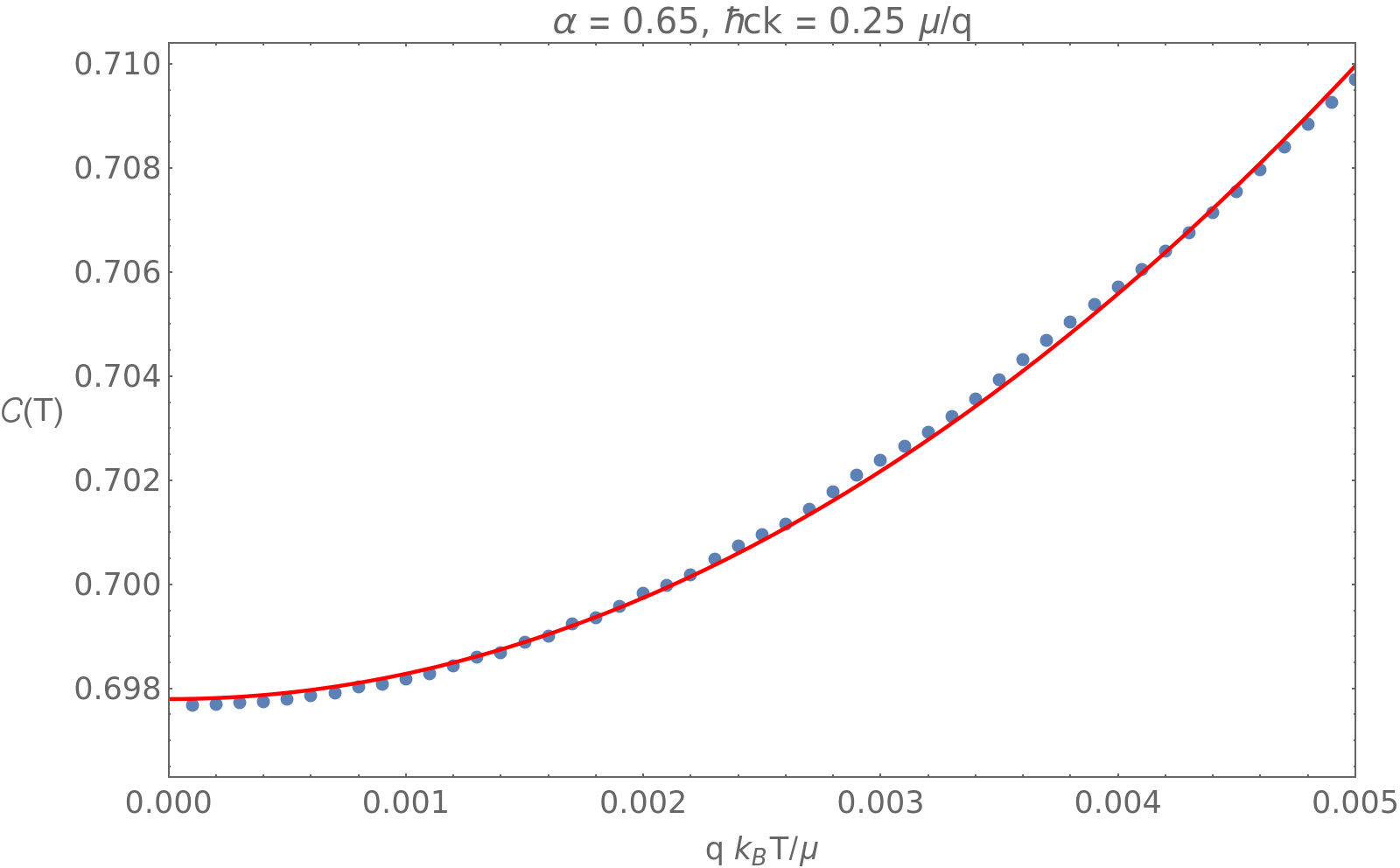}
  \caption{Temperature dependence of the coefficient $\mathcal{C}$ with a quadratic fit (red line)}
  \label{fig:c_t}
\end{subfigure}
\caption{Example of the momentum dependence of the coefficient $\mathcal{C}$ resulting from a fit of the numerical solution at fixed temperature $T/\mu = 0.005$, with a function of the form of Eq.\ \eqref{eq:near_behavior} with an energy-independent coefficient $\mathcal{C}(\omega, k, T) \simeq \mathcal{C}(0, k, T)$. It shows that the latter coefficient is linear in $k/\mu$ as expected from the argument in the main text, while temperature corrections are quadratic and thus negligible at sufficiently low temperatures where $\mathcal{C}(\omega, k, T) \simeq \mathcal{C}(0, k, 0)$.}
\label{fig:c_dependence}
\end{figure}

In this section, we briefly remind the reader of the relationship between the infrared Green's function and the full one for the boundary fermionic operator $\mathcal{O}$.
The basic idea is that we divide the spacetime into two regions, the inner region for small $r/\mu$ where we solved the Dirac equation as explained in the previous section, and an outer region where $r \gg \omega, T$ and we can solve the equations in a series expansion in these two quantities. For $\omega, T \ll \mu$, this two regions overlap when $\omega, T \ll r$ and $r \ll \mu$. In this overlap region, we can then match the two solutions. The details of this matching procedure between the solutions to the Dirac equation in the inner and outer region of the spacetime are nicely outlined in, for example, Refs. \cite{Faulkner2011, Iqbal2011}. We ultimately have that the full holographic Green's function can be written as
\begin{align}\label{eq:hol_green_matching}
    G_{H}(\omega, k; T)\mu^{-2 m} =  \frac{b_+(\omega, k, T) + b_-(\omega, k, T) \mathcal{G}_k(\omega; T)\mu^{-2\nu_k}}{a_+(\omega, k, T) + a_-(\omega, k, T) \mathcal{G}_k(\omega; T)\mu^{-2\nu_k}} \text{ ,}
\end{align}
where the coefficients $b_{\pm}$, $a_{\pm}$ depend on the full spacetime and, in most cases, as it is in this paper, they can only be computed numerically. They are all real and have an analytic expansion in terms of $\omega$ and $T$ 
\begin{align}
    a_{\pm}(\omega, k, T) = a^{(0,0)}_{\pm}(k) + \omega a^{(1,0)}_{\pm}(k) + T a^{(0,1)}_{\pm}(k)  + \cdots \text{ ,}
\end{align}
for $\omega$, $T \ll \mu$ and similarly for $b_\pm$. The expansion coefficients depend on momentum and can be Taylor expanded as well. Notice that, from the structure of Eq.\ \eqref{eq:hol_green_matching}, an analogous relation holds for the Green's function in alternative quantization. In particular, we therefore expect that for small energies and temperatures, we can accurately describe the dissipative physics, encoded in the imaginary part of the inverse Green's function with a form 
\begin{align}\label{eq:near_behavior}
    {\text{Im}[G_H(\omega, k; T)^{-1}\mu^{2 m}]} \simeq  - \mathcal{C}(\omega, k, T)\text{Im}\left[\mathcal{G}_{k}(\omega; T)\mu^{-2\nu_k}\right] \text{ ,}
\end{align}
with $\mathcal{G}_{k}(\omega; T)$ given in Eq.\ \eqref{eq:ir_green_2d_nonzero_T} and $\mathcal{C}$ admitting an analytical expansion in its variables. For $q k/\mu$ in the range of interest to us, we expect it to be well described by $\mathcal{C} = \mathcal{C}^{(0)} + k \mathcal{C}^{(k)} + T \mathcal{C}^{(T)} + \omega \mathcal{C}^{(\omega)} + \cdots$, with the dominant correction coming from the $k$ dependence and only small energy and temperature-dependent corrections, as we are interested in describing low-energy excitations near the Fermi surface where $k/\mu \gg \omega/\mu, T/\mu$.
We verified this numerically for different values of the holographic fermion charge $q$ in the range needed to describe the ARPES data and show this in Fig.\ \ref{fig:fit_exact}.
We see that the imaginary part of the full solution is very accurately described by the IR results at all values of temperatures considered, with the coefficient $\mathcal{C}(\omega, k, T)$ linear in $k$ as shown, for example, in Fig.\ \ref{fig:c_dependence} for a value of the exponent $\alpha = 0.65$ (from Eq.\ \eqref{eq:self_energy_hol}) and $q T/\mu = 0.005$. 
The real part of the Green's function, on the contrary, cannot be understood in terms of the emergent geometry only, as it depends on the full solution and it has to be analyzed numerically.

We can further see from Eq.\ \eqref{eq:hol_green_matching} that the Green's function for $\mathcal{O}$ shows a Fermi surface whenever there exists a value of momentum $k \equiv k_F$ such that $a_+^{(0,0)}(k = k_F) = 0$. Expanding the Green's function near this point we see that it can be written as
\begin{align}\label{eq:near_kF_expansion}
    G_H(\omega, k; T)/\mu^{2m} \simeq \frac{b_+^{(0,0)}(k_F)}{\partial_k a_{+}^{(0,0)}(k_F) (k - k_F) + \omega a_{+}^{(1,0)}(k_F) + a_{-}^{(0,0)}(k_F) \mathcal{G}_{k_F}(\omega; T) \mu^{-2\nu_{k_F}}}  \text{ ,}
\end{align}
justifying the form used in Eq.\ \eqref{eq:holographic_green_fermi} in the previous section. We now understand that $k_F/\mu$, as well as $Z$ and $v_H$, require the full solutions and depend therefore on the UV completion of the theory, while $\Sigma''$ is, up to a prefactor, fully determined  by the imaginary part of $\mathcal{G}_{k}$. In particular then, near the Fermi surface, we recover the PLL self-energy $\Sigma'' \propto (\omega^2)^\alpha$, proposed as a model of the cuprates strange metal, by fine-tuning the charge of the bulk fermion such that $\nu_{k_F} = 2 q k_F/\mu \equiv \alpha$,  with $\alpha = 1/2$ at optimal doping and increasing approximately linearly with doping towards the Fermi-liquid value of $1$ \cite{Smit2021}. 

The full solution starts to deviate from the IR analytical behavior for small values of $\hbar c k/\mu$, reintroducing for a moment dimensionful units. However, we are only interested in describing the MDC peaks near the Fermi momentum with $q \hbar c k_F/\mu \equiv \alpha/2 \in [0.25, 0.5)$. As noted in Sec. \ref{sec:fermions}, we consider $c$ as a velocity of the same order of the Fermi velocity. We can verify that this interpretation is consistent with the infrared regime $\hbar\omega, k_B T \ll \mu$ corresponding to an energy and temperature range of experimental interest. Since we have $q \hbar \omega/\mu = \alpha \hbar \omega/2 \hbar c k_F$, choosing for estimation purposes a velocity $\hbar c = 4 \text{eV\AA}$ and $k_F = 0.5 \text{\AA}^{-1}$, we have that, choosing, for example, the largest values shown in Fig.\ \ref{fig:fit_exact}, $q \hbar \omega/\mu = 0.05$ corresponds to an energy $\hbar \omega \in (0.2 \text{eV}, 0.4 \text{eV}]$ depending on the value of $\alpha \in [0.25, 0.5)$. This is also where the range of validity of the linear approximation in the dispersion observed in the ARPES measurements starts to break down, and we hence do not expect the low-energy holographic description to be valid anymore. Accordingly, the highest temperature shown is $q k_B T/\mu = 0.005$ which corresponds to the high temperatures $T \in (232 K, 465 K]$. 

\section{Semi-holography and effective low-energy response}\label{sec:semi-holography}
As we have shown, holography provides us with a way to compute the low-energy response function of strongly interacting systems at non-zero temperature and chemical potential and for any dimensionality, a task not easily achievable with such a generality within other frameworks. However, when comparing with experimental measurements, we have to keep in mind that the holographic model considered does not fully correspond, at all energies, to the theory of an electron system, but it has to be regarded as yielding an effective theory capturing the low-energy behavior of a fermionic massless composite operator, whose UV behavior might be far from what is expected from a metal in the laboratory. This is also clear from the fact that the holographic Green's function presented above does not satisfy the electronic sum rule and obeys
\begin{align}\label{eq:sum_rule}
    \frac{1}{\pi}\int_{-\infty}^{+\infty} \dif \omega\, \text{Im}[G_H(\omega, k)] \ne 1 \text{ ,} \quad \forall k \text{ .}
\end{align}
We want, hence, to write down a low-energy effective action to decouple the dissipative physics related to the interior of the spacetime from the ultraviolet conformal field theory of the asymptotically AdS spacetime. 
This is done in an approach known as \textit{semi-holography} \cite{Faulkner:2010tq, Gursoy2012}, that considers the effective action as arising from a fermionic field $\chi$, living on the boundary of the spacetime whose decay is controlled by the interaction with the strongly coupled sector described by the holographic fermionic operator $\mathcal{O}$, so that the effective action takes the form 
\begin{align}\label{eq:effective_action}
    S^{\text{eff}} = \int \frac{\dif \omega \dif^d k}{(2\pi)^{d + 1}} (\chi^\dagger (-\hbar\omega + \epsilon(k) - \mu) \chi + g_k \chi^\dagger \mathcal{O} + g_k \mathcal{O}^\dagger \chi) + S^\text{strong}(\mathcal{O}) \text{ ,}
\end{align}
with $g_k$ a momentum-dependent (assumed real) coupling constant, and $S^\text{strong}$ the action from the near-horizon holographic sector. The Green's function for the fermion field then becomes 
\begin{align}\label{eq:free_fermion_green}
    G_{\chi\chi}(\omega, k) = \frac{\hbar}{-\hbar\omega + \epsilon(k) - \mu - g_k^2 \mathcal{G}_{k}(\omega)} \equiv \frac{1}{G_0(\omega, k)^{-1} + \Sigma(\omega, k)}  \text{ ,}
\end{align}
with $-g_k^2 \mathcal{G}_{k}(\omega)/\hbar$ assuming the role of the electron self-energy ($\mathcal{G}_{k}(\omega)$ being the IR Green's function presented above).

It is this semi-holographic construction that we considered in what follows, as we are, in the first instance, primarily interested in understanding if there is evidence in the cuprates of the peculiar physics arising from the universal infrared behavior described by the gauge-gravity duality. Only after this question is answered affirmatively can we undertake the more challenging problem of a complete description of the theory. We, therefore, start with the simplest approach of modeling the Green's function near the Fermi surface by 
\begin{align}\label{eq:semi_holo_green}
     G_{\chi\chi}(\omega,k) =  \frac{\hbar}{-\hbar\omega + \hbar v_F (k - k_F) - i g_k^2 \text{Im}[\mathcal{G}_{k}(\omega)]} \text{ ,}
\end{align}
where we now consider the renormalized Fermi velocity $v_F$ to be a $k$-independent adjustable parameter, incorporating the real contribution of the self-energy that renormalizes the bare velocity $v_B$, and the freedom in $g_k^2>0$ is set to match the measured peak width near the Fermi surface. In this way, the holographic input is all in the imaginary part of the electron self-energy $\hbar\Sigma''(\omega, k) \equiv  g_k^2 \text{Im}[\mathcal{G}_{k}(\omega)]$. 

It is interesting to notice that Gursoy et al. in Ref.\ \cite{Gursoy2012} proposed a semi-holography approach where the fermion is coupled to the full holographic theory (and not only to the IR emergent sector), as they showed that such theory satisfies the ARPES sum rule in Eq.\ \eqref{eq:sum_rule}. However, as we show in the appendix, this does not allow for a description of the observed peak width.

\section{Comparing holography to ARPES data}\label{sec:comparison_with_experiment}
The purpose of this paper is to show that the simple semi-holographic model presented above is able to describe the energy and momentum dependence of the cuprate spectral function measured in ARPES experiments, and, in particular, we argue that it is a step forward compared to the momentum-independent self-energy of the power-law liquid. While the latter is accurate near the Fermi surface, the former provides a better comparison with the data over a larger range of energies. We, hence, explain here how to compare the holographic prediction to the experimental data from Ref.\ \cite{Smit2021}. 
We start by looking at the zero-temperature solution and leave the discussion about temperature corrections, only relevant as $k_B T \simeq \omega$, to a later section.

The analysis of the electronic MDCs is based on a fit of the form of Eq.\ \eqref{eq:fit_function_pll}
\begin{align}\label{eq:spec_generic}
    \mathcal{A}(\omega, k) = \frac{W(\omega)}{\pi}\frac{\Gamma(\omega, k)/2}{(k-k_{*}(\omega))^{2}+(\Gamma(\omega, k)/2)^{2}} \text{ ,}
\end{align}
where $\Gamma(\omega, k) = 2 \Sigma''(\omega, k)/v_F + G_0(\omega)/v_F$, with $G_0(\omega)$ a momentum-independent contribution usually attributed to the electron-phonon interaction and disorder. The ARPES measurements on the cuprate strange metal have generally been analyzed in a framework in which the self-energy is dominated by the frequency dependence. It has then been assumed that it could be modeled with a momentum-independent ansatz, the power-law liquid:
\begin{align}
    \hbar\Sigma''_{\text{PLL}} = \hbar\Sigma''_{\text{PLL}}(\omega, T) = \frac{((\hbar\omega)^2 + (\beta k_B T)^2)^\alpha}{(\hbar\omega_N)^{2 \alpha - 1}} \text{ ,} 
\end{align}
with $\hbar \omega_N = 0.5 \text{eV}$ 
a fixed energy scale, $\beta$ a constant with an experimentally determined value around $\pi$ \cite{Reber2019}, and $\alpha$ a doping-dependent exponent with $\alpha = 1/2$ at optimal doping and $1/2 < \alpha < 1$ as we move into the overdoped region of the phase diagram.
As a first-order test of the semi-holographic prediction, according to Eq.\ \eqref{eq:semi_holo_green}, we simply replace the PLL's imaginary part of the self-energy with the semi-holographic result $\hbar\Sigma''(\omega, k) \equiv g_k^2 \text{Im}[\mathcal{G}_{k}(\omega)]$. Notice that, by neglecting the real part of the self-energy, this approach is independent of the UV completion of our holographic theory. One could argue that, given a non-linear momentum dependence, the real part of the self-energy should introduce some non-linearity in the dispersion relation, as well as momentum dependence in $W = W(\omega, k)$ in Eq.\ \eqref{eq:spec_generic}. From the experimental fact that the peak position near the Fermi surface is well described by a linear dispersion with a phonon kink, we consider it a reasonable first step to only assume that the real part of the self-energy simply renormalizes the Fermi velocity. It is nonetheless a very interesting problem to consider the effect of a full holographic self-energy on the dispersion and normalization, but that is left for future work.

\subsection{Particle-hole symmetry}
\begin{figure}
    \centering
    \includegraphics[width=1\linewidth]{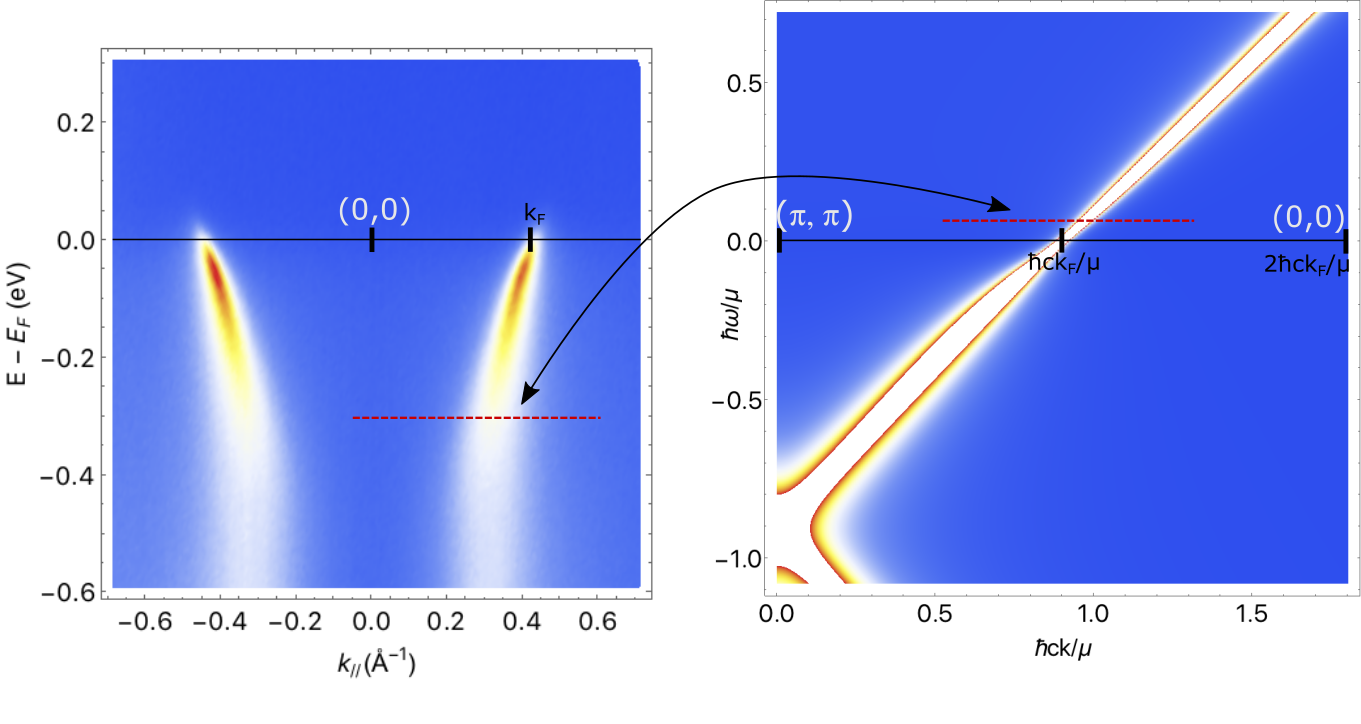}
    \caption{With (semi-)holography we obtain the spectral function for the holes (right panel). In an emergent particle-hole symmetry near the Fermi surface, we relate the ARPES MDC along the nodal line for the electrons in a cuprate at $E < E_F$, with momentum origin at the (0,0) point with the holographic distribution at the corresponding $\hbar\omega/\mu = (E_F - E)/\mu > 0$ (\textcolor{red}{dashed red line}). The origin of the momentum axis is at $2 \hbar c k_F/\mu$}
    \label{fig:particle_hole_spectra}
\end{figure}

\begin{figure}[h!]
    \centering
    \includegraphics[width=0.45\linewidth]{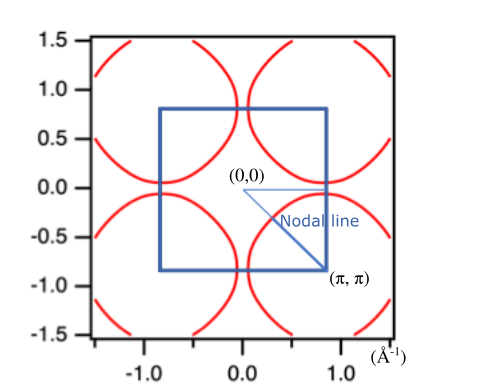}
    \caption{Structure of the Fermi surface. The origin of the momentum axis is set at the $(0,0)$ point, with measurements performed along the nodal line. The holographic model describes the response of the hole system from the $(\pi, \pi)$ point, within an emergent particle-hole symmetry where the $(\pi,\pi)$ point is assumed at $2 k_F$ along the nodal direction.}
    \label{fig:fermi_surface_structure}
\end{figure}

  Near the Fermi energy, the MDC peaks are sharp and the holographic prediction for the imaginary part of the self-energy is well approximated by the momentum-independent form of the PLL, i.e., $\Sigma'' \propto ( \omega^2)^{\nu_{k_F}}$ with the identification $\alpha \equiv \nu_{k_F} = 2 q \abs{k_F}/\mu$. In particular, we see that the value of the charge $q$ in the bulk Dirac equation can be used to describe the doping dependence of the power-law exponent, with $\alpha$ approximately linear in $q$ since the value of $k_F/\mu$ varies very little as a function of $q$ (see Fig.\ \ref{fig:fit_green_parameters}). As we move away from the Fermi level, however, the observed peaks become broad enough (the FWHM is of the same order of magnitude as the Fermi momentum) that we expect to be able to notice deviations from the symmetric Lorentzian shape due to the effect of the $k$-dependent exponent as $\Sigma''(\omega, k, T = 0) \propto (\omega^2)^{\alpha(1 + (k - k_F)/k_F)}$. It is easy to see that this implies more spectral weight in the tail of the peak for $\abs{k} < \abs{k_*}$ as shown in an example in Fig.\ \ref{fig:fermi_surface}. This is in contrast with experimental findings along the nodal line, where the peaks appear as if they are ``tilted'' in the other direction, with more spectral weight for $\abs{k} > \abs{k_*}$. We believe that this is due to the holographic fermion in the bulk being dual to a fermionic operator for the hole, $\mathcal{O}(\omega, k) \equiv \mathcal{O}_h(\omega, k)$, hence the semi-holographic effective theory, Eq.\ \eqref{eq:effective_action}, describes the response of a hole system, $\chi(\omega, k) \equiv \chi_h(\omega, k)$. On the other hand, ARPES measures the response of the electrons in cuprates, associated with $\chi_e(\omega, k)$. In practice, this implies that ARPES measurements below the Fermi surface correspond to positive values of $\hbar \omega$ in the semi-holographic spectral function, as we depict in Fig.\ \ref{fig:particle_hole_spectra}. The semi-holographic spectral function, in the right panel of Fig.\ \ref{fig:particle_hole_spectra} then describes the hole response in the nodal direction. The left-hand panel in Fig.\ \ref{fig:particle_hole_spectra} represents photoemission data and thus necessarily is in an electron picture. Since the semi-holographic spectral function in the right-hand panel is in the hole picture, the red line shows the unoccupied hole states, which corresponds to the red line in the occupied electron states in the left-hand panel. We thus need to identify the zero of the momentum axis with the $(\pi,\pi)$ point, and the rotationally invariant Fermi surface at $\omega = 0$ of the semi-holographic spectral function, then, corresponds to the ``round'' Fermi surface around the $(\pi,\pi)$ point of Fig.\ \ref{fig:fermi_surface_structure}. Assuming an emergent particle-hole symmetry near the Fermi surface, we hence identify the position of the $(0,0)$ point along the nodal direction at $2 k_F$. Performing a particle-hole conjugation on the fermionic composite operator $\mathcal{O}_h(\omega, k) \rightarrow \mathcal{O}_e^\dagger(-\omega, 2k_F - k)$, transforms the self-energy as
  \begin{align} \label{eq:electron_self}
   \begin{split}
      \hbar \Sigma_h(\omega, k) &= - g_k^{2} \mathcal{G}_k(\omega) \rightarrow + g_{2 k_F - k}^2 \mathcal{G}_{2 k_F - k}^*(-\omega) = \hbar \Sigma_e(\omega, k)\\
      &\propto -g_{2 k_F - k}^2 (-\omega) (-\omega^2)^{\alpha(1-(k - k_F)/k_F)-1/2} \text{ ,}
    \end{split}
  \end{align}
  where we used the fact that $g_k^2$ is real and we dropped the absolute value in the exponent as we always consider $0 \le k < 2 k_F$. 
  In summary, the electronic spectral function at $E < E_F$, as measured in ARPES MDCs, is described in our semi-holographic framework by the peaks for positive frequencies, as depicted in Fig.\ \ref{fig:particle_hole_spectra}, with the momentum $k$ measured along the nodal direction from the $(0,0)$ point. From now on, we simply use the notation $\hbar \Sigma(\omega, k) = - g_k^2 \mathcal{G}_k(\omega)$ to refer to the self-energy for the electron as in Eq.\ \eqref{eq:electron_self}, with the scaling exponent $\alpha(k) \equiv \alpha(1 - (k - k_F)/k_F)$.

\subsection{Fit-function and the phonon}
\begin{figure}
    \centering
    \includegraphics[width=0.5\linewidth]{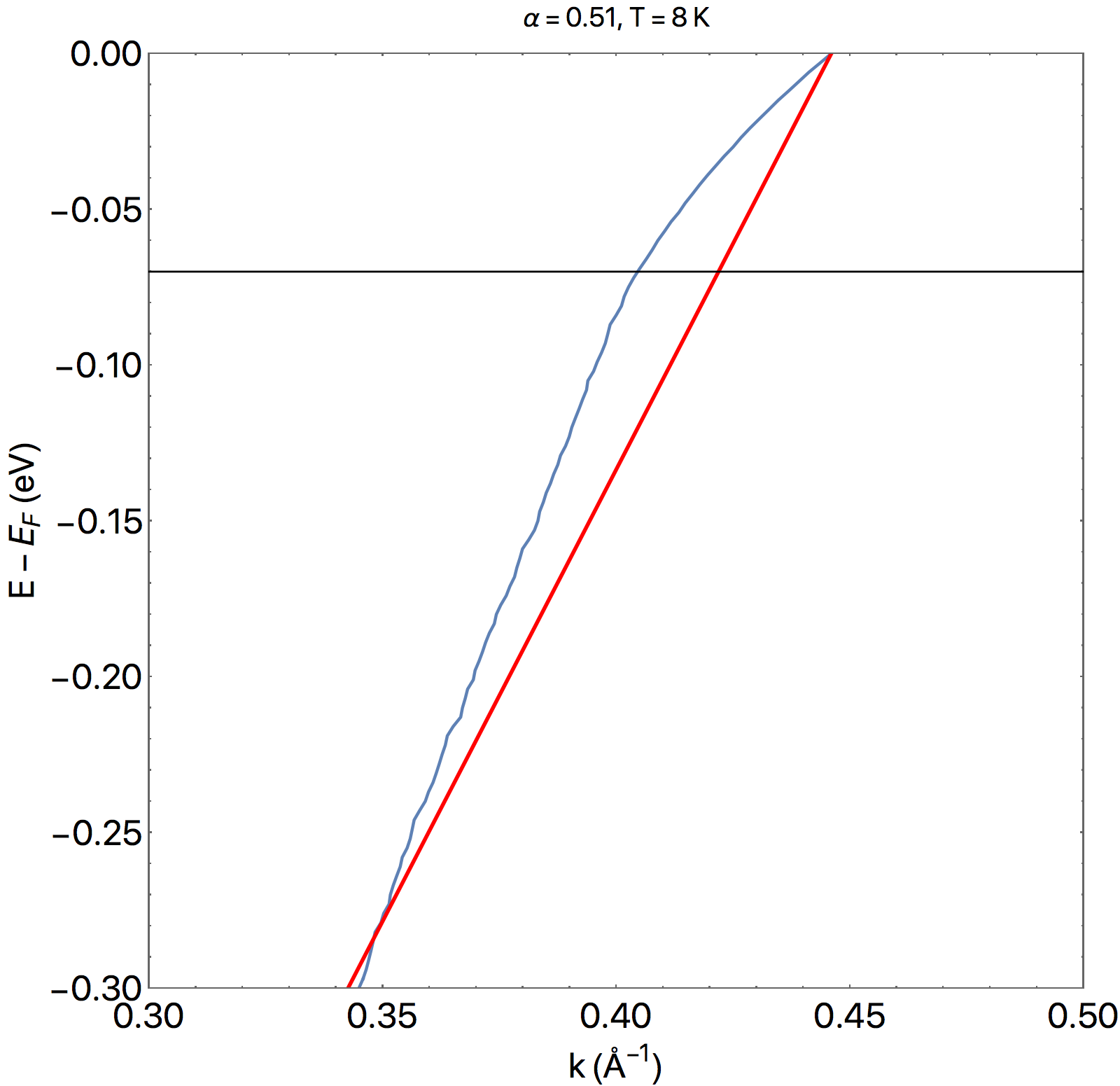}
    \caption{Plot of the dispersion relation from ARPES measurements along the nodal direction at optimal doping and low-temperature (\textcolor{blue}{blue line}), and comparison with a linear dispersion $E - E_F = \hbar v_F (k - k_F)$, with $v_F = 2.9\text{eV\AA}$ (\textcolor{red}{red line}). We can see a kink in the linear dispersion around $E - E_F = -0.07\text{ eV}$.}
    \label{fig:ARPES_dispersion}
\end{figure}

\begin{figure}[h!]\centering
  \begin{subfigure}{.43\textwidth}
    \centering
    \includegraphics[width=\linewidth]{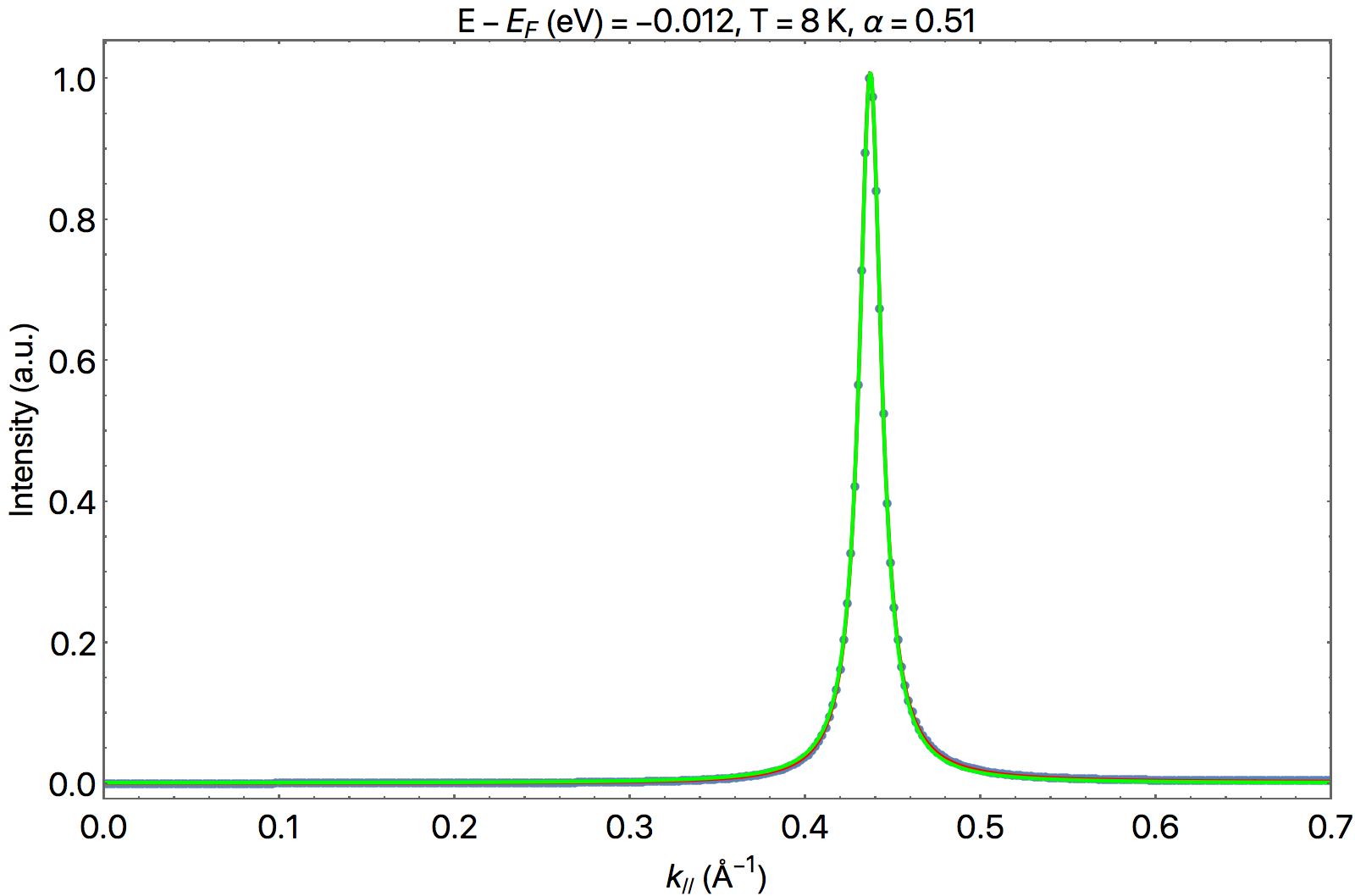}
  \end{subfigure}%
\begin{subfigure}{.43\textwidth}
  \centering
  \includegraphics[width=\linewidth]{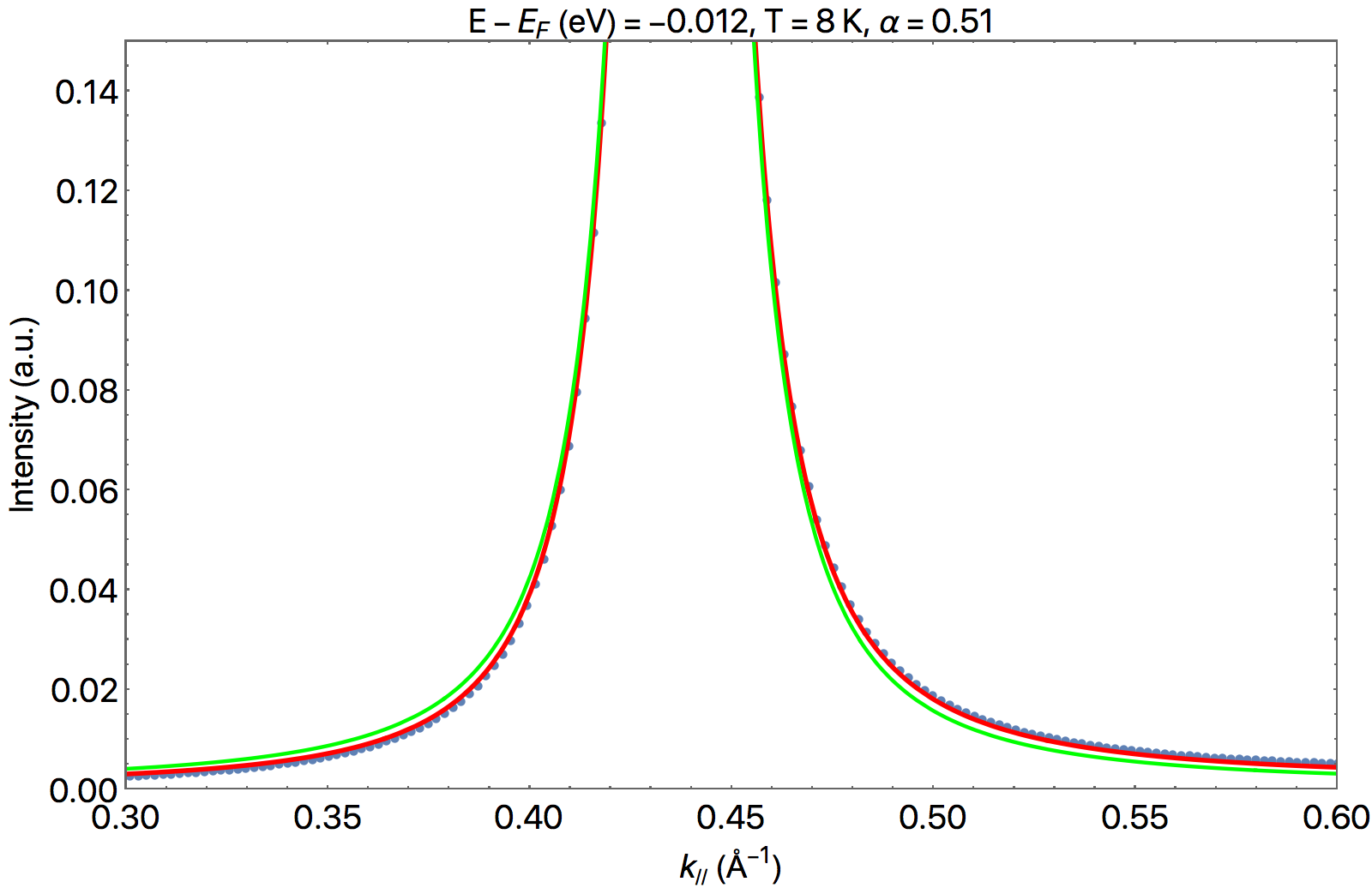}
\end{subfigure}
\vskip\baselineskip
 \begin{subfigure}{.43\textwidth}
    \centering
    \includegraphics[width=\linewidth]{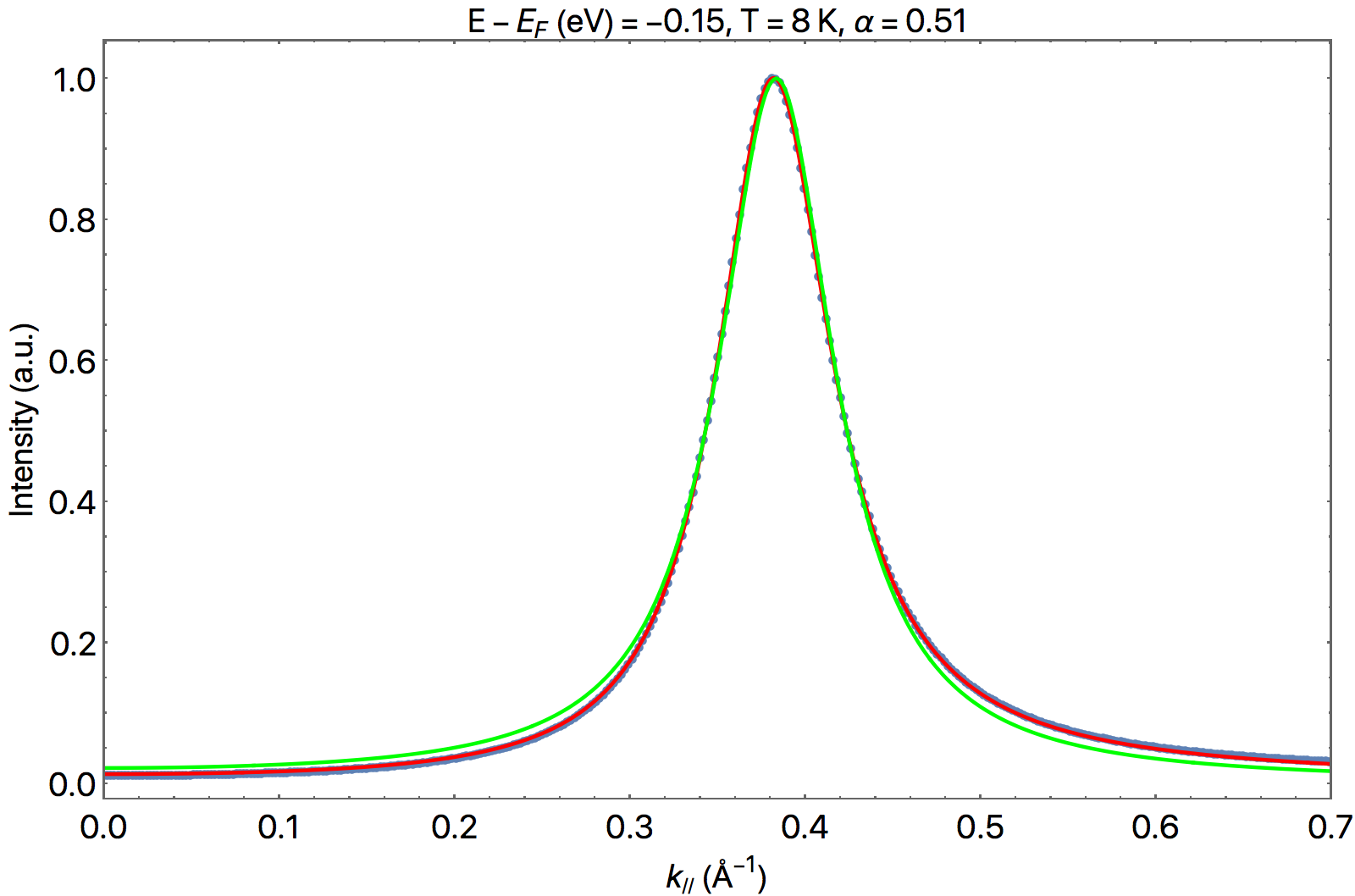}
  \end{subfigure}%
 \begin{subfigure}{.43\textwidth}
    \centering
    \includegraphics[width=\linewidth]{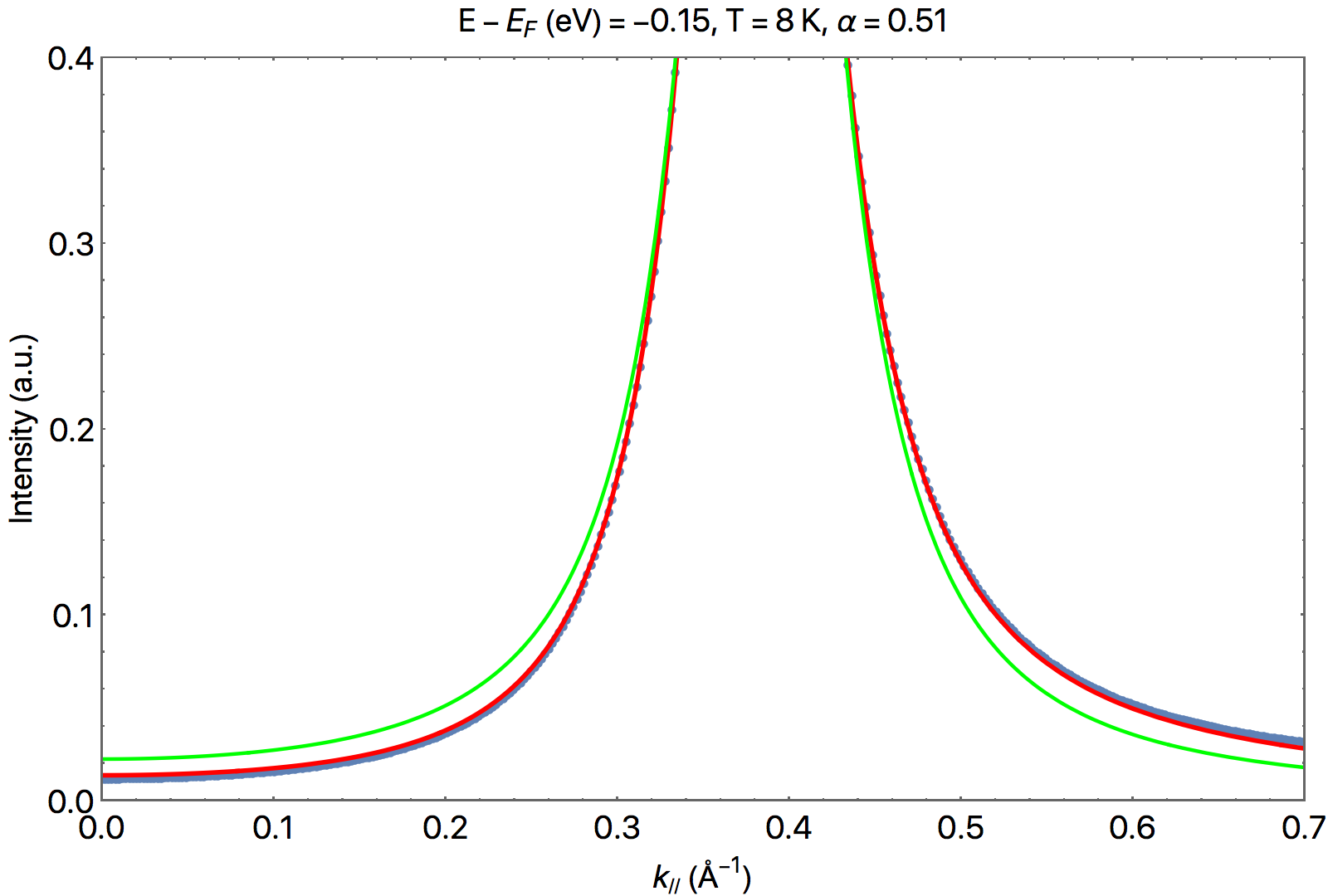}
  \end{subfigure}%
  \vskip\baselineskip
 \begin{subfigure}{.43\textwidth}
    \centering
    \includegraphics[width=\linewidth]{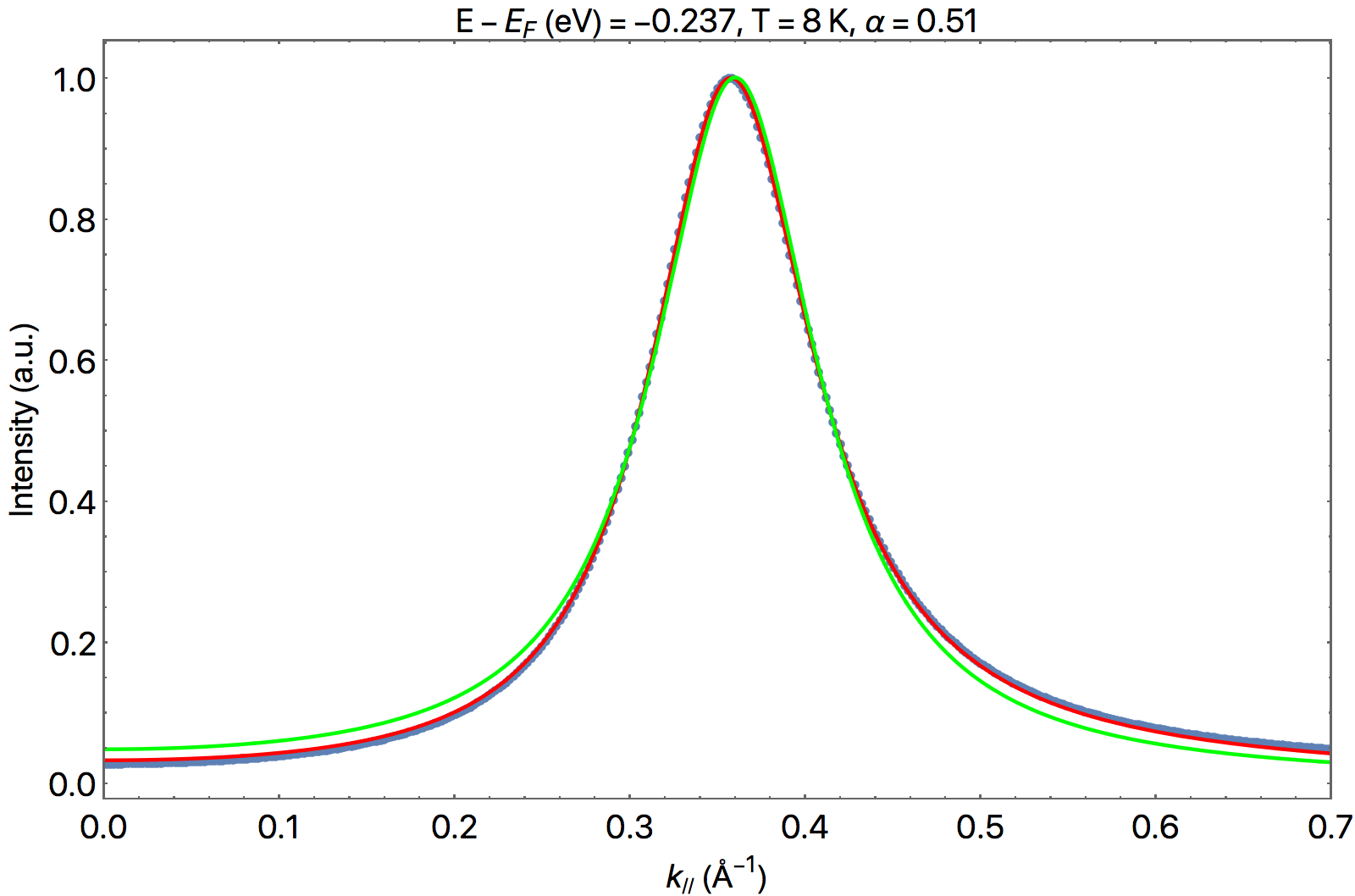}
  \end{subfigure}%
 \begin{subfigure}{.43\textwidth}
    \centering
    \includegraphics[width=\linewidth]{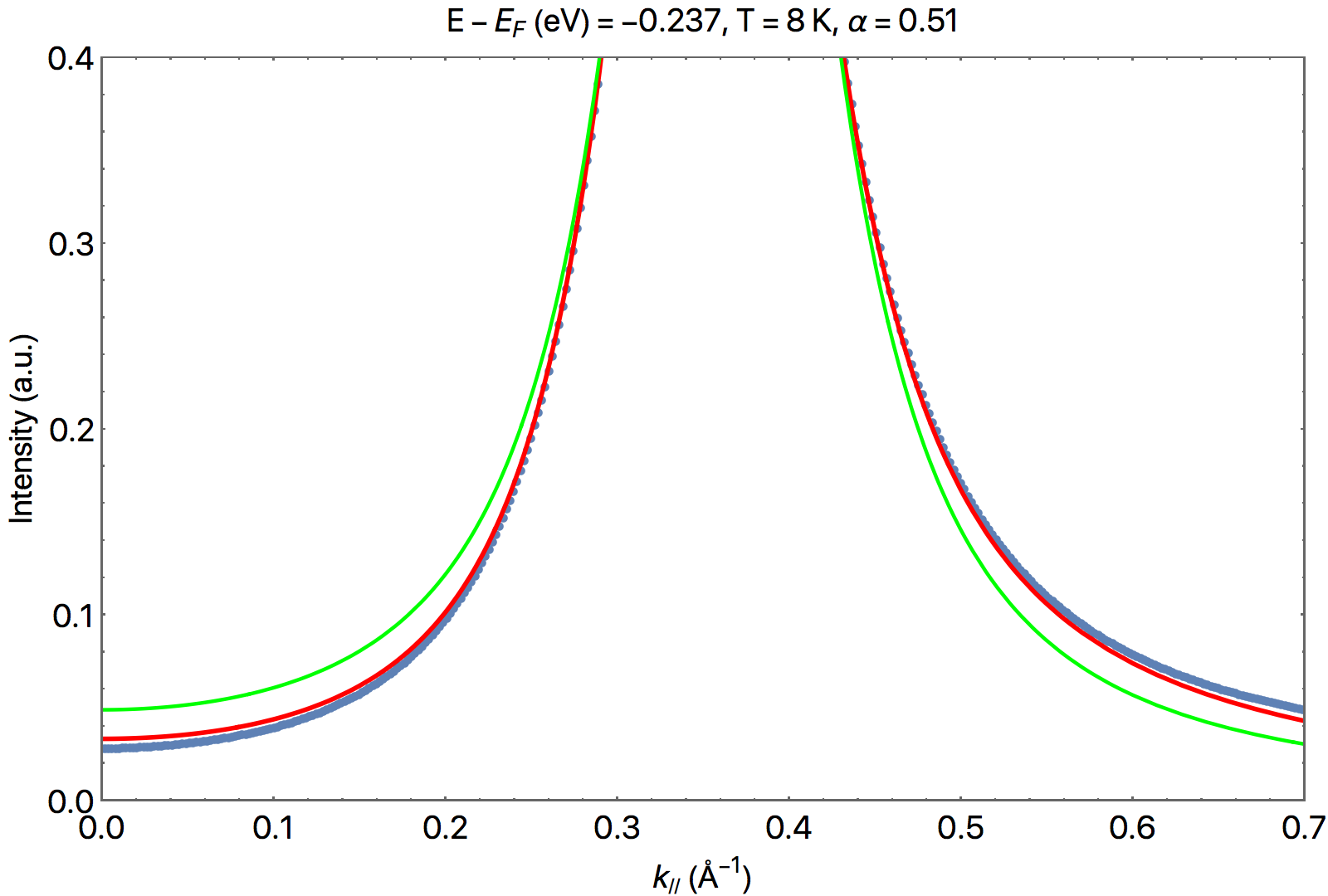}
  \end{subfigure}%
\caption{\label{fig:spectral_opD_8K} Comparison of the (smoothed) experimental MDC (\textcolor{blue}{blue dots}) for a sample at optimal doping $\alpha = 0.51$ at $T = 8 K$, with a Lorentzian fit based on the PLL (\textcolor{green}{green line}), and a fit based on the semi-holographic prediction for the self-energy in Eq.\ \eqref{eq:self-energy_holo_fit} (\textcolor{red}{red line}). Near the Fermi surface both models provide a good fit, but as we move away from the Fermi surface the holographic model provides a better fit, being able to capture the asymmetry in the peak. Deep below the Fermi surface, for $E -E_F \lesssim -0.25$, we start to see deviations even for the semi-holographic model. This is, however, where we expect the low-energy approximation to break down, as is also indicated by the dispersion becoming non-linear in Fig.\ \ref{fig:phonon_opD_8K}.}
\end{figure}

In data collected from ARPES measurements, there is a kink in the dispersion relation, as shown in Fig.\ \ref{fig:ARPES_dispersion}, at about $E - E_F \simeq 0.07 \text{eV}$. This kink is associated with a jump in the lifetime of the excitations and it is generally attributed to the electron-phonon interaction. This gives an additional contribution to the imaginary part of the self-energy that is accounted for in the experimental fitting procedure by the addition of a momentum-independent parameter $G_0(\omega)$. As explained in more details in Ref.\ \cite{Smit2021}, the analysis of the experimental data is done by fitting each momentum-dependent curve, i.e., a slice at constant energy as for example in Fig.\ \ref{fig:spectral_opD_8K}), with the ansatz
\begin{align}\label{eq:full_fit_function}
    \mathcal{A}(\omega, k) = \frac{\textcolor{red}{W(\omega)}}{\pi} \frac{\overbrace{\frac{\textcolor{blue}{\lambda} f_k \omega_N}{2}\left[\left(\frac{\omega}{\omega_N}\right)^2\right]^{\alpha(k)} + \frac{\textcolor{red}{G_0(\omega)}}{2} }^{\Gamma(\omega, k)/2}}{(k - \textcolor{red}{k_*(\omega)})^2 + \left(\underbrace{\frac{\textcolor{blue}{\lambda} f_k \omega_N}{2}\left[\left(\frac{\omega}{\omega_N}\right)^2\right]^{\alpha(k)}}_{\Sigma''(\omega, k)/v_F} + \frac{\textcolor{red}{G_0(\omega)}}{2}\right)^2 } ~,
\end{align}
where \textcolor{blue}{$\lambda$} is a doping-independent constant that is determined by a 2-dimensional fit for momenta and energies near the Fermi level, and is then fixed and does not enter as a fit parameter in the MDC fits. It is also kept fixed independently of the model used for the self-energy. In addition, $\hbar \omega_N = 0.5\text{eV}$ is an energy scale related to the
microscopic details of the $\text{CuO}_2$ layer \cite{Reber2019}. Finally, \textcolor{red}{$W(\omega)$}, \textcolor{red}{$k_*(\omega)$} and
\textcolor{red}{$G_0(\omega)$} are fit functions. $W(\omega)$ is the overall normalization, that we won't discuss here, $k_*(\omega)$ the dispersion that we expect to take the form $k_*(\omega) = k_F - \omega/v_F$ plus a renormalization contribution from the phonon as explained below. Finally, $G_0(\omega)$ accounts for the contribution to the width of the peak that does not come from the electron self-energy, and we expect it to be well described by a model of the electron-phonon interaction. 
In order to check this, we compare the results for $G_0(\omega)$ with a simple approximation for a dispersionless phonon, with characteristic frequency obeying $\hbar\omega_\text{ph} = 0.07 \text{eV}$, given by
\begin{align}\label{eq:phonon}
  \frac{\Sigma_{\text{ph}}(\omega)}{v_F} = \frac{G_{\text{ph}}}{2 \pi} \log\left(\frac{\omega - \omega_\text{ph} - i \Omega}{\omega + \omega_\text{ph} + i \Omega}\right) \text{ ,}
\end{align}
with, as we will shortly see, $\Omega > 0$ even at $T = 0$,  likely due to the smearing of the Fermi surface due to strong interactions in the system. This has the effect of smoothing out the step-function in the imaginary part of the phonon self-energy, as well as the kink in the dispersion. 

For the PLL, $f_k = 1$ and $\alpha(k) = \alpha$, while for the semi-holographic model presented above $\alpha(k) = \alpha (1 - (k - k_F)/k_F)$ from Eq.\ \eqref{eq:electron_self}, and from the imaginary part in Eq.\ \eqref{eq:ir_green_2d} we have
\begin{align}\label{eq:f_k_semiholo}
    f_k = g_k^2 \left(\frac{\omega_N \alpha}{3^{3/4} 2 k_F} \right)^{2 \alpha(k)}\frac{\Gamma[1/2 - \alpha(k)]}{\Gamma[1/2 + \alpha(k)]} \cos(\pi \alpha(k)) \text{ ,}
\end{align}
where we used the fact that $\alpha \equiv 2 q k_F/\mu$ in the first prefactor. 
Remember that, upon reintroducing dimensionful units, this prefactor reads $\alpha \omega_N/2 \hbar c k_F$. This then contains an unknown velocity that, however, can be adsorbed into the coupling constant $g_k^2$. 
Specifically, as we would like to keep the model as simple as possible, we consider a coupling $g_k^2 = \mathcal{C}_0(1/\beta_0)^{2(\alpha(k))}$, with $\mathcal{C}_0$ a momentum-independent constant that is uniquely determined by requiring that at the Fermi momentum we recover the form of the PLL, that is $f_{k_F} = 1$. Then the self-energy in Eq.\ \eqref{eq:full_fit_function} takes the form (temporarily reintroducing dimensionful quantities for clarity)
\begin{align}\label{eq:self-energy_holo_fit}
 \begin{split}
    \frac{\Sigma''(\omega, k)}{v_F} =&\frac{\textcolor{blue}{\lambda} \hbar \omega_N}{2} \left(\frac{\alpha \hbar \omega_N}{3^{3/4} 2 \hbar c \beta_0 k_F} \right)^{-\alpha\frac{k - k_F}{k_F}}\frac{\Gamma[1/2 + \alpha]}{\Gamma[1/2 - \alpha]}\frac{\Gamma[1/2 - \alpha(k)]}{\Gamma[1/2 - \alpha(k)]} \frac{\cos(\pi \alpha(k))}{\cos(\pi \alpha)}\left(\frac{\omega^2}{\omega_N^2}\right)^{\alpha(k)} \text{ ,}\\
    \alpha(k) =& \alpha \left(1 - \frac{k - k_F}{k_F} \right) \text{ .}
 \end{split}
\end{align}
Here $\beta_0$ is such that $\hbar c \beta_0 \simeq 0.21  \text{eV\AA}$, chosen by comparison with experimental data. It is a fixed constant, independent of doping and temperature, and it does thus not introduce any extra fitting parameter compared to the PLL model. 

\begin{figure}[h]\centering
  \begin{subfigure}{.5\textwidth}
    \centering
    \includegraphics[width=\linewidth]{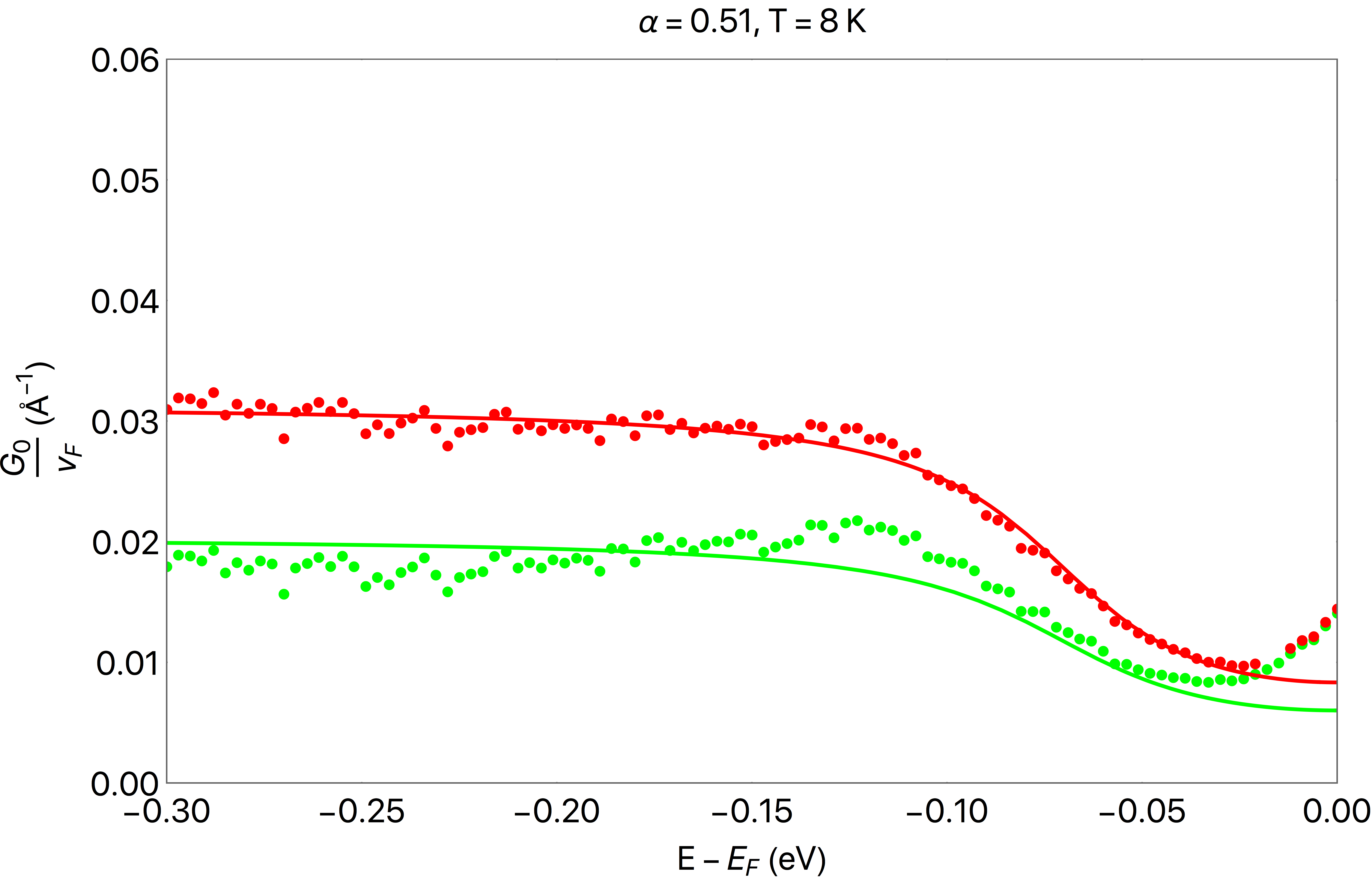}
  \end{subfigure}%
\begin{subfigure}{.5\textwidth}
  \centering
  \includegraphics[width=\linewidth]{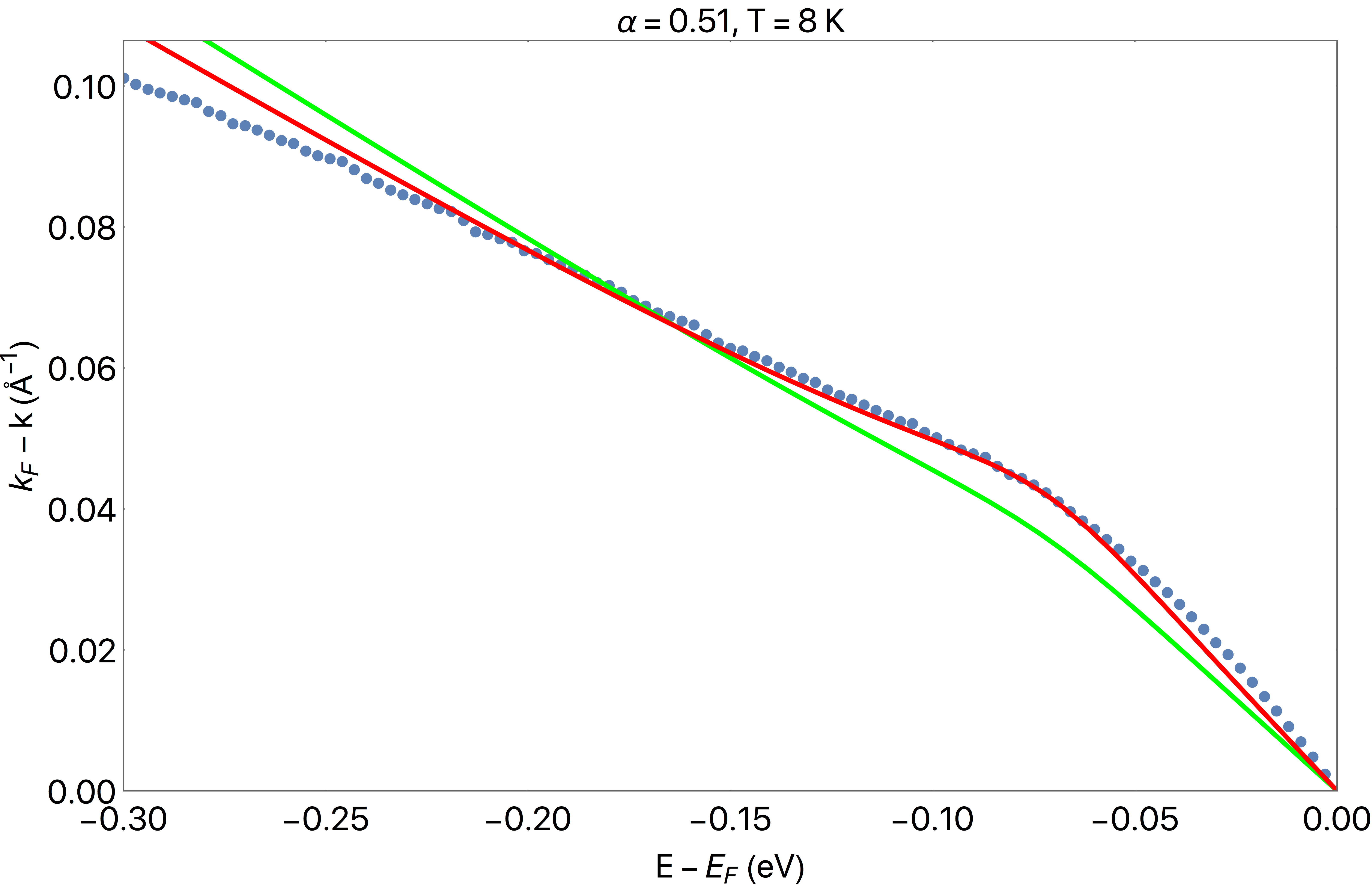}
\end{subfigure}%
\caption{\label{fig:phonon_opD_8K} (Left) Fit of $G_0(\omega)$ in Eq.\ \eqref{eq:full_fit_function} to the electron-phonon model in Eq.\ \eqref{eq:phonon} for a PLL fit function (\textcolor{green}{green line}), giving $G_{\text{ph}} \simeq 0.020$ and $\Omega \simeq 0.034$, and the semi-holographic one (\textcolor{red}{red line}), that gives $G_{\text{ph}} \simeq 0.031$ and $\Omega \simeq 0.030$ (it can be seen that there is a deviation at very small energy,  probably related to some leftover disorder). (Right) Comparison between the experimentally observed dispersion relation (\textcolor{blue}{blue dots}) with the dispersion as expected from the electron-phonon model given the parameters $G_{\text{ph}}$ and $\Omega$ obtained from the fit. We see that the semi-holographic model provides a better fit to the dispersions with $\hbar v_F \simeq 2.9\text{eV\AA}$, while the PLL fit does not properly capture the phonon kink and gives $\hbar v_F \simeq 2.7 \text{eV\AA}$.}
\end{figure}

Below, in Fig.\ \ref{fig:spectral_opD_8K}, we present the comparison of the fit to a ``cleaned-up'' version of data from Ref.\ \cite{Smit2021} at low-temperature and near optimal doping ($T = 8 K$, $\alpha = 0.51$), where the width due to the instrumental resolution and fluctuations in the data have been removed by a combination of deconvolution and smoothing respectively. We show one branch of the dispersion, but the fit is working on both branches, so as to correctly account for the overlap of the tail of each branch on the lineshape of the other. Near the Fermi surface, both models provide a good fit to the data, but as we move away from $E_F$ the semi-holographic model (red line in the figures) does a much better job in describing the data by capturing extremely well the peak asymmetry over the entire energy range where the dispersion is approximately linear. We start to see deviations from the holographic model at $\hbar\omega = E - E_F < -0.25 \text{ eV}$. However, from the right panel of Fig.\ \ref{fig:phonon_opD_8K}, one can see that this energy is one at which the approximation of a linear dispersion seems to break down. We stress again here that, while we cannot claim with certainty the origin of this observed asymmetry, many other possible simple explanations---that are not rooted in the momentum-dependence of the electron self-energy---have been analyzed and ruled out in Ref.\ \cite{Smit2021}. 
To validate the fit procedure, we have to check that what we obtain for $G_0(\omega)$ can have a reasonable physical explanation and, in particular, whether it is consistent with the contribution from the electron-phonon interaction as per Eq.\ \eqref{eq:phonon}. Notice that in the way the fit is performed, the contribution of the phonon to $\Sigma''(\omega, k)$ and the dispersion relation are two independent parameters, but, within the validity of the phonon-model approximation, we expect them to be related. We therefore perform a fit of $G_0(\omega)$ for both the power-law liquid and the semi-holographic model to determine $\Omega$ and $G_{\text{ph}}$ from Eq.\ \eqref{eq:phonon}, and check them against the dispersion $k_*(\omega)$:
\begin{align}
    k_*(\omega) - k_F \overset{\text{?}}{\approx} \frac{\omega}{\textcolor{red}{v_F}} + \frac{\textcolor{blue}{G_{\text{ph}}}}{2\pi}\text{Re}\left[ \log\left(\frac{\omega - \omega_\text{ph} - i \textcolor{blue}{\Omega}}{\omega + \omega_\text{ph} + i \textcolor{blue}{\Omega}}\right) \right] \text{ .}
\end{align}
The renormalized velocity \textcolor{red}{$v_F$} is left as a parameter in the fit for the dispersion, but we expect a value $v_F \simeq 3 \text{eV\AA}$ \cite{Kordyuk2005}. In Fig.\ \ref{fig:phonon_opD_8K} we show the result of the fit of this model (continuous line) to the extracted values of $G_0(\omega)$ (dots) for the PLL (green) and the semi-holographic (red) fit function. The semi-holographic fit function, in contrast with the PLL, seems to give a description for $G_0(\omega)$ consistent with an electron-phonon model, nicely reproducing the dispersion (including the kink at $\hbar\omega_\text{ph} = 0.07 \text{eV}$) with $\hbar v_F = 2.9 \text{eV\AA}$, up to energies where it ceases to be linear. While we should be careful in considering this as further evidence of the validity of the model as there could be other experimental factors influencing $G_0(\omega)$ that we might be overlooking, the success of such a simple semi-holographic model with a phonon in accurately depicting the results of ARPES measurements on a single-layer cuprate provides a simple and useful benchmark to compare to other theoretical models at low-temperature.
A further test that is left for future work could be to check if the value found for $\Omega$ can be related to the density distribution arising from the semi-holographic Green's function $N(k) \propto \int_{-\infty}^{0} \dif \omega \text{Im}[\text{Tr}[G_{\chi\chi}]]$.

\subsection{Doping dependence}
\begin{figure}[h!]\centering
\begin{subfigure}{.5\textwidth}
    \centering
    \includegraphics[width=\linewidth]{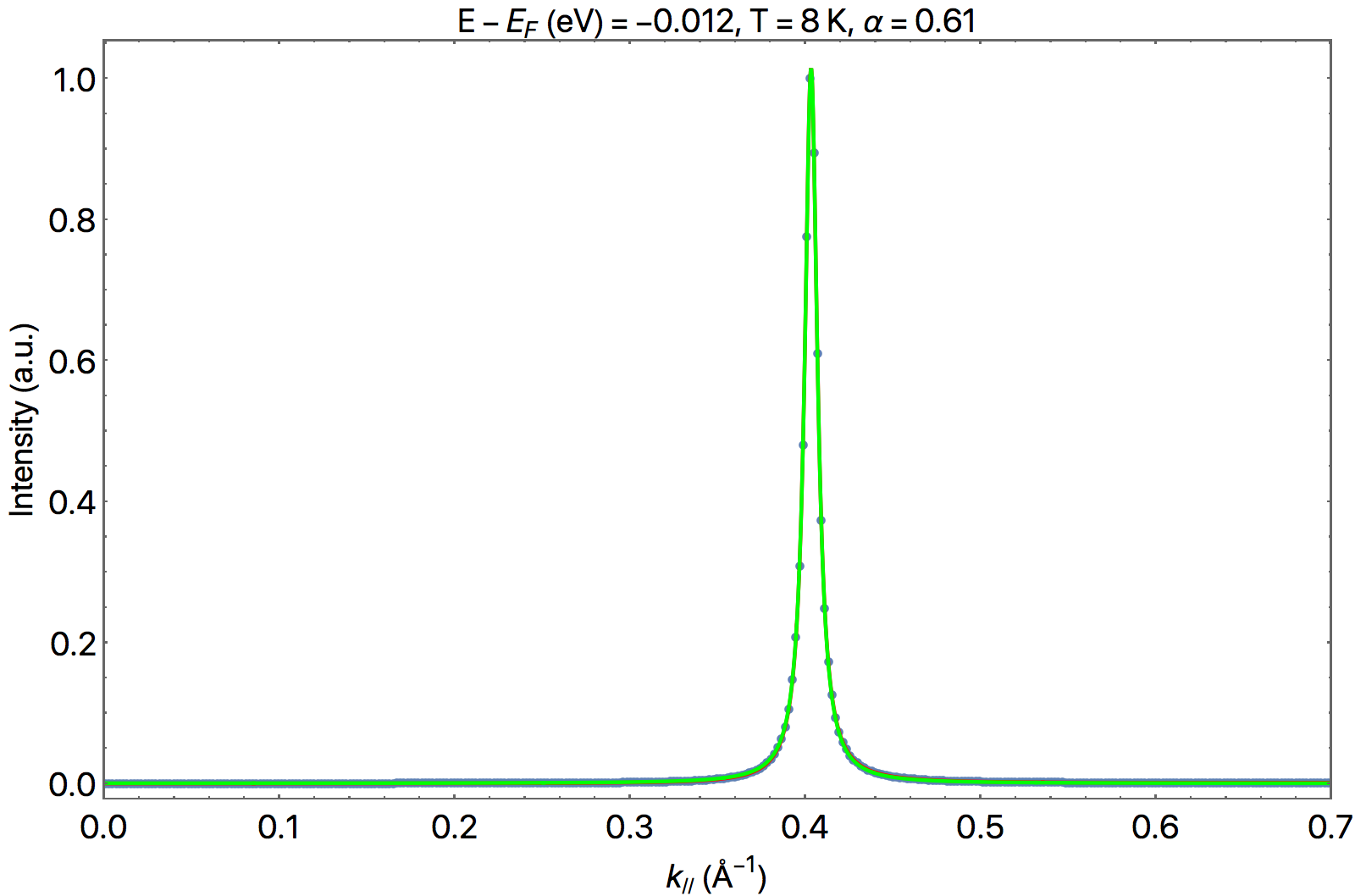}
  \end{subfigure}%
 \begin{subfigure}{.5\textwidth}
    \centering
    \includegraphics[width=\linewidth]{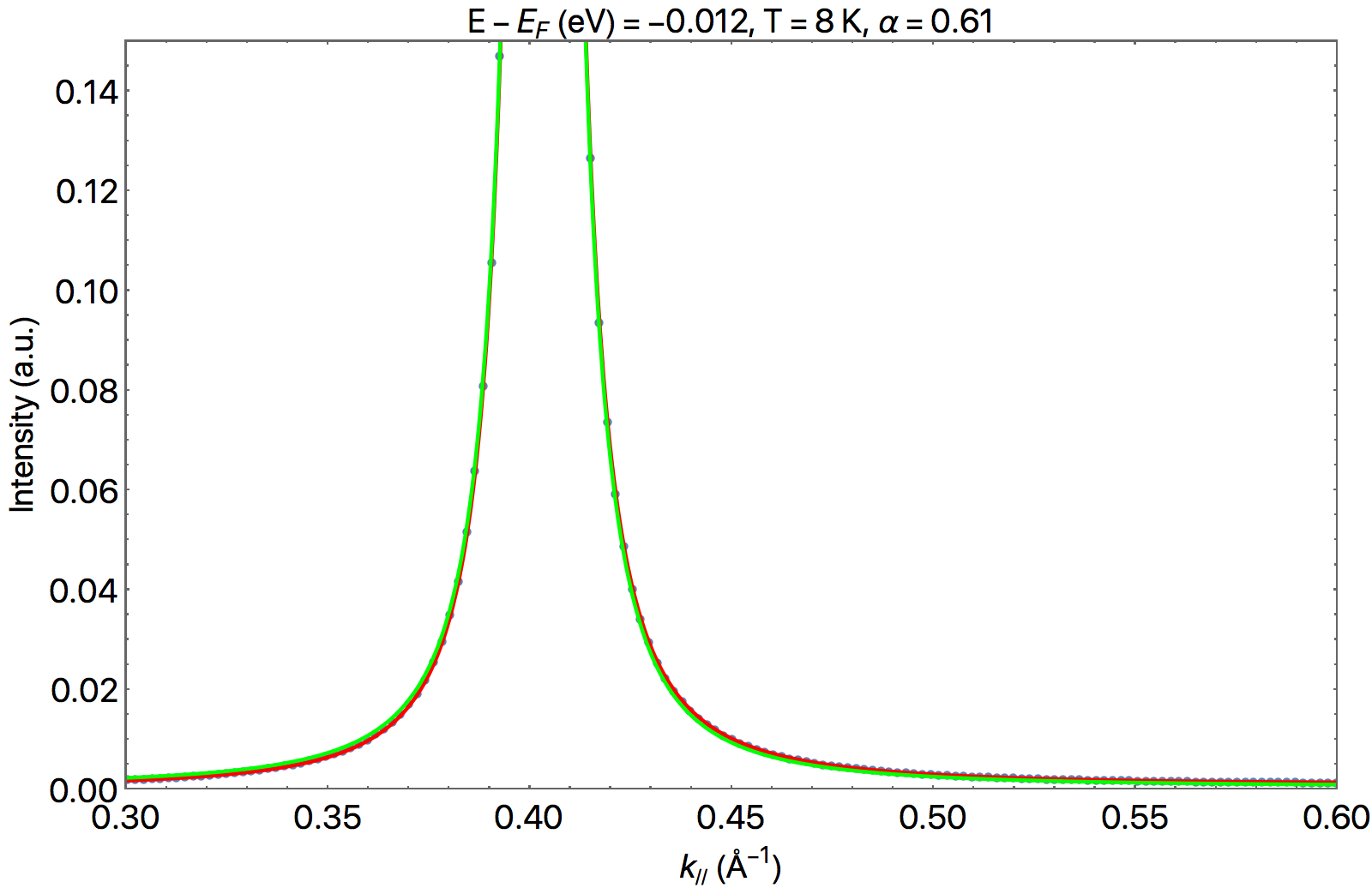}
  \end{subfigure}%
  \vskip\baselineskip
  \begin{subfigure}{.5\textwidth}
    \centering
    \includegraphics[width=\linewidth]{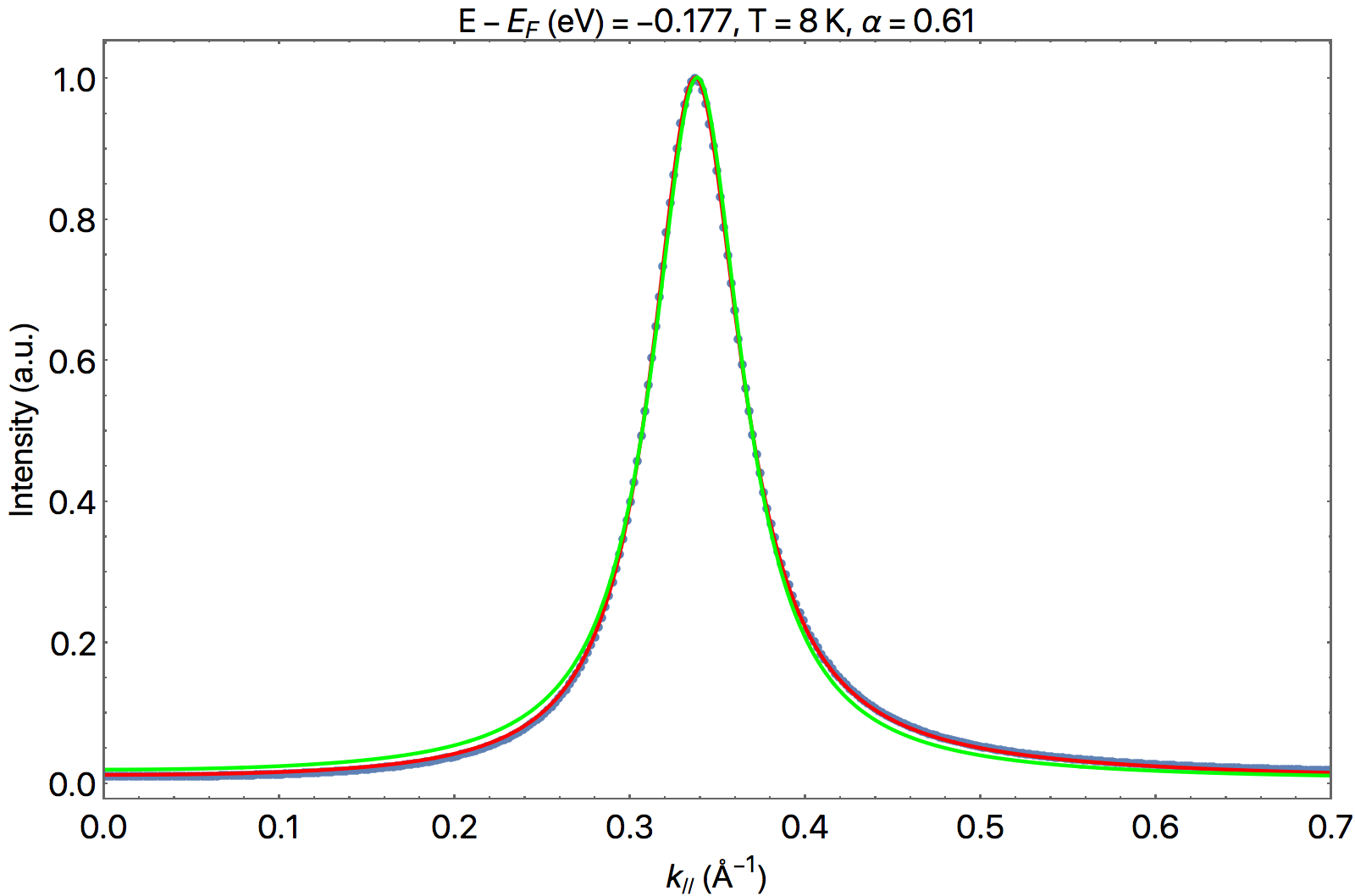}
  \end{subfigure}%
\begin{subfigure}{.5\textwidth}
  \centering
  \includegraphics[width=\linewidth]{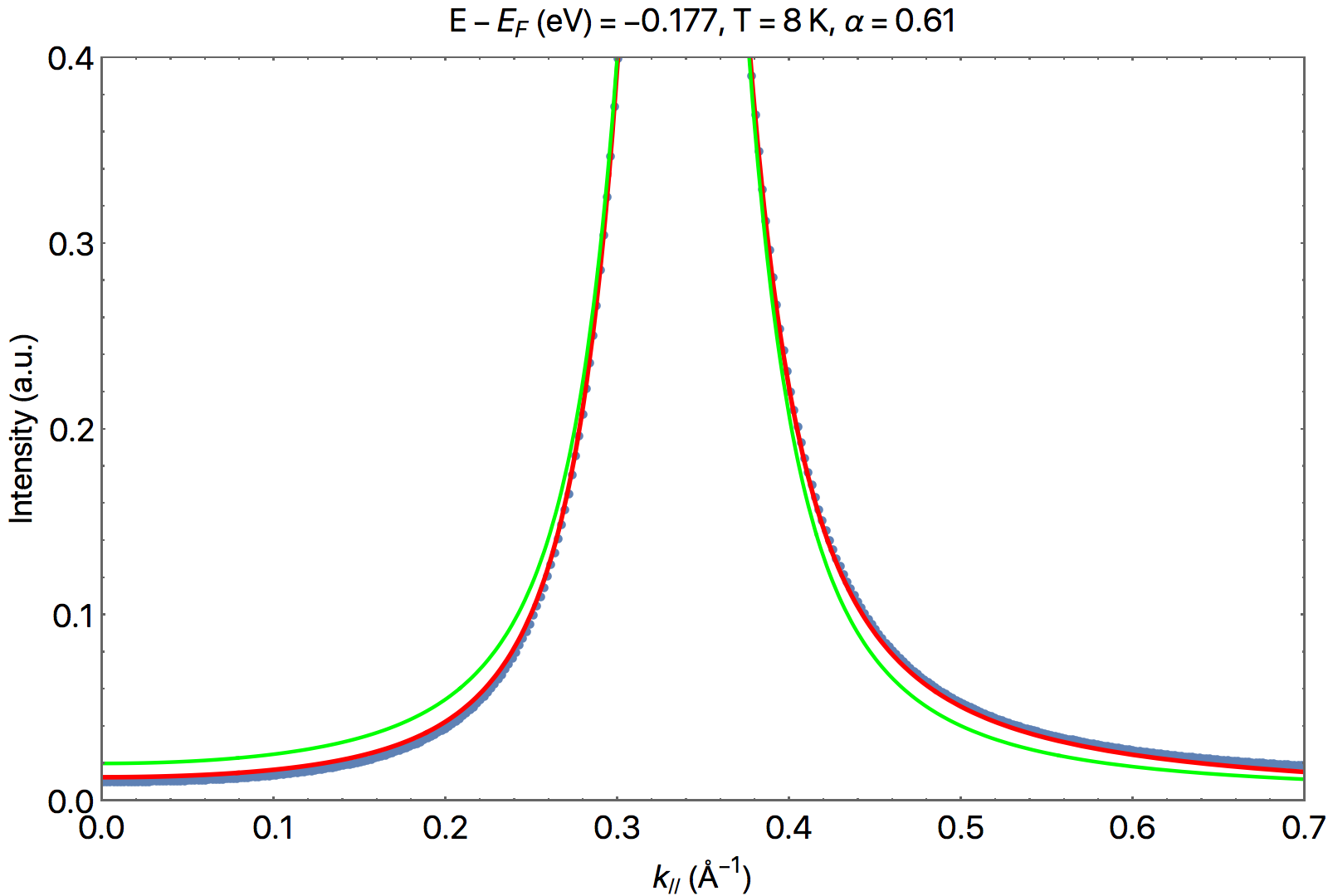}
\end{subfigure}
\caption{\label{fig:spectral_OD23_8K} Comparison of the PLL fit (\textcolor{green}{green line}) and semi-holographic fit (\textcolor{red}{red line}) to MDC data (\textcolor{blue}{blue dots}) for an overdoped sample with $\alpha = 0.61$ at $T = 8K$. Near the Fermi surface, both models provide a good fit to the data, while further away from it the semi-holographic model accounts for the asymmetry in the peak.}
\end{figure}

\begin{figure}[h!]\centering
  \begin{subfigure}{.5\textwidth}
    \centering
    \includegraphics[width=\linewidth]{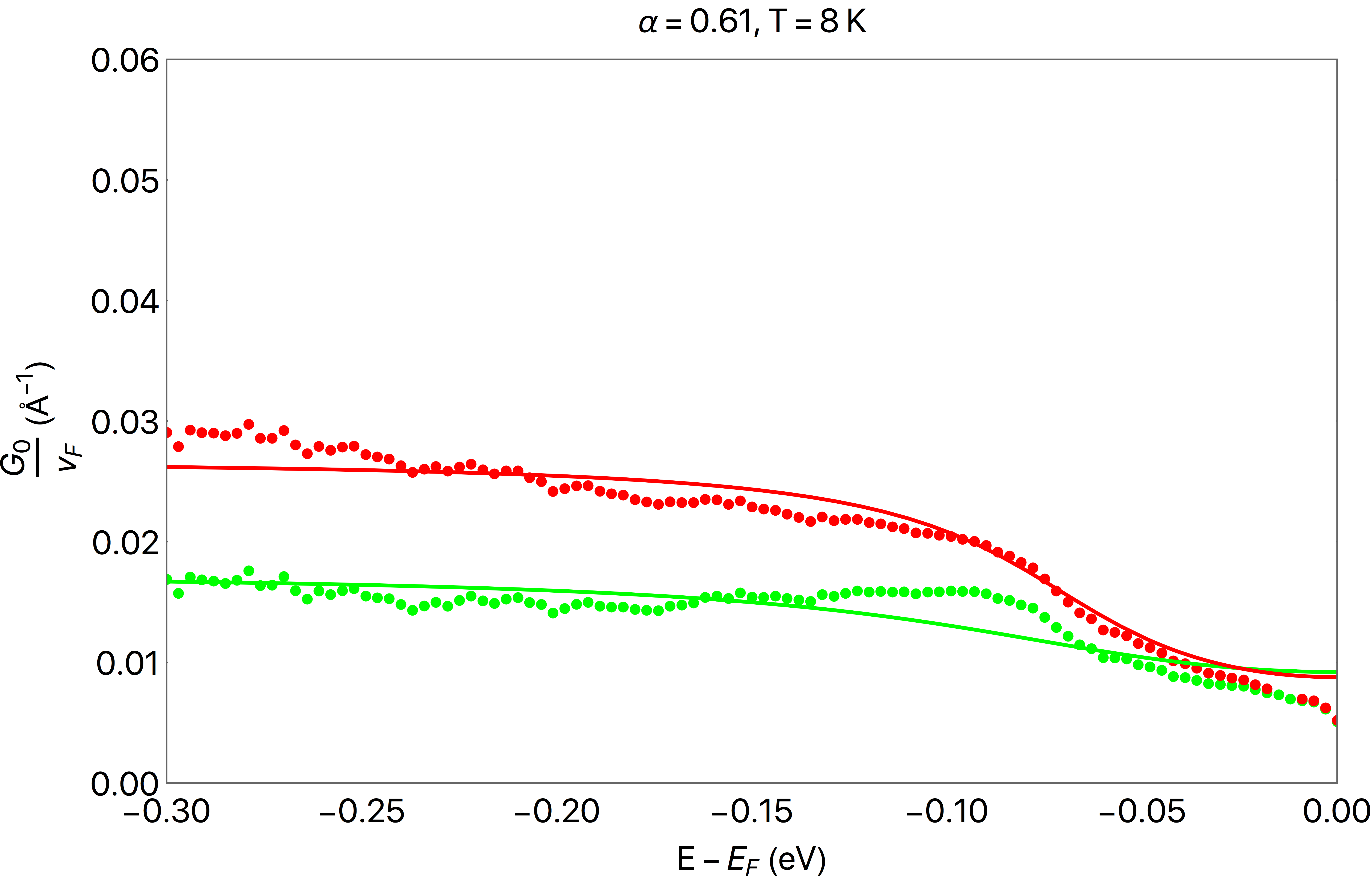}
  \end{subfigure}%
\begin{subfigure}{.5\textwidth}
  \centering
  \includegraphics[width=\linewidth]{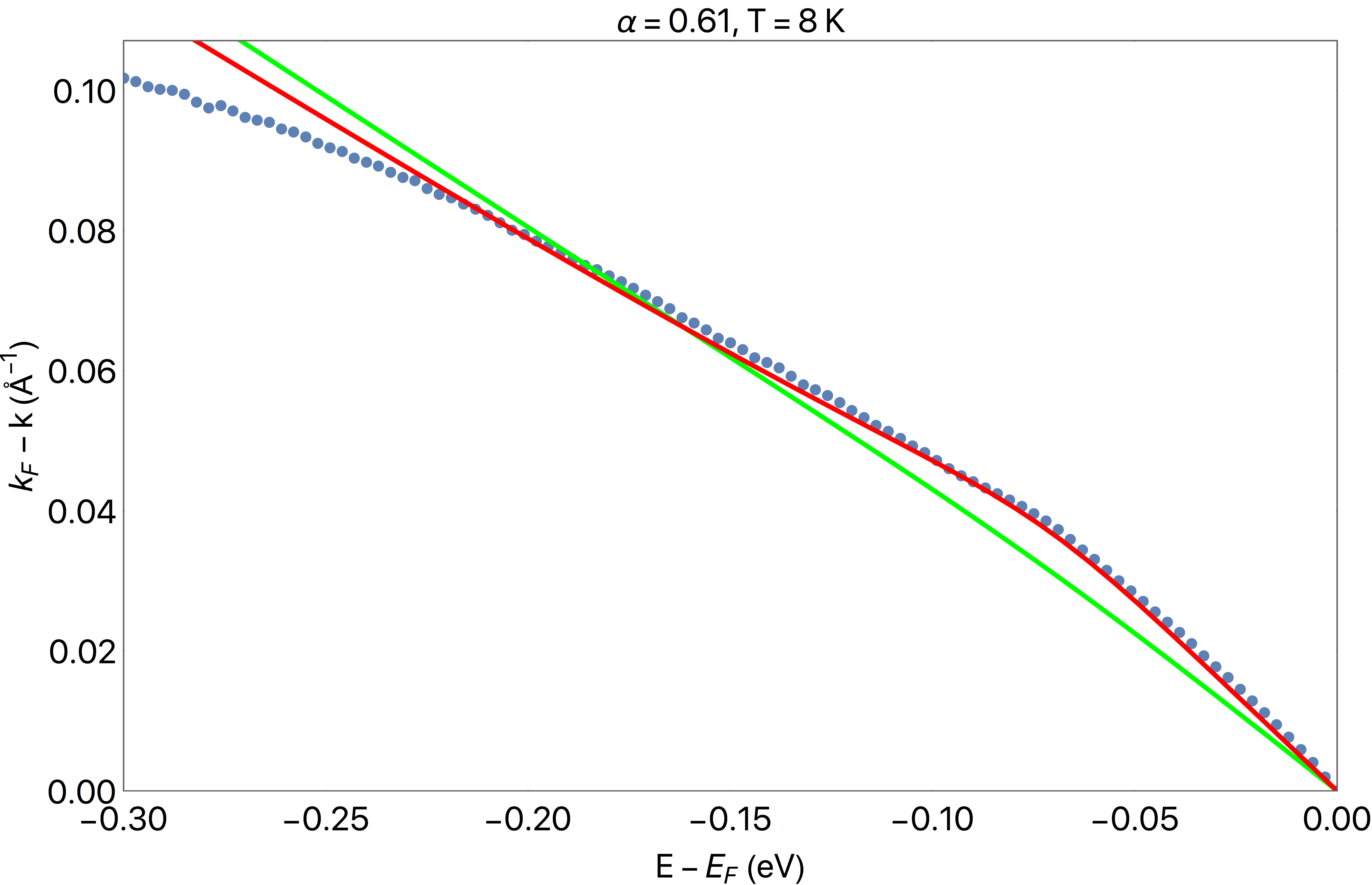}
\end{subfigure}
\caption{\label{fig:phonon_OD23_8K} (Left) Fit to $G_0(\omega)$ with the electron-phonon model and comparison with the dispersion relation (right), for the overdoped sample with $\alpha = 0.65$. We see here as well that the semi-holographic fit function provides a description consistent with the electron-phonon model.}
\end{figure}

\begin{figure}[h!]\centering
  \begin{subfigure}{.5\textwidth}
    \centering
    \includegraphics[width=\linewidth]{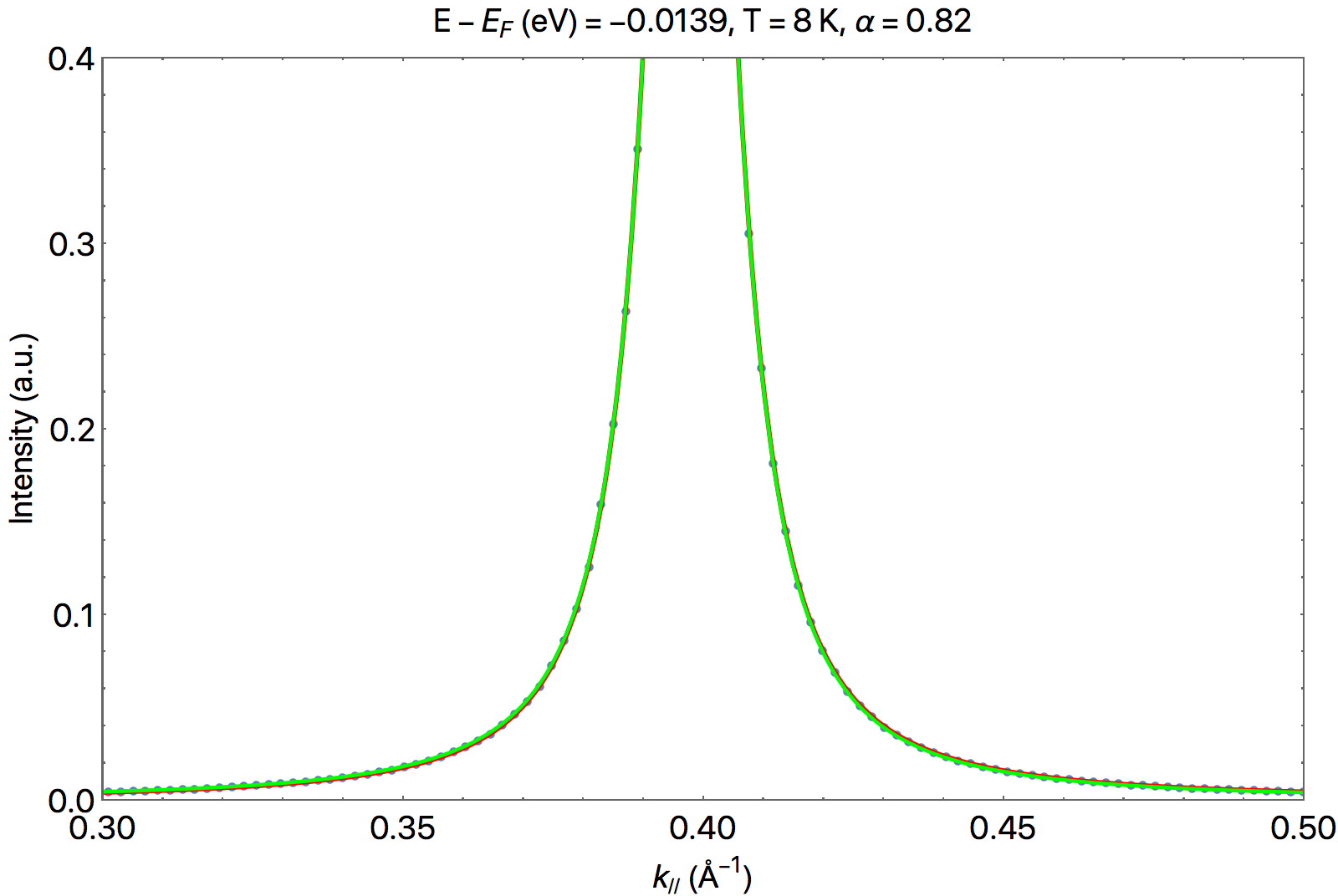}
  \end{subfigure}%
\begin{subfigure}{.5\textwidth}
  \centering
  \includegraphics[width=\linewidth]{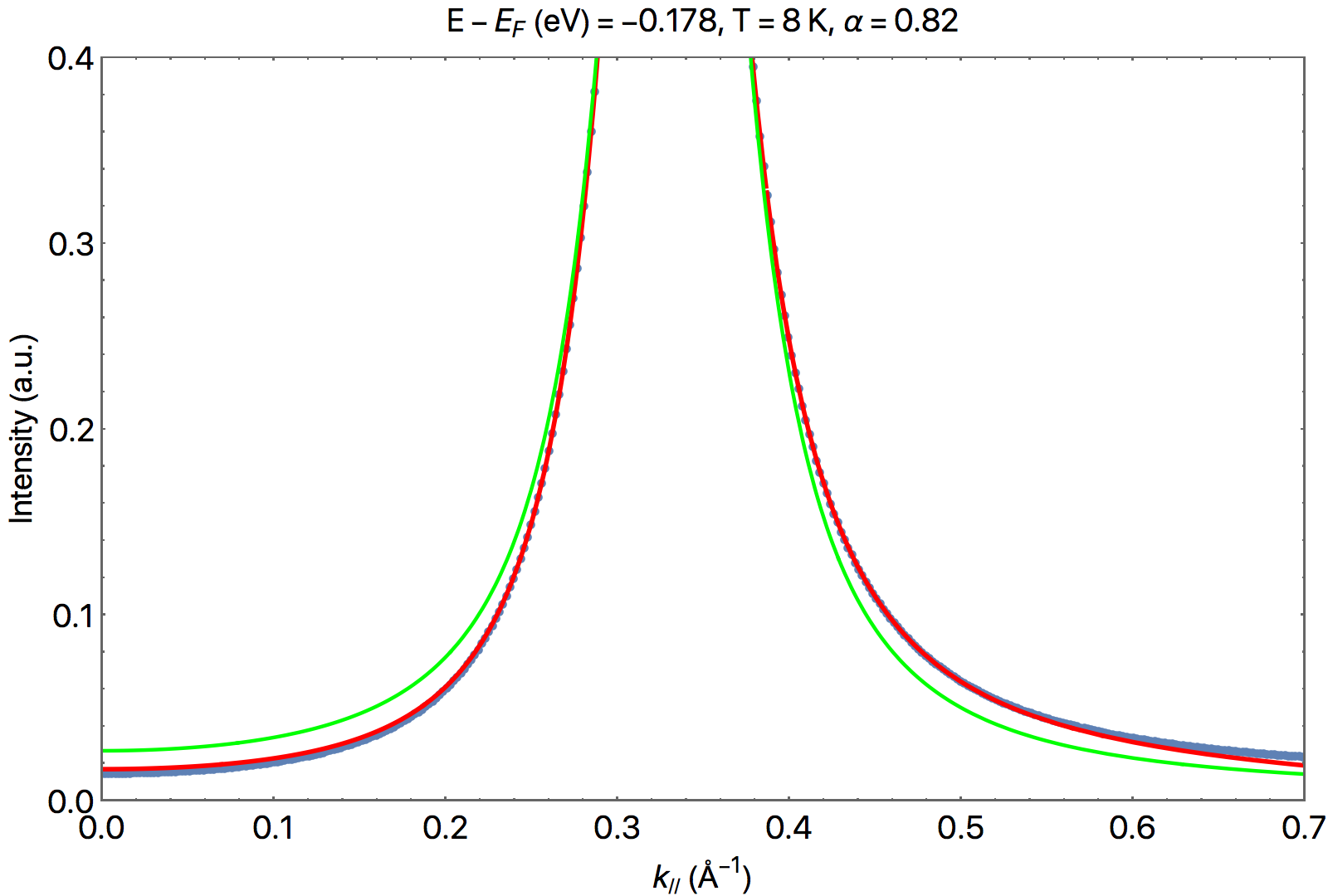}
\end{subfigure}
\caption{\label{fig:spectral_OD3_8K} Details of the PLL fit (\textcolor{green}{green line}) and semi-holographic fit (\textcolor{red}{red line}) to MDC data \textcolor{blue}{blue dots} for an overdoped sample with $\alpha = 0.82$ at $T = 8K$, both near and deep below the Fermi surface.}
\end{figure}

\begin{figure}[h!]\centering
  \begin{subfigure}{.5\textwidth}
    \centering
    \includegraphics[width=\linewidth]{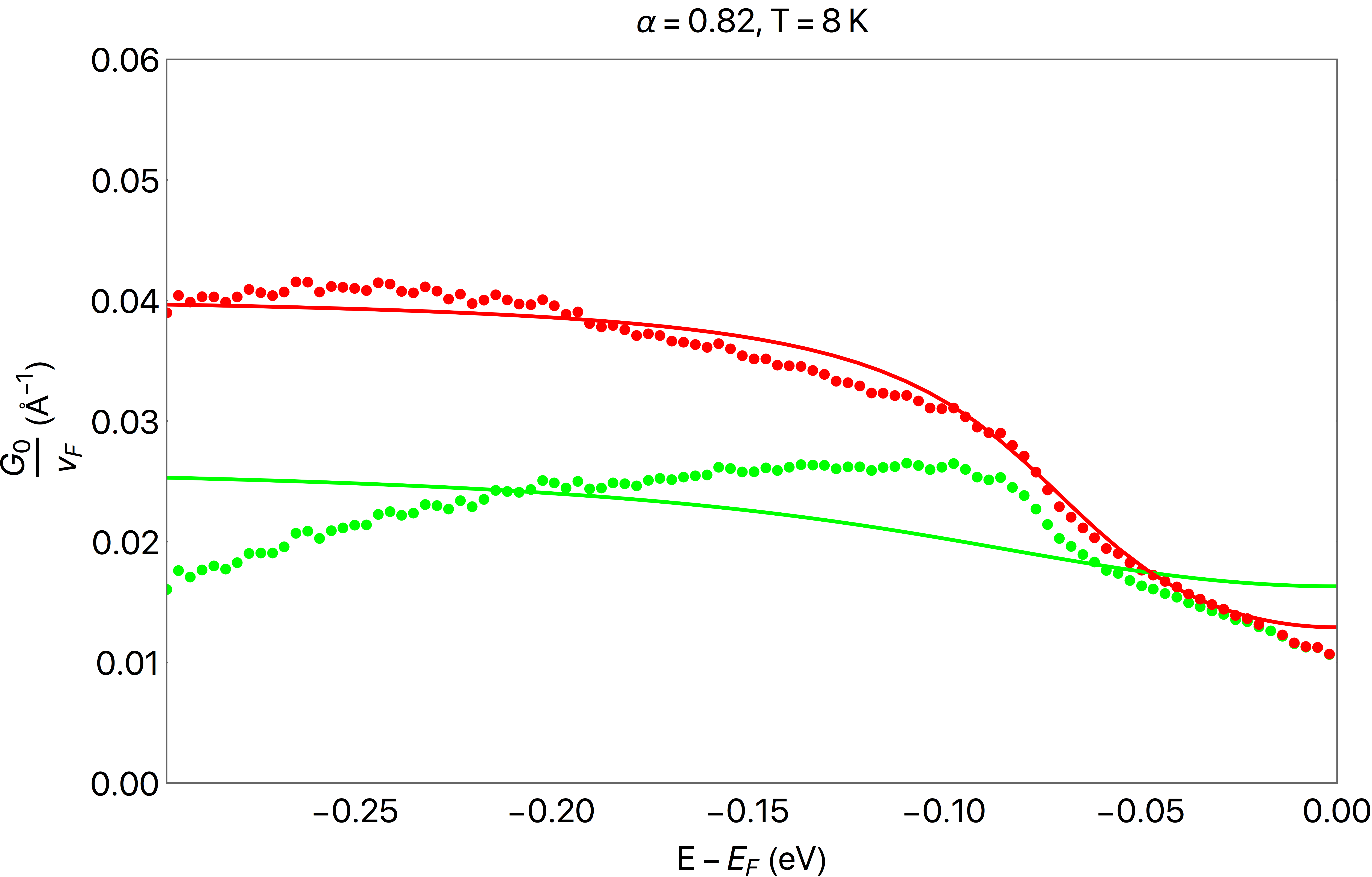}
  \end{subfigure}%
\begin{subfigure}{.5\textwidth}
  \centering
  \includegraphics[width=\linewidth]{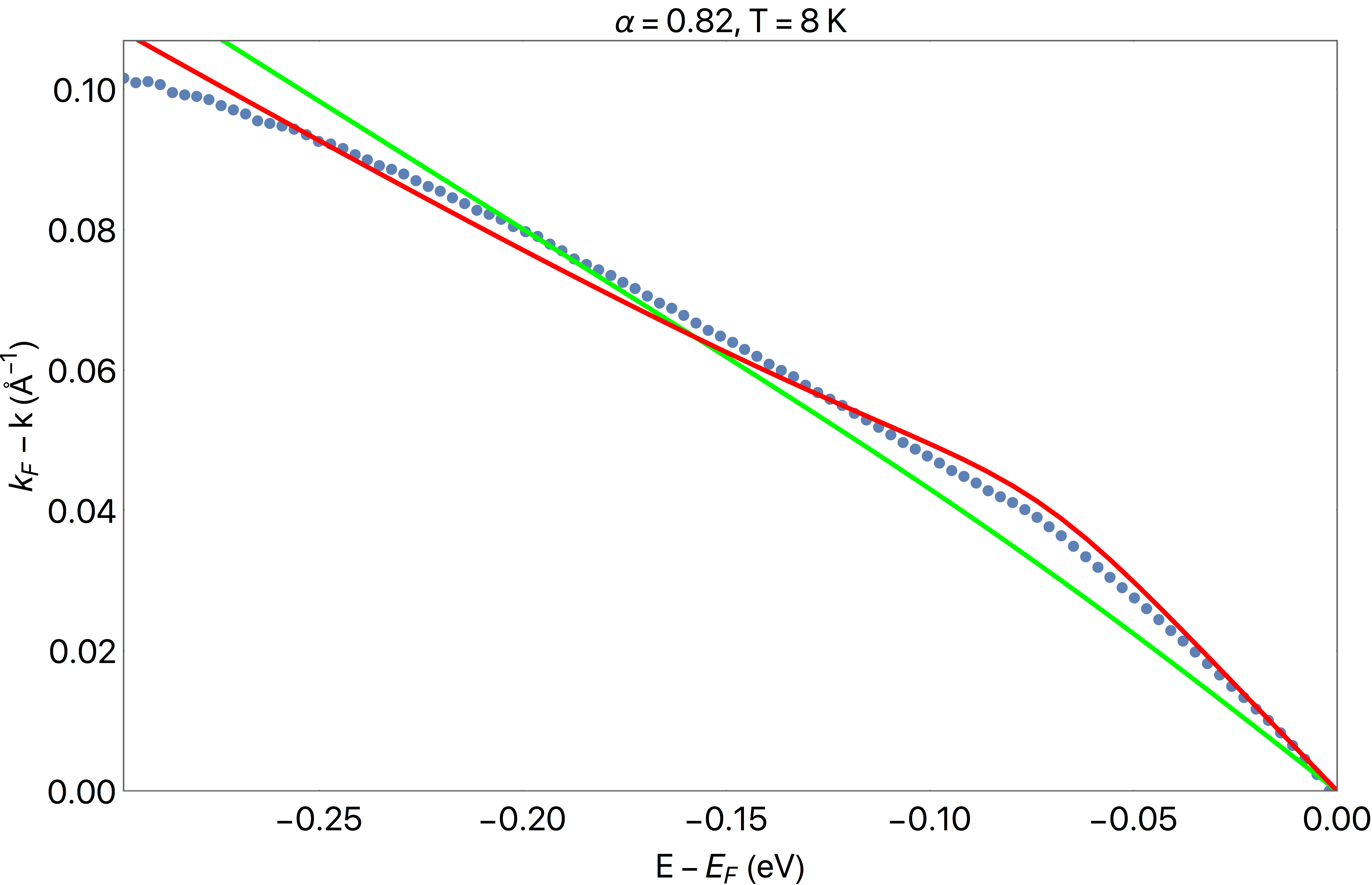}
\end{subfigure}
\caption{\label{fig:phonon_OD3_8K} (Left) Fit to $G_0(\omega)$ with the electron-phonon model and comparison with the dispersion relation (right), for the overdoped sample with $\alpha = 0.81$.}
\end{figure}

We now repeat the analysis presented above for a pair of overdoped samples at the same low-temperature $T = 8 K$, and show the results in Figs. \ref{fig:spectral_OD3_8K}  and \ref{fig:phonon_OD3_8K} for $\alpha = 0.61$ and Figs. \ref{fig:spectral_OD23_8K}  and \ref{fig:phonon_OD23_8K} for $\alpha = 0.82$. Note that the only quantities that change with doping in the fit function Eq.\ \eqref{eq:full_fit_function} are the scaling exponent $\alpha$ and $k_F = k_F(\alpha)$---both determined by a PLL fit near the Fermi surface---while all the other parameters are kept fixed. We see again, also in both overdoped cases, that the semi-holographic model provides a better description of the data at higher energies, as well as of the phonon contribution and dispersion relation \footnote{for the phonon, we considered the Fermi velocity doping independent while we allowed the coupling $G_\text{ph}$ and the factor $\Omega$ to change with doping.}. The slight disagreement near the Fermi surface between $G_0(\omega)$ and the phonon model could simply arise from the fact that near $E_F$ the approximation used for the phonon self-energy in Eq.\ \eqref{eq:phonon} deviates the most from the Fermi-Dirac distribution with a smeared out Fermi surface (see also Fig.\ \ref{fig:phonon_fermi_dirac_comparison}).

\subsection{Temperature corrections}\label{subsec:non_zero_T}
\begin{figure}
    \centering
    \includegraphics[width=0.5\linewidth]{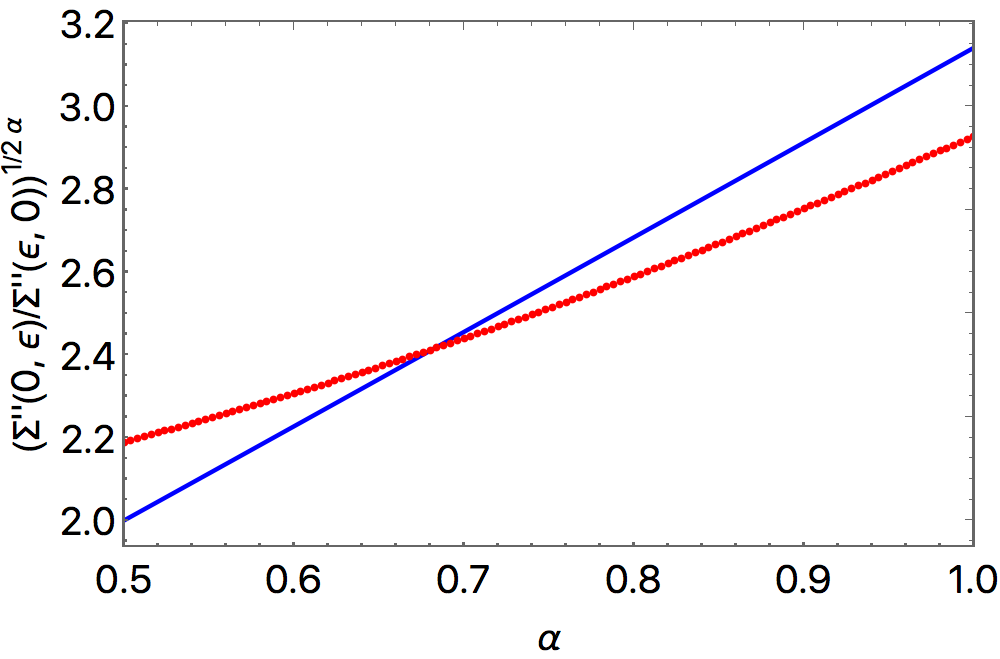}
    \caption{Plot of the ratio $(\Sigma''(\omega = 0, k_B T = \epsilon, k_F)/\Sigma''(\omega = \epsilon, k_B T = 0, k_F))^{1/2\alpha}$ from the analytical semi-holographic model (\textcolor{blue}{blue line}), and from the full numerical solution (\textcolor{red}{red line}), where there are corrections dependent on the mass $m$ in the bulk Dirac equation. In the PLL this ratio is the fit parameter $\beta$, with experimentally-determined values between 3 and 4.}
    \label{fig:asymptotic_ratio}
\end{figure}

As we have seen above, the zero-temperature IR Green's function provides a compelling model to describe ARPES data at low temperatures and in the range of frequencies of interest. Moving closer to room temperatures, we want to compare experimental data with the non-zero-temperature semi-holographic prediction from Eq.\ \eqref{eq:ir_green_2d_nonzero_T}. 
The PLL instead generalizes to 
\begin{align}
    \Sigma''_{\text{PLL}} \propto ((\hbar\omega)^2 + (\beta k_B T)^2)^{\alpha} \text{ ,}
\end{align}
that has been shown to well capture the temperature behavior with a value of the parameter $\beta$ believed to be $\pi$, and found to lie between 3 and 4 depending on doping, see Ref.\ \cite{Smit2021}. There, ARPES data were also compared with a ``semi-holographic inspired'' generalization, where the self-energy from the power-law liquid model was simply generalized by making the exponent momentum-dependent $((\hbar\omega)^2 + (\beta k_B T)^2)^\alpha/(\hbar\omega_N)^{2\alpha} \to ((\hbar\omega)^2 + (\beta k_B T)^2)^{\alpha(k)}/(\hbar\omega_N)^{2\alpha(k)}$, with $\alpha(k)$ as in Eq.\ \eqref{eq:self-energy_holo_fit} and all the other parameters kept as in the PLL. This provided an accurate description of the temperature dependence. 
On the other hand, the semi-holographic model seems to present a rather different temperature behavior as per Eq.\ \eqref{eq:ir_green_2d_nonzero_T}. It is first interesting to notice that, when $\alpha(k) = 1$, the latter equation simplifies to the Fermi-liquid form
\begin{align}
    \mathcal{G}_k^{2D}/\mu = q \frac{4 i}{3 \sqrt{3} \mu^2}((\hbar\omega)^2 + (\pi k_B T)^2) \text{ ,}
\end{align}
consistent with the PLL temperature behavior. However, the factor in front of the temperature term becomes smaller for $\alpha(k) < 1$.
In Fig.\ \ref{fig:asymptotic_ratio} we show the ratio $(\Sigma''(\hbar\omega = 0, k_B T = \epsilon, k_F)/\Sigma''(\hbar\omega = \epsilon, k_B T = 0, k_F))^{1/2\alpha}$ for the semi-holographic prediction from Eq.\ \eqref{eq:ir_green_2d_nonzero_T}, where $\epsilon \ll 1$. This ratio corresponds to $\beta$ for the PLL liquid. We also checked this relation numerically, where at small frequencies and especially for lower values of $\nu_k$ there are corrections to the analytical formula coming from the $m$ and $q$ terms in the bulk Dirac equation. We thus explored the mass parameter space between $(-1/2, 1/2)$ to find the values of $m$ that bring the ratio closest to the expected value of $\pi$ even at lower $\nu_k$. We find that this happens as we approach the lower limit of allowed mass $m = -1/2$. However, even in this limit, the temperature prefactor is still lower than the expected value for all dopings, as seen from the red line in Fig.\ \ref{fig:asymptotic_ratio}.

\begin{figure}[h]\centering
  \begin{subfigure}{.5\textwidth}
    \centering
    \includegraphics[width=\linewidth]{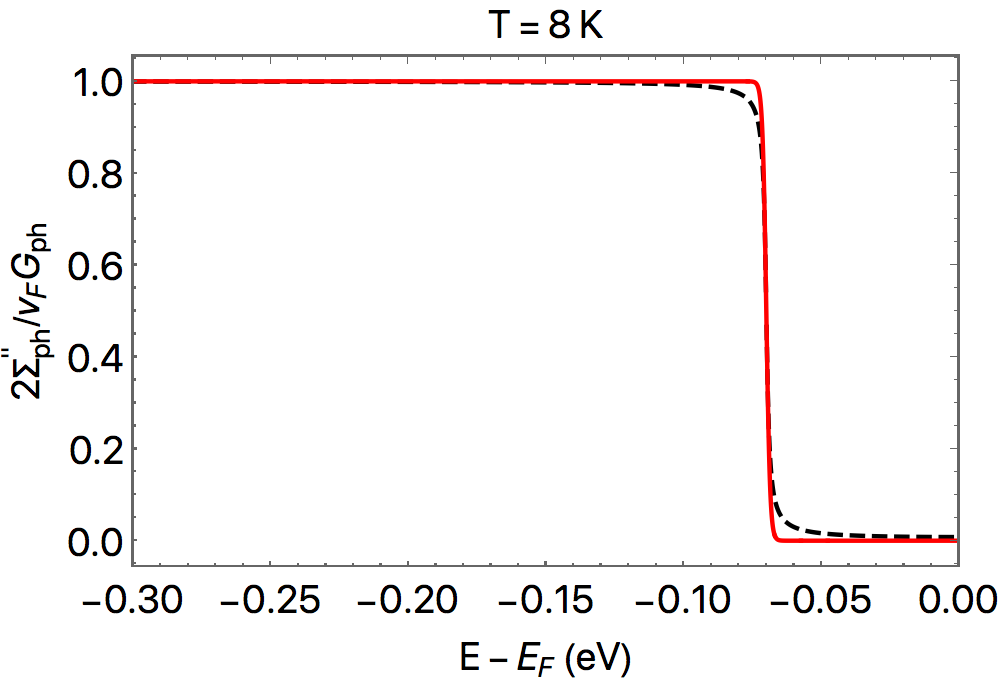}
  \end{subfigure}%
\begin{subfigure}{.5\textwidth}
  \centering
  \includegraphics[width=\linewidth]{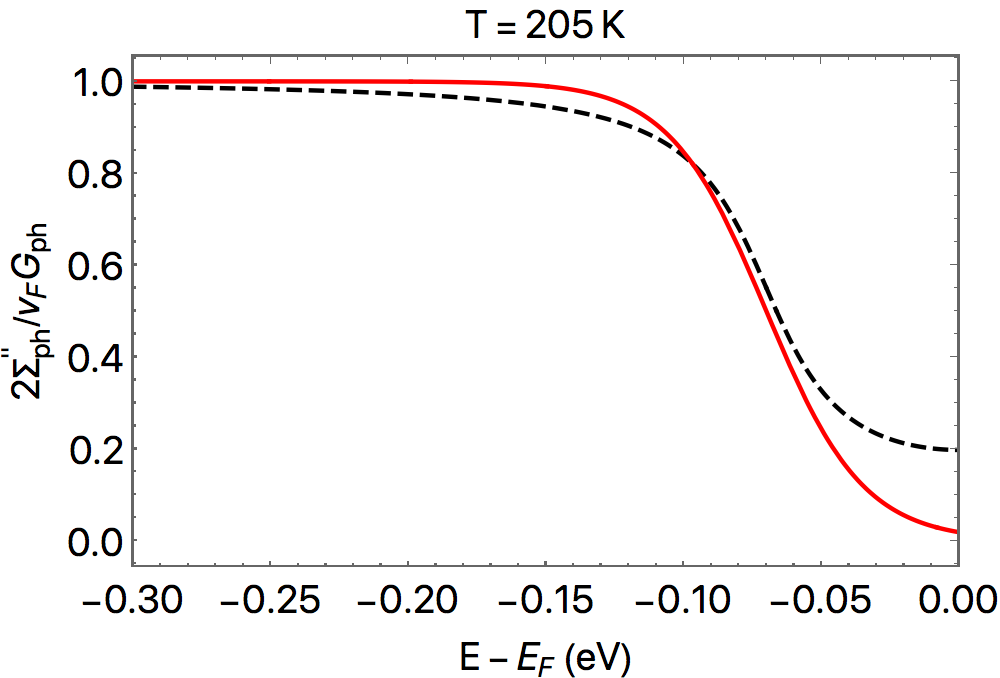}
\end{subfigure}
\caption{\label{fig:phonon_fermi_dirac_comparison} Comparison between the Fermi-Dirac distribution (\textcolor{red}{red solid line}) and the imaginary part of the approximation in Eq.\ \eqref{eq:phonon_nonzero_t} (black dashed line) with $\Omega = 0$.}
\end{figure}

\begin{figure}[h]\centering
    \includegraphics[width=0.7\linewidth]{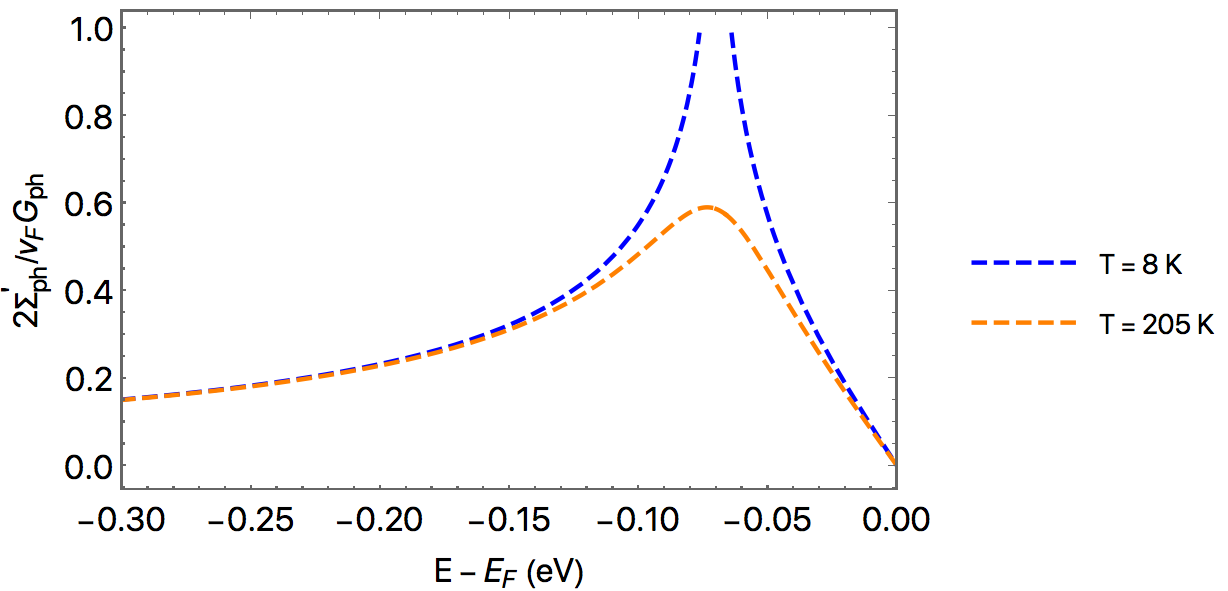}
\caption{\label{fig:phonon_real} Plot of the real part of the approximation to the electron-phonon contribution in Eq.\ \eqref{eq:phonon_nonzero_t}, with $\Omega = 0$. We can see a divergence at the phonon frequency $\hbar\omega_\text{ph} = 0.07\text{eV}$ as we approach zero temperature.}
\end{figure}

\begin{figure}[h!]\centering
  \begin{subfigure}{.5\textwidth}
    \centering
    \includegraphics[width=\linewidth]{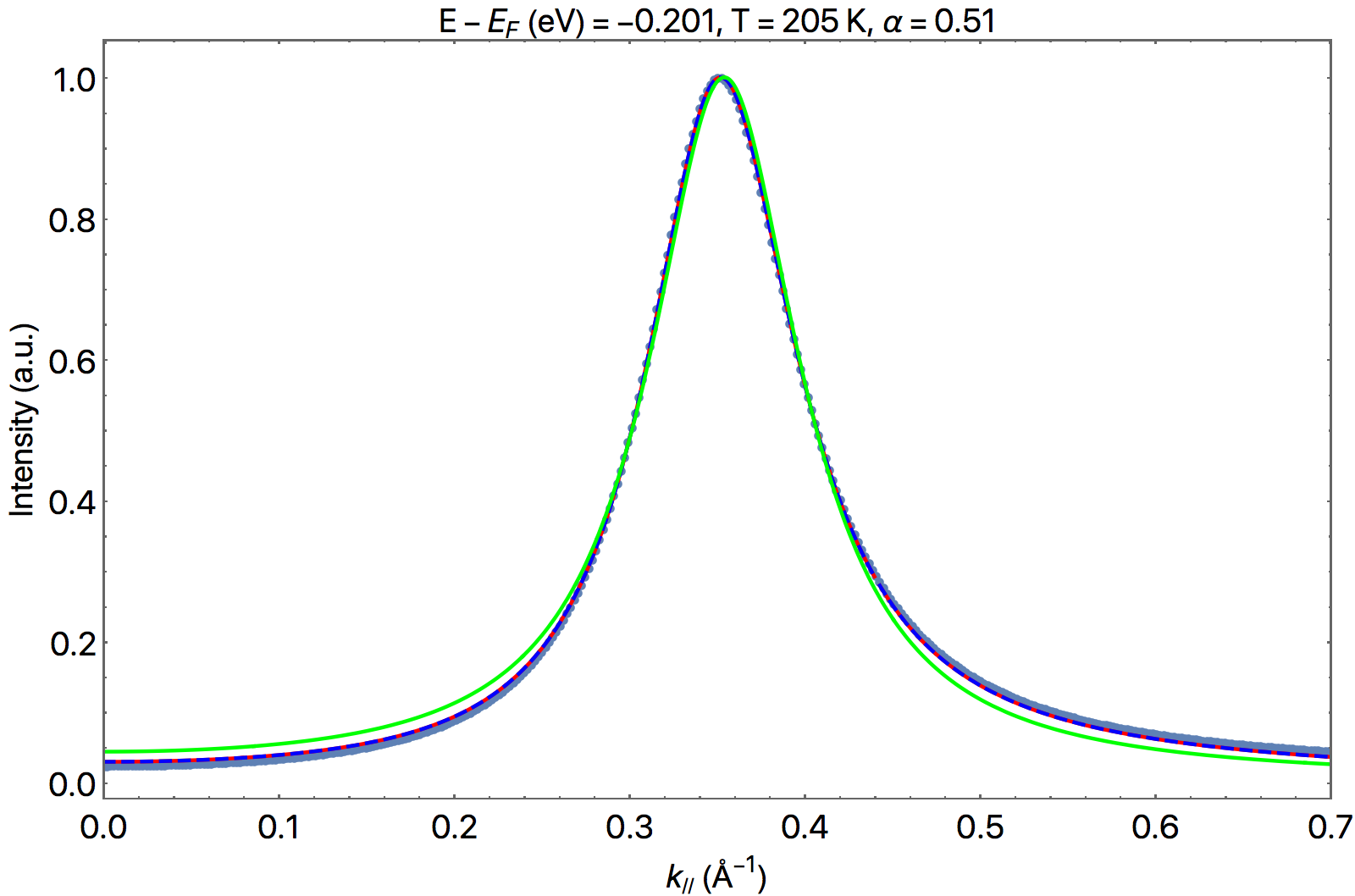}
  \end{subfigure}%
\begin{subfigure}{.5\textwidth}
  \centering
  \includegraphics[width=\linewidth]{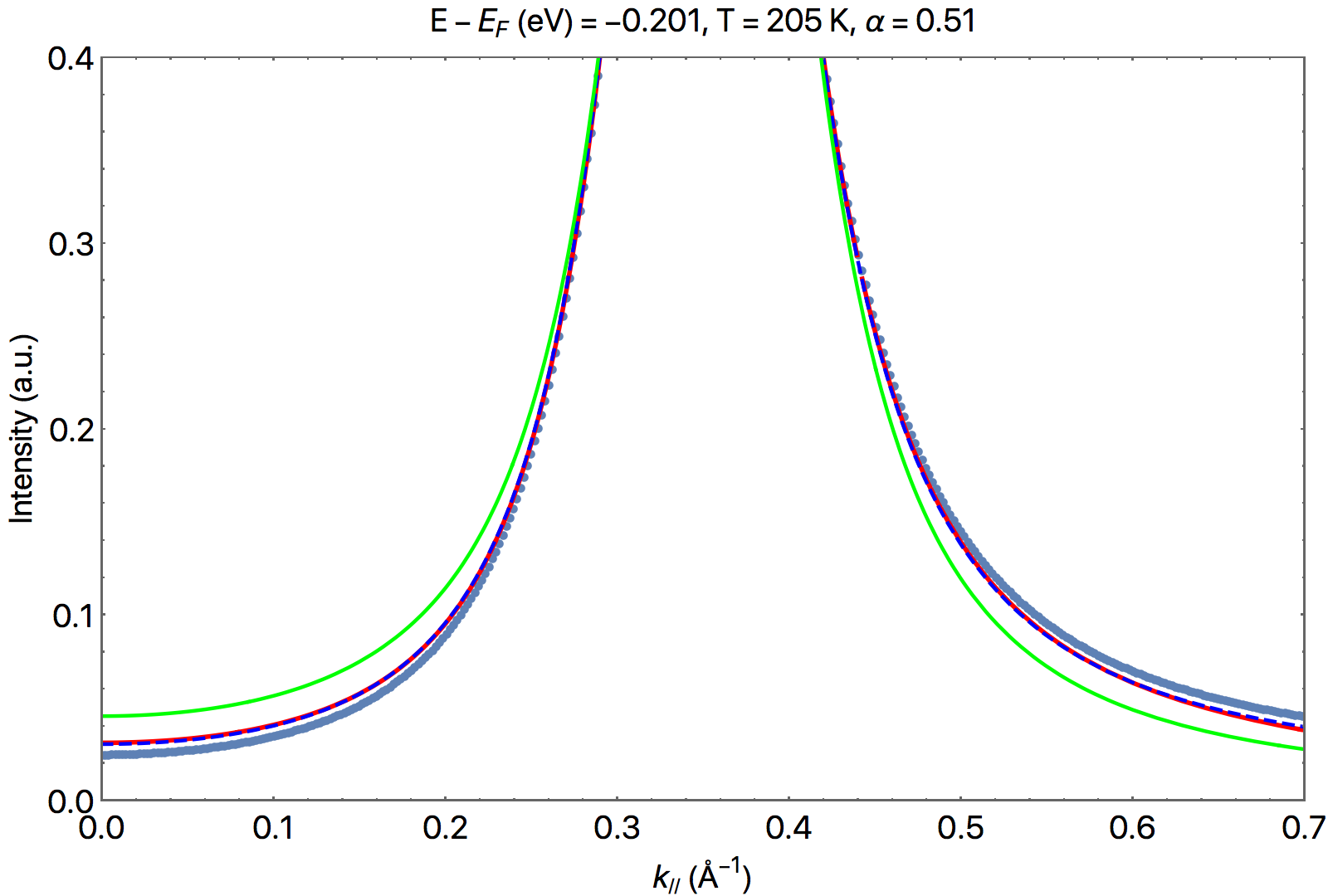}
\end{subfigure}
\caption{\label{fig:spectral_opD_205K} Comparison of the PLL fit (\textcolor{green}{green line}) and semi-holographic fit (\textcolor{red}{red line}) to MDC data (\textcolor{blue}{blue dots}) for an optimally-doped sample $\alpha = 0.51$ at high temperature $T = 205K$. Here, also added for comparison, is the ``holographic inspired'' model of Ref.\ \cite{Smit2021} (\textcolor{blue}{blue line}). We see that the semi-holographic model still provides a better fit than the PLL to the asymmetric peak far from the Fermi surface, however, this implies a $G_0(\omega)$ that cannot be simply described by the electron-phonon model as shown in Fig.\ \ref{fig:dispersion_205K} below.}
\end{figure}

\begin{figure}[h!]\centering
  \begin{subfigure}{.5\textwidth}
    \centering
    \includegraphics[width=\linewidth]{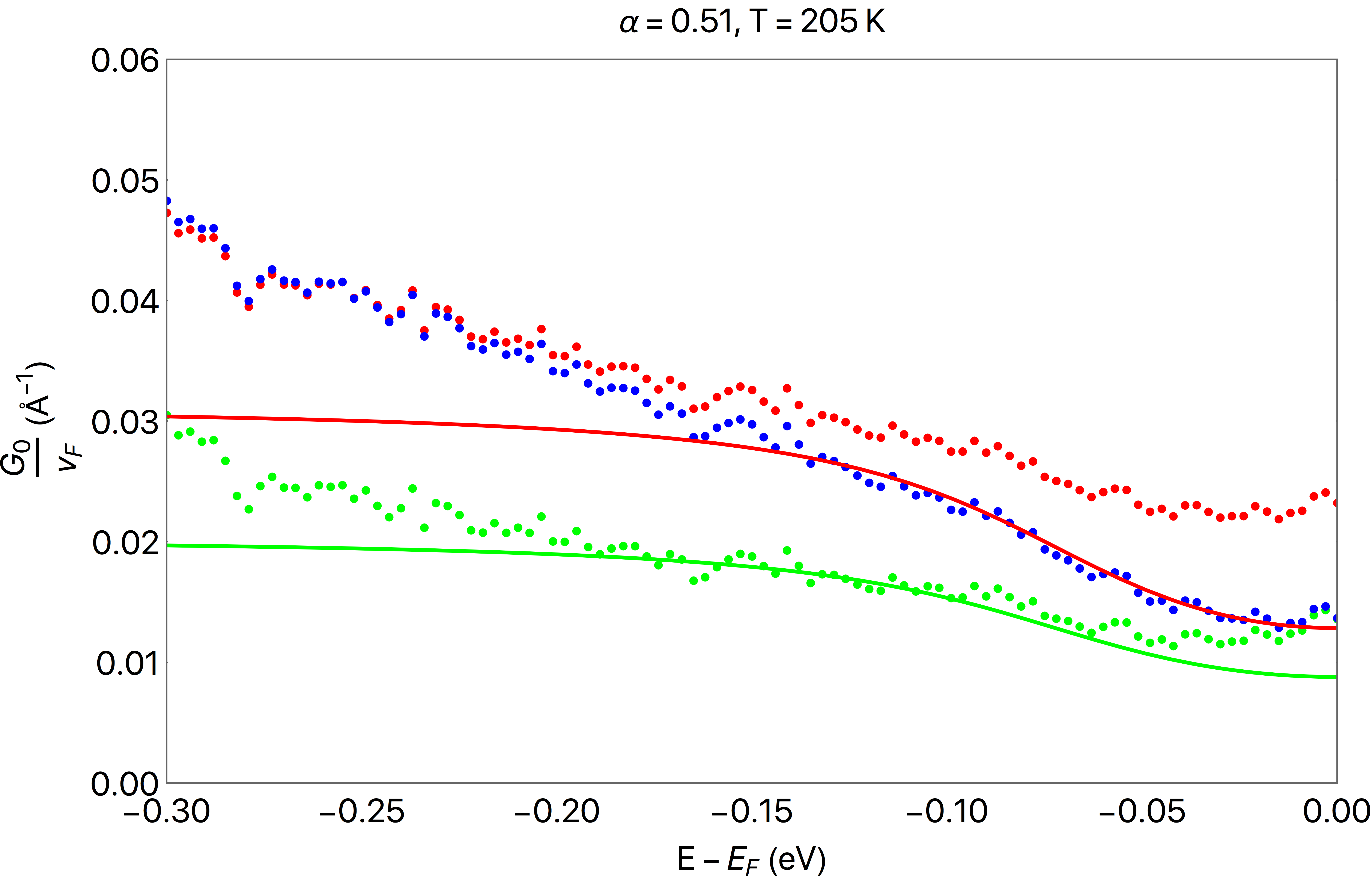}
  \end{subfigure}%
\begin{subfigure}{.5\textwidth}
  \centering
  \includegraphics[width=\linewidth]{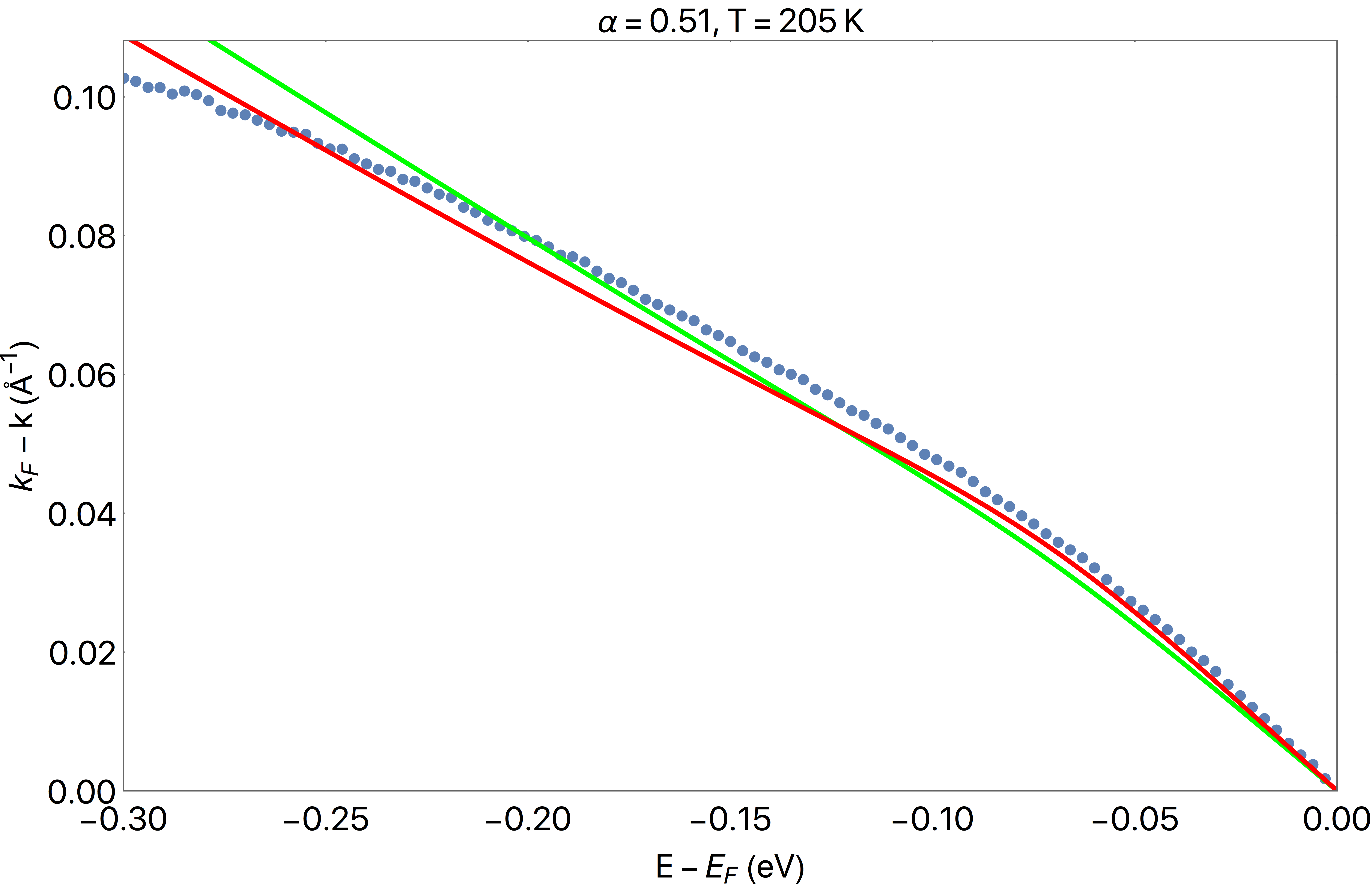}
\end{subfigure}
\caption{\label{fig:dispersion_205K} (Left) Results for  $G_0(\omega)$ in Eq.\ \eqref{eq:full_fit_function} from the fit of the MDC data to a PLL fit function (\textcolor{green}{green dots}), the ``semi-holographic inspired model'' of Ref.\ \cite{Smit2021}, (\textcolor{blue}{blue dots}), and the semi-holographic model (\textcolor{red}{red dots}) and comparison with the prediction from the high-temperature generalization of the electron-phonon model (solid lines) as in Eq.\ \eqref{eq:phonon_nonzero_t}, with the parameter $G_{\text{ph}}$ and $\Omega$ as obtained from the fit at low-temperature. While we see that in all three cases away from the Fermi surface there are contributions to $G_0(\omega)$ that are not explained by the simple model of electron-phonon interactions, the semi-holographic model also shows deviations near the Fermi surface, as expected, due to the different prediction of the temperature behavior of the electron self-energy. The deviations from the electron-phonon model are also evident in the discrepancy between the expected and measured dispersion (right).}
\end{figure}

Given this premise, we then do not expect the semi-holographic model to properly describe the temperature behavior, in fact, if we keep the coupling $g_k$ fixed to the value found for the low-temperature case, we should find that the model underestimates the width of the ARPES peak for $\hbar\omega \lesssim k_B T$ compared to the prediction from the PLL. By repeating the fitting procedure explained above, we find that the semi-holographic model still accurately describes the asymmetric peak shape in the MDCs, as shown in Fig.\ \ref{fig:spectral_opD_205K}. However, $G_0(\omega)$ departs from the expected high-temperature generalization of the electron-phonon self-energy. 
We generalize Eq.\ \eqref{eq:phonon} to non-zero temperature by requiring that in the limit $\Omega \to 0$ the imaginary part of the model gives a good approximation of the Fermi-Dirac distribution, that is 
\begin{align}\label{eq:phonon_nonzero_t}
  \frac{\Sigma_{\text{ph}}}{v_F} = \frac{G_{\text{ph}}}{2\pi} \log\left(\frac{\hbar\omega - \hbar\omega_\text{ph} - i (\hbar\Omega + 4 k_B T/\pi)}{\hbar\omega - \hbar\omega_\text{ph} + i (\hbar\Omega + 4 k_B T/\pi)}\right) \text{ ,}
\end{align}
where the values for $G_{\text{ph}}$ and $\Omega$ are kept fixed to the ones at $T = 0$.
In Fig.\ \ref{fig:phonon_fermi_dirac_comparison} we show the comparison between the imaginary part of this approximation and a Fermi-Dirac distribution, while in Fig.\ \ref{fig:phonon_real} we show the effect of temperature on the divergence in the real part at the phonon frequency. This generalization gives a good description of the dispersion at higher temperatures as it can be seen in the right panel of Fig.\ \ref{fig:dispersion_205K}, which compares to the experimental data at $T = 205 K$.
The contribution to $\Sigma''(\omega, k)$ that does not come from the electron-electron contribution to the self-energy is, however, larger than what is predicted by the simple addition of the electron-phonon interaction, as shown in the left panel of Fig.\ \ref{fig:dispersion_205K}. 
Here, we also make the comparison with the ``semi-holographic inspired'' model mentioned above and presented in Ref.\ \cite{Smit2021} (blue line in Fig.\ \ref{fig:spectral_opD_205K} and blue dots in \ref{fig:dispersion_205K}). 
This difference between $G_0(\omega)$ and the expected electron-phonon contribution might signal a shortcoming of the semi-holographic model considered in this paper in describing the non-zero temperature behavior of the cuprate strange metal, hinting at the necessity of searching for other models in the large class of $z = \infty$ holographic theory with a temperature behavior that more closely resembles the experimental behavior. 
On the other hand, it might also be that a proper description at non-zero temperature must take into account contributions to the self-energy, other than the phonon, that are activated at non-zero temperature, contributing to $G_0(\omega)$, or corrections to the coupling $g_k \to g_k (1 + c_1 T + \dots)$, that we did not consider here. While we believe in the importance of pointing out a possible shortcoming here, we leave a deeper analysis and identification of viable resolutions to future studies. 

\section{Conclusions}

The main aim of this paper is to provide a simple phenomenological model that can describe the momentum distribution curves measured in high-resolution angle-resolved photoemission spectroscopy experiments on the strange metal phase on optimally-doped and overdoped single-layer cuprates. In particular, we propose a model coming from a semi-holographic theory based on the Gubser-Rocha model of a non-Fermi liquid, and test it on recent experimental data \cite{Smit2021}. We find that such a model plus an electron-phonon contribution very accurately describes the behavior of the observed spectral functions in the form of the momentum distribution curves (MDCs), and provides an improvement over the previously proposed power-law-liquid (PLL) model in the low-temperature limit, since:
\begin{itemize}
    \item it gives a better fit to the MDC peaks over a much larger range of energy below the Fermi surface. This is because it captures an asymmetry observed in the experimental MDCs that shifts more spectral weight towards the tail of the distribution for $\abs{k} > \abs{k_*}$, and it reduces to the PLL result near the Fermi energy where this asymmetry is too small to be observed. While there could be some alternative, non-conventional explanation of this asymmetry, we argue in favor of it coming from properties of the electron self-energy, as other possible simple causes have been analyzed and ruled out in Ref.\ \cite{Smit2021}.
    
    \item it is compatible with a simple model of the electron-phonon interaction that well describes---to a better degree than in the power-law liquid---both the observed dispersion relation and the contributions to the self-energy other than that from electron-electron interactions. 
\end{itemize}

The most practically important implication of these results is that, with just a few fit parameters, we can provide a phenomenological description, across a large range of energies and dopings, of low-temperature momentum distribution curves in cuprates along the nodal direction. This description can be used as a benchmark to compare various theories on the self-energy in the strongly interacting cuprates. Up to a normalization we, in fact, have that 
\begin{align}\label{eq:full_fit_function_conclusions}
    \mathcal{A}(\omega, k) \propto \text{Im}\left[\frac{1}{k - (k_F - \frac{\omega}{\textcolor{red}{v_F}}) - i \frac{\lambda f_k \omega_N}{2}\left[\left(\frac{\omega}{\omega_N}\right)^2\right]^{\alpha(1 - (k - k_F)/k_F)} +  \frac{\textcolor{red}{G_{\text{ph}}}}{2 \pi} \log\left(\frac{\omega - \omega_\text{ph} - i \textcolor{red}{\Omega}}{\omega + \omega_\text{ph} + i \textcolor{red}{\Omega}}\right) } \right] \text{ ,}
\end{align}
where $f_k$ is given in Eq.\ \eqref{eq:f_k_semiholo}. The parameter $\lambda$ and the exponent $\alpha$ are fixed to the PLL values at low energies, and the only fit parameters for each MDC are highlighted in red. Remember that these are constant \textbf{parameters}, in contrast with the fit \textbf{functions} $k_*(\omega)$ and $G_0(\omega)$ as in Eq.\ \eqref{eq:full_fit_function}. 
We stress especially that---while our model is rooted in the holographic duality and we believe these results provide good arguments in favor of the use of this technique in the description of the non-Fermi liquid---the validity of the model is independent of its holographic origin. We note that momentum-dependent scaling exponents have also been predicted in a one-dimensional nonlinear-Luttinger liquid model \cite{Jin2019}, hinting, perhaps, at the emergence of one-dimensional physics governing the electron response along the nodal direction in copper-oxide layers.

We have also shown the shortcomings of the semi-holographic model adopted in this paper, in the fact that it underestimates the temperature contribution to the self-energy. However, our model is based on the specific choice of the Gubser-Rocha dual gravitational theory, which was simply dictated by the fact that it is perhaps the most simple model among the ones proposed for the description of the strange metal phase, and it allows for an analytical solution of the gravitational background. We find it already remarkable that it gives such an accurate description at low temperatures, and a not-so-large discrepancy in the temperature dependence might simply point to the need for a more refined dual gravitational theory among the large class of conformal-to-AdS$_2$ metals. The differences observed at elevated temperatures might also be due to the need for a temperature-dependent coupling $g_k^2(T)$ in Eq.\ \eqref{eq:f_k_semiholo}. This could be the topic of further research, although this would go against our main goal of providing a simple phenomenological model with as few adjustable parameters as possible. 
Moreover, as we have shown in Fig.\ \ref{fig:dispersion_205K}, at high temperatures it appears that there are contributions to $G_0(\omega)$ that cannot simply be explained within the electron-phonon interaction, pointing to the fact that we are possibly ignoring some other effects that might change the analysis of the results if included. This is also a matter that requires further investigation in future studies. 

Finally, we are careful not to claim that the momentum-dependence in the scaling exponent of the self-energy is the only possible explanation of the observed asymmetry, as there could be other factors that might generate this shift in the spectral weight, some of which, though, are carefully analyzed and ruled out in Ref.\ \cite{Smit2021}. Moreover, it is important to understand that, while the asymmetry could be accounted for by simply adding an additional momentum dependence as an extra parameter in the phenomenological PLL model of the electron self-energy, the model presented here \textbf{predicts} the form of this momentum dependence, and it is far from trivial that MDC experimental results across a large range of energies can be so well described by this function, without adding any additional adjustable parameter to the PLL model.

\section*{Acknowledgments}
This work is supported by the Stichting voor Fundamenteel Onderzoek der Materie (FOM) and is part of the D-ITP consortium, a program of the Netherlands Organization for Scientific
Research (NWO) that is funded by the Dutch Ministry
of Education, Culture and Science (OCW).

\appendix*
\section{Limitations of the semi-holographic approach for a consistent electronic sum rule}
In this appendix, we show that the construction proposed in Ref.\ \cite{Gursoy2012} to compute the single-particle fermionic Green's function from holography, does not allow to model the data within the approach used in this paper. In Ref.\ \cite{Gursoy2012}, it was shown that by writing down an action coupling the fermion with the full holographic theory, and not only the IR emergent sector, we find
\begin{align}\label{eq:umutsemiholography}
    G_{\chi\chi}(\omega,k) = \frac{\hbar}{-\hbar\omega + \epsilon(k) - \mu + g_k^2 G_H^{-1}(\omega,k)} \underset{\omega \ll \mu}{\approx} \frac{\hbar}{-\hbar\omega + \tilde \epsilon(k) - \mu - \tilde g_k^2 \mathcal{G}_{k}(\omega,k)} \text{ ,}
\end{align}
and we, hence, obtain a spectral function that satisfies the electronic sum rule for masses in the range $(-1/2, 1/2)$. In fact, in this mass range, the contribution from the holographic Green's function is subleading at high energies where Eq.\ \eqref{eq:umutsemiholography} reduces to that of a free fermion so that the electronic sum rule for $G_{\chi\chi}$ is satisfied, pointing to the possibility  that this construction could be used to obtain in a consistent way the spectral function for a single electron in a strongly interacting theory at all energies. Notice, however, that in this way the non-universal ultraviolet physics of the holographic theory enters into the spectral function, modifying the dispersion relation and, as we will shortly show, imposing a lower limit on the lifetime of the excitations near the Fermi surface. Combining Eqs. \eqref{eq:holographic_green_fermi} and \eqref{eq:umutsemiholography} we can see in fact that the Green's function takes the low-energy form
\begin{align}\label{eq:low_energy_fermion}
\begin{split}
   G_{\chi\chi}(\omega,k) &\simeq  \frac{\hbar}{-\hbar\omega(1 + g_k^2/Z) + (\hbar v_B + g_k^2 \hbar v_H/Z)(k - k_F) - i g_k^2 \mathcal{C} \text{Im}[\mathcal{G}_{k}(\omega)]}\\
   \text{Im}[G_{\chi\chi}(\omega,k)] &\simeq \frac{\tilde Z}{v_F} \frac{\Gamma(\omega, k)/2}{(k - k_*(\omega))^2 + \Gamma(\omega, k)^2/4}\text{ ,}
\end{split}
\end{align}
where $\tilde Z = Z/(Z + g_k^2)$, $k_*(\omega) = k_F + \omega/v_F$, $v_F(k) = v_B \frac{1 + g_k^2 v_B/Z v_H}{1 + g_k^2/Z}$, $v_B$ is the bare Fermi velocity, and $\Gamma(\omega, k) = \frac{g_k^2}{v_F (1 + g_k^2/Z)} \mathcal{C} \text{Im}[\mathcal{G}_{k}(\omega)]$.
We thus see that the lifetime of the electronic excitations as a function of the adjustable parameter $g_k^2$ is bounded from below, with a minimum as we approach the holographic result for $\lim_{g_k \to \infty} G_{\chi\chi}(\omega,k) \propto G_H(\omega,k)$. In this limit, given the small value of $Z$ as shown in Fig.\ \ref{fig:fit_green_parameters}, the holographic prediction gives very sharp peaks near the Fermi surface, and cannot then offer a quantitative description of the much broader peaks measured in ARPES experiments. This is an indication that the holographic theory considered here does not provide a proper description of the electronic excitations at all energies and the UV completion of the theory, influencing the low-energy behavior through the values of $Z$, $k_F$, and $v_F$ has to be modified if we want to write down a full theory of the cuprate strange metal valid across the entire energy range observed. This is, however, outside the scope of our present work.

\bibliography{apssamp}

\providecommand{\noopsort}[1]{}\providecommand{\singleletter}[1]{#1}%
\begin{thebibliography}{52}%
\makeatletter
\providecommand \@ifxundefined [1]{%
 \@ifx{#1\undefined}
}%
\providecommand \@ifnum [1]{%
 \ifnum #1\expandafter \@firstoftwo
 \else \expandafter \@secondoftwo
 \fi
}%
\providecommand \@ifx [1]{%
 \ifx #1\expandafter \@firstoftwo
 \else \expandafter \@secondoftwo
 \fi
}%
\providecommand \natexlab [1]{#1}%
\providecommand \enquote  [1]{``#1''}%
\providecommand \bibnamefont  [1]{#1}%
\providecommand \bibfnamefont [1]{#1}%
\providecommand \citenamefont [1]{#1}%
\providecommand \href@noop [0]{\@secondoftwo}%
\providecommand \href [0]{\begingroup \@sanitize@url \@href}%
\providecommand \@href[1]{\@@startlink{#1}\@@href}%
\providecommand \@@href[1]{\endgroup#1\@@endlink}%
\providecommand \@sanitize@url [0]{\catcode `\\12\catcode `\$12\catcode `\&12\catcode `\#12\catcode `\^12\catcode `\_12\catcode `\%12\relax}%
\providecommand \@@startlink[1]{}%
\providecommand \@@endlink[0]{}%
\providecommand \url  [0]{\begingroup\@sanitize@url \@url }%
\providecommand \@url [1]{\endgroup\@href {#1}{\urlprefix }}%
\providecommand \urlprefix  [0]{URL }%
\providecommand \Eprint [0]{\href }%
\providecommand \doibase [0]{https://doi.org/}%
\providecommand \selectlanguage [0]{\@gobble}%
\providecommand \bibinfo  [0]{\@secondoftwo}%
\providecommand \bibfield  [0]{\@secondoftwo}%
\providecommand \translation [1]{[#1]}%
\providecommand \BibitemOpen [0]{}%
\providecommand \bibitemStop [0]{}%
\providecommand \bibitemNoStop [0]{.\EOS\space}%
\providecommand \EOS [0]{\spacefactor3000\relax}%
\providecommand \BibitemShut  [1]{\csname bibitem#1\endcsname}%
\let\auto@bib@innerbib\@empty
\bibitem [{\citenamefont {Bednorz}\ and\ \citenamefont {M{\"u}ller}(1986)}]{Bednorz1986}%
  \BibitemOpen
  \bibfield  {author} {\bibinfo {author} {\bibfnamefont {J.~G.}\ \bibnamefont {Bednorz}}\ and\ \bibinfo {author} {\bibfnamefont {K.~A.}\ \bibnamefont {M{\"u}ller}},\ }\bibfield  {title} {\bibinfo {title} {Possible {high-$T_c$} superconductivity in the {Ba{\textendash}La{\textendash}Cu{\textendash}O} system},\ }\href {https://doi.org/10.1007/BF01303701} {\bibfield  {journal} {\bibinfo  {journal} {Zeitschrift f{\"u}r Physik B Condensed Matter}\ }\textbf {\bibinfo {volume} {64}},\ \bibinfo {pages} {189} (\bibinfo {year} {1986})}\BibitemShut {NoStop}%
\bibitem [{\citenamefont {Bardeen}\ \emph {et~al.}(1957)\citenamefont {Bardeen}, \citenamefont {Cooper},\ and\ \citenamefont {Schrieffer}}]{Bardeen1957}%
  \BibitemOpen
  \bibfield  {author} {\bibinfo {author} {\bibfnamefont {J.}~\bibnamefont {Bardeen}}, \bibinfo {author} {\bibfnamefont {L.~N.}\ \bibnamefont {Cooper}},\ and\ \bibinfo {author} {\bibfnamefont {J.~R.}\ \bibnamefont {Schrieffer}},\ }\bibfield  {title} {\bibinfo {title} {Theory of superconductivity},\ }\href {https://doi.org/10.1103/PhysRev.108.1175} {\bibfield  {journal} {\bibinfo  {journal} {Phys. Rev.}\ }\textbf {\bibinfo {volume} {108}},\ \bibinfo {pages} {1175} (\bibinfo {year} {1957})}\BibitemShut {NoStop}%
\bibitem [{\citenamefont {Keimer}\ \emph {et~al.}(2015)\citenamefont {Keimer}, \citenamefont {Kivelson}, \citenamefont {Norman}, \citenamefont {Uchida},\ and\ \citenamefont {Zaanen}}]{Keimer2015}%
  \BibitemOpen
  \bibfield  {author} {\bibinfo {author} {\bibfnamefont {B.}~\bibnamefont {Keimer}}, \bibinfo {author} {\bibfnamefont {S.~A.}\ \bibnamefont {Kivelson}}, \bibinfo {author} {\bibfnamefont {M.~R.}\ \bibnamefont {Norman}}, \bibinfo {author} {\bibfnamefont {S.}~\bibnamefont {Uchida}},\ and\ \bibinfo {author} {\bibfnamefont {J.}~\bibnamefont {Zaanen}},\ }\bibfield  {title} {\bibinfo {title} {From quantum matter to high-temperature superconductivity in copper oxides},\ }\href {https://doi.org/10.1038/nature14165} {\bibfield  {journal} {\bibinfo  {journal} {Nature}\ }\textbf {\bibinfo {volume} {518}},\ \bibinfo {pages} {179} (\bibinfo {year} {2015})}\BibitemShut {NoStop}%
\bibitem [{\citenamefont {Greene}\ \emph {et~al.}(2020)\citenamefont {Greene}, \citenamefont {Mandal}, \citenamefont {Poniatowski},\ and\ \citenamefont {Sarkar}}]{Greene2020}%
  \BibitemOpen
  \bibfield  {author} {\bibinfo {author} {\bibfnamefont {R.~L.}\ \bibnamefont {Greene}}, \bibinfo {author} {\bibfnamefont {P.~R.}\ \bibnamefont {Mandal}}, \bibinfo {author} {\bibfnamefont {N.~R.}\ \bibnamefont {Poniatowski}},\ and\ \bibinfo {author} {\bibfnamefont {T.}~\bibnamefont {Sarkar}},\ }\bibfield  {title} {\bibinfo {title} {The strange metal state of the electron-doped cuprates},\ }\href {https://doi.org/10.1146/annurev-conmatphys-031119-050558} {\bibfield  {journal} {\bibinfo  {journal} {Annual Review of Condensed Matter Physics}\ }\textbf {\bibinfo {volume} {11}},\ \bibinfo {pages} {213} (\bibinfo {year} {2020})},\ \Eprint {https://arxiv.org/abs/https://doi.org/10.1146/annurev-conmatphys-031119-050558} {https://doi.org/10.1146/annurev-conmatphys-031119-050558} \BibitemShut {NoStop}%
\bibitem [{\citenamefont {Lee}\ \emph {et~al.}(2006)\citenamefont {Lee}, \citenamefont {Nagaosa},\ and\ \citenamefont {Wen}}]{Lee2006}%
  \BibitemOpen
  \bibfield  {author} {\bibinfo {author} {\bibfnamefont {P.~A.}\ \bibnamefont {Lee}}, \bibinfo {author} {\bibfnamefont {N.}~\bibnamefont {Nagaosa}},\ and\ \bibinfo {author} {\bibfnamefont {X.-G.}\ \bibnamefont {Wen}},\ }\bibfield  {title} {\bibinfo {title} {Doping a mott insulator: Physics of high-temperature superconductivity},\ }\href {https://doi.org/10.1103/RevModPhys.78.17} {\bibfield  {journal} {\bibinfo  {journal} {Rev. Mod. Phys.}\ }\textbf {\bibinfo {volume} {78}},\ \bibinfo {pages} {17} (\bibinfo {year} {2006})}\BibitemShut {NoStop}%
\bibitem [{\citenamefont {Scalapino}(2012)}]{Scalapino2012}%
  \BibitemOpen
  \bibfield  {author} {\bibinfo {author} {\bibfnamefont {D.~J.}\ \bibnamefont {Scalapino}},\ }\bibfield  {title} {\bibinfo {title} {A common thread: The pairing interaction for unconventional superconductors},\ }\href {https://doi.org/10.1103/RevModPhys.84.1383} {\bibfield  {journal} {\bibinfo  {journal} {Rev. Mod. Phys.}\ }\textbf {\bibinfo {volume} {84}},\ \bibinfo {pages} {1383} (\bibinfo {year} {2012})}\BibitemShut {NoStop}%
\bibitem [{\citenamefont {Chien}\ \emph {et~al.}(1991)\citenamefont {Chien}, \citenamefont {Wang},\ and\ \citenamefont {Ong}}]{Chen1991}%
  \BibitemOpen
  \bibfield  {author} {\bibinfo {author} {\bibfnamefont {T.~R.}\ \bibnamefont {Chien}}, \bibinfo {author} {\bibfnamefont {Z.~Z.}\ \bibnamefont {Wang}},\ and\ \bibinfo {author} {\bibfnamefont {N.~P.}\ \bibnamefont {Ong}},\ }\bibfield  {title} {\bibinfo {title} {Effect of zn impurities on the normal-state hall angle in single-crystal ${\mathrm{yba}}_{2}$${\mathrm{cu}}_{3\mathrm{\ensuremath{-}}\mathit{x}}$${\mathrm{zn}}_{\mathit{x}}$${\mathrm{o}}_{7\mathrm{\ensuremath{-}}\mathrm{\ensuremath{\delta}}}$},\ }\href {https://doi.org/10.1103/PhysRevLett.67.2088} {\bibfield  {journal} {\bibinfo  {journal} {Phys. Rev. Lett.}\ }\textbf {\bibinfo {volume} {67}},\ \bibinfo {pages} {2088} (\bibinfo {year} {1991})}\BibitemShut {NoStop}%
\bibitem [{\citenamefont {Custers}\ \emph {et~al.}(2003)\citenamefont {Custers}, \citenamefont {Gegenwart}, \citenamefont {Wilhelm}, \citenamefont {Neumaier}, \citenamefont {Tokiwa}, \citenamefont {Trovarelli}, \citenamefont {Geibel}, \citenamefont {Steglich}, \citenamefont {P{\'{e}}pin},\ and\ \citenamefont {Coleman}}]{Custers2003}%
  \BibitemOpen
  \bibfield  {author} {\bibinfo {author} {\bibfnamefont {J.}~\bibnamefont {Custers}}, \bibinfo {author} {\bibfnamefont {P.}~\bibnamefont {Gegenwart}}, \bibinfo {author} {\bibfnamefont {H.}~\bibnamefont {Wilhelm}}, \bibinfo {author} {\bibfnamefont {K.}~\bibnamefont {Neumaier}}, \bibinfo {author} {\bibfnamefont {Y.}~\bibnamefont {Tokiwa}}, \bibinfo {author} {\bibfnamefont {O.}~\bibnamefont {Trovarelli}}, \bibinfo {author} {\bibfnamefont {C.}~\bibnamefont {Geibel}}, \bibinfo {author} {\bibfnamefont {F.}~\bibnamefont {Steglich}}, \bibinfo {author} {\bibfnamefont {C.}~\bibnamefont {P{\'{e}}pin}},\ and\ \bibinfo {author} {\bibfnamefont {P.}~\bibnamefont {Coleman}},\ }\bibfield  {title} {\bibinfo {title} {The break-up of heavy electrons at a quantum critical point},\ }\href {https://doi.org/10.1038/nature01774} {\bibfield  {journal} {\bibinfo  {journal} {Nature}\ }\textbf {\bibinfo {volume} {424}},\ \bibinfo {pages} {524} (\bibinfo {year} {2003})}\BibitemShut {NoStop}%
\bibitem [{\citenamefont {Cooper}\ \emph {et~al.}(2009)\citenamefont {Cooper}, \citenamefont {Wang}, \citenamefont {Vignolle}, \citenamefont {Lipscombe}, \citenamefont {Hayden}, \citenamefont {Tanabe}, \citenamefont {Adachi}, \citenamefont {Koike}, \citenamefont {Nohara}, \citenamefont {Takagi}, \citenamefont {Proust},\ and\ \citenamefont {Hussey}}]{Cooper603}%
  \BibitemOpen
  \bibfield  {author} {\bibinfo {author} {\bibfnamefont {R.~A.}\ \bibnamefont {Cooper}}, \bibinfo {author} {\bibfnamefont {Y.}~\bibnamefont {Wang}}, \bibinfo {author} {\bibfnamefont {B.}~\bibnamefont {Vignolle}}, \bibinfo {author} {\bibfnamefont {O.~J.}\ \bibnamefont {Lipscombe}}, \bibinfo {author} {\bibfnamefont {S.~M.}\ \bibnamefont {Hayden}}, \bibinfo {author} {\bibfnamefont {Y.}~\bibnamefont {Tanabe}}, \bibinfo {author} {\bibfnamefont {T.}~\bibnamefont {Adachi}}, \bibinfo {author} {\bibfnamefont {Y.}~\bibnamefont {Koike}}, \bibinfo {author} {\bibfnamefont {M.}~\bibnamefont {Nohara}}, \bibinfo {author} {\bibfnamefont {H.}~\bibnamefont {Takagi}}, \bibinfo {author} {\bibfnamefont {C.}~\bibnamefont {Proust}},\ and\ \bibinfo {author} {\bibfnamefont {N.~E.}\ \bibnamefont {Hussey}},\ }\bibfield  {title} {\bibinfo {title} {Anomalous criticality in the electrical resistivity of {La2{\textendash}xSrxCuO4}},\ }\href {https://doi.org/10.1126/science.1165015} {\bibfield  {journal} {\bibinfo  {journal} {Science}\
  }\textbf {\bibinfo {volume} {323}},\ \bibinfo {pages} {603} (\bibinfo {year} {2009})}\BibitemShut {NoStop}%
\bibitem [{\citenamefont {Bruin}\ \emph {et~al.}(2013)\citenamefont {Bruin}, \citenamefont {Sakai}, \citenamefont {Perry},\ and\ \citenamefont {Mackenzie}}]{bruin_similarity_2013}%
  \BibitemOpen
  \bibfield  {author} {\bibinfo {author} {\bibfnamefont {J.~A.~N.}\ \bibnamefont {Bruin}}, \bibinfo {author} {\bibfnamefont {H.}~\bibnamefont {Sakai}}, \bibinfo {author} {\bibfnamefont {R.~S.}\ \bibnamefont {Perry}},\ and\ \bibinfo {author} {\bibfnamefont {A.~P.}\ \bibnamefont {Mackenzie}},\ }\bibfield  {title} {\bibinfo {title} {Similarity of {Scattering} {Rates} in {Metals} {Showing} {T}-{Linear} {Resistivity}},\ }\href {https://doi.org/10.1126/science.1227612} {\bibfield  {journal} {\bibinfo  {journal} {Science}\ }\textbf {\bibinfo {volume} {339}},\ \bibinfo {pages} {804} (\bibinfo {year} {2013})}\BibitemShut {NoStop}%
\bibitem [{\citenamefont {Analytis}\ \emph {et~al.}(2014)\citenamefont {Analytis}, \citenamefont {Kuo}, \citenamefont {McDonald}, \citenamefont {Wartenbe}, \citenamefont {Rourke}, \citenamefont {Hussey},\ and\ \citenamefont {Fisher}}]{Analytis2014}%
  \BibitemOpen
  \bibfield  {author} {\bibinfo {author} {\bibfnamefont {J.~G.}\ \bibnamefont {Analytis}}, \bibinfo {author} {\bibfnamefont {H.-H.}\ \bibnamefont {Kuo}}, \bibinfo {author} {\bibfnamefont {R.~D.}\ \bibnamefont {McDonald}}, \bibinfo {author} {\bibfnamefont {M.}~\bibnamefont {Wartenbe}}, \bibinfo {author} {\bibfnamefont {P.~M.~C.}\ \bibnamefont {Rourke}}, \bibinfo {author} {\bibfnamefont {N.~E.}\ \bibnamefont {Hussey}},\ and\ \bibinfo {author} {\bibfnamefont {I.~R.}\ \bibnamefont {Fisher}},\ }\bibfield  {title} {\bibinfo {title} {Transport near a quantum critical point in {BaFe}2({As1}-{xPx})2},\ }\href {https://doi.org/10.1038/nphys2869} {\bibfield  {journal} {\bibinfo  {journal} {Nature Physics}\ }\textbf {\bibinfo {volume} {10}},\ \bibinfo {pages} {194} (\bibinfo {year} {2014})}\BibitemShut {NoStop}%
\bibitem [{\citenamefont {Legros}\ \emph {et~al.}(2018)\citenamefont {Legros}, \citenamefont {Benhabib}, \citenamefont {Tabis}, \citenamefont {Lalibert{\'{e}}}, \citenamefont {Dion}, \citenamefont {Lizaire}, \citenamefont {Vignolle}, \citenamefont {Vignolles}, \citenamefont {Raffy}, \citenamefont {Li}, \citenamefont {Auban-Senzier}, \citenamefont {Doiron-Leyraud}, \citenamefont {Fournier}, \citenamefont {Colson}, \citenamefont {Taillefer},\ and\ \citenamefont {Proust}}]{Legros2018}%
  \BibitemOpen
  \bibfield  {author} {\bibinfo {author} {\bibfnamefont {A.}~\bibnamefont {Legros}}, \bibinfo {author} {\bibfnamefont {S.}~\bibnamefont {Benhabib}}, \bibinfo {author} {\bibfnamefont {W.}~\bibnamefont {Tabis}}, \bibinfo {author} {\bibfnamefont {F.}~\bibnamefont {Lalibert{\'{e}}}}, \bibinfo {author} {\bibfnamefont {M.}~\bibnamefont {Dion}}, \bibinfo {author} {\bibfnamefont {M.}~\bibnamefont {Lizaire}}, \bibinfo {author} {\bibfnamefont {B.}~\bibnamefont {Vignolle}}, \bibinfo {author} {\bibfnamefont {D.}~\bibnamefont {Vignolles}}, \bibinfo {author} {\bibfnamefont {H.}~\bibnamefont {Raffy}}, \bibinfo {author} {\bibfnamefont {Z.~Z.}\ \bibnamefont {Li}}, \bibinfo {author} {\bibfnamefont {P.}~\bibnamefont {Auban-Senzier}}, \bibinfo {author} {\bibfnamefont {N.}~\bibnamefont {Doiron-Leyraud}}, \bibinfo {author} {\bibfnamefont {P.}~\bibnamefont {Fournier}}, \bibinfo {author} {\bibfnamefont {D.}~\bibnamefont {Colson}}, \bibinfo {author} {\bibfnamefont {L.}~\bibnamefont {Taillefer}},\ and\ \bibinfo {author} {\bibfnamefont
  {C.}~\bibnamefont {Proust}},\ }\bibfield  {title} {\bibinfo {title} {Universal t-linear resistivity and planckian dissipation in overdoped cuprates},\ }\href {https://doi.org/10.1038/s41567-018-0334-2} {\bibfield  {journal} {\bibinfo  {journal} {Nature Physics}\ }\textbf {\bibinfo {volume} {15}},\ \bibinfo {pages} {142} (\bibinfo {year} {2018})}\BibitemShut {NoStop}%
\bibitem [{\citenamefont {Licciardello}\ \emph {et~al.}(2019)\citenamefont {Licciardello}, \citenamefont {Buhot}, \citenamefont {Lu}, \citenamefont {Ayres}, \citenamefont {Kasahara}, \citenamefont {Matsuda}, \citenamefont {Shibauchi},\ and\ \citenamefont {Hussey}}]{Licciardello2019}%
  \BibitemOpen
  \bibfield  {author} {\bibinfo {author} {\bibfnamefont {S.}~\bibnamefont {Licciardello}}, \bibinfo {author} {\bibfnamefont {J.}~\bibnamefont {Buhot}}, \bibinfo {author} {\bibfnamefont {J.}~\bibnamefont {Lu}}, \bibinfo {author} {\bibfnamefont {J.}~\bibnamefont {Ayres}}, \bibinfo {author} {\bibfnamefont {S.}~\bibnamefont {Kasahara}}, \bibinfo {author} {\bibfnamefont {Y.}~\bibnamefont {Matsuda}}, \bibinfo {author} {\bibfnamefont {T.}~\bibnamefont {Shibauchi}},\ and\ \bibinfo {author} {\bibfnamefont {N.~E.}\ \bibnamefont {Hussey}},\ }\bibfield  {title} {\bibinfo {title} {Electrical resistivity across a nematic quantum critical point},\ }\href {https://doi.org/10.1038/s41586-019-0923-y} {\bibfield  {journal} {\bibinfo  {journal} {Nature}\ }\textbf {\bibinfo {volume} {567}},\ \bibinfo {pages} {213} (\bibinfo {year} {2019})}\BibitemShut {NoStop}%
\bibitem [{\citenamefont {Hussey}\ \emph {et~al.}(2018)\citenamefont {Hussey}, \citenamefont {Buhot},\ and\ \citenamefont {Licciardello}}]{Hussey_2018}%
  \BibitemOpen
  \bibfield  {author} {\bibinfo {author} {\bibfnamefont {N.~E.}\ \bibnamefont {Hussey}}, \bibinfo {author} {\bibfnamefont {J.}~\bibnamefont {Buhot}},\ and\ \bibinfo {author} {\bibfnamefont {S.}~\bibnamefont {Licciardello}},\ }\bibfield  {title} {\bibinfo {title} {A tale of two metals: contrasting criticalities in the pnictides and hole-doped cuprates},\ }\href {https://doi.org/10.1088/1361-6633/aaa97c} {\bibfield  {journal} {\bibinfo  {journal} {Reports on Progress in Physics}\ }\textbf {\bibinfo {volume} {81}},\ \bibinfo {pages} {052501} (\bibinfo {year} {2018})}\BibitemShut {NoStop}%
\bibitem [{\citenamefont {Ogata}\ and\ \citenamefont {Fukuyama}(2008)}]{tJ_review}%
  \BibitemOpen
  \bibfield  {author} {\bibinfo {author} {\bibfnamefont {M.}~\bibnamefont {Ogata}}\ and\ \bibinfo {author} {\bibfnamefont {H.}~\bibnamefont {Fukuyama}},\ }\bibfield  {title} {\bibinfo {title} {The {$t–J$} model for the oxide high-{$T_c$} superconductors},\ }\href {https://doi.org/10.1088/0034-4885/71/3/036501} {\bibfield  {journal} {\bibinfo  {journal} {Reports on Progress in Physics}\ }\textbf {\bibinfo {volume} {71}},\ \bibinfo {pages} {036501} (\bibinfo {year} {2008})}\BibitemShut {NoStop}%
\bibitem [{\citenamefont {Mukuda}\ \emph {et~al.}(2012)\citenamefont {Mukuda}, \citenamefont {Shimizu}, \citenamefont {Iyo},\ and\ \citenamefont {Kitaoka}}]{AFM_multilayer}%
  \BibitemOpen
  \bibfield  {author} {\bibinfo {author} {\bibfnamefont {H.}~\bibnamefont {Mukuda}}, \bibinfo {author} {\bibfnamefont {S.}~\bibnamefont {Shimizu}}, \bibinfo {author} {\bibfnamefont {A.}~\bibnamefont {Iyo}},\ and\ \bibinfo {author} {\bibfnamefont {Y.}~\bibnamefont {Kitaoka}},\ }\bibfield  {title} {\bibinfo {title} {High-{$T_c$} superconductivity and antiferromagnetism in multilayered copper oxides –a new paradigm of superconducting mechanism–},\ }\href {https://doi.org/10.1143/JPSJ.81.011008} {\bibfield  {journal} {\bibinfo  {journal} {Journal of the Physical Society of Japan}\ }\textbf {\bibinfo {volume} {81}},\ \bibinfo {pages} {011008} (\bibinfo {year} {2012})}\BibitemShut {NoStop}%
\bibitem [{\citenamefont {Kakehashi}\ and\ \citenamefont {Fulde}(2005)}]{marginal_FL}%
  \BibitemOpen
  \bibfield  {author} {\bibinfo {author} {\bibfnamefont {Y.}~\bibnamefont {Kakehashi}}\ and\ \bibinfo {author} {\bibfnamefont {P.}~\bibnamefont {Fulde}},\ }\bibfield  {title} {\bibinfo {title} {Marginal fermi liquid theory in the hubbard model},\ }\href {https://doi.org/10.1103/PhysRevLett.94.156401} {\bibfield  {journal} {\bibinfo  {journal} {Phys. Rev. Lett.}\ }\textbf {\bibinfo {volume} {94}},\ \bibinfo {pages} {156401} (\bibinfo {year} {2005})}\BibitemShut {NoStop}%
\bibitem [{\citenamefont {Kokalj}\ \emph {et~al.}(2012)\citenamefont {Kokalj}, \citenamefont {Hussey},\ and\ \citenamefont {McKenzie}}]{marginal_FM_2}%
  \BibitemOpen
  \bibfield  {author} {\bibinfo {author} {\bibfnamefont {J.}~\bibnamefont {Kokalj}}, \bibinfo {author} {\bibfnamefont {N.~E.}\ \bibnamefont {Hussey}},\ and\ \bibinfo {author} {\bibfnamefont {R.~H.}\ \bibnamefont {McKenzie}},\ }\bibfield  {title} {\bibinfo {title} {Transport properties of the metallic state of overdoped cuprate superconductors from an anisotropic marginal fermi liquid model},\ }\href {https://doi.org/10.1103/PhysRevB.86.045132} {\bibfield  {journal} {\bibinfo  {journal} {Phys. Rev. B}\ }\textbf {\bibinfo {volume} {86}},\ \bibinfo {pages} {045132} (\bibinfo {year} {2012})}\BibitemShut {NoStop}%
\bibitem [{\citenamefont {Zaanen}\ \emph {et~al.}(1996)\citenamefont {Zaanen}, \citenamefont {Osman}, \citenamefont {Eskes},\ and\ \citenamefont {van Saarloos}}]{Zaanen1996}%
  \BibitemOpen
  \bibfield  {author} {\bibinfo {author} {\bibfnamefont {J.}~\bibnamefont {Zaanen}}, \bibinfo {author} {\bibfnamefont {O.~Y.}\ \bibnamefont {Osman}}, \bibinfo {author} {\bibfnamefont {H.}~\bibnamefont {Eskes}},\ and\ \bibinfo {author} {\bibfnamefont {W.}~\bibnamefont {van Saarloos}},\ }\bibfield  {title} {\bibinfo {title} {Dynamical stripe correlations in cuprate superconductors},\ }\href {https://doi.org/10.1007/BF00768446} {\bibfield  {journal} {\bibinfo  {journal} {Journal of Low Temperature Physics}\ }\textbf {\bibinfo {volume} {105}},\ \bibinfo {pages} {569} (\bibinfo {year} {1996})}\BibitemShut {NoStop}%
\bibitem [{\citenamefont {Emery}\ \emph {et~al.}(1999)\citenamefont {Emery}, \citenamefont {Kivelson},\ and\ \citenamefont {Tranquada}}]{Emery1999}%
  \BibitemOpen
  \bibfield  {author} {\bibinfo {author} {\bibfnamefont {V.~J.}\ \bibnamefont {Emery}}, \bibinfo {author} {\bibfnamefont {S.~A.}\ \bibnamefont {Kivelson}},\ and\ \bibinfo {author} {\bibfnamefont {J.~M.}\ \bibnamefont {Tranquada}},\ }\bibfield  {title} {\bibinfo {title} {Stripe phases in high-temperature superconductors},\ }\href {https://doi.org/10.1073/pnas.96.16.8814} {\bibfield  {journal} {\bibinfo  {journal} {Proceedings of the National Academy of Sciences}\ }\textbf {\bibinfo {volume} {96}},\ \bibinfo {pages} {8814} (\bibinfo {year} {1999})}\BibitemShut {NoStop}%
\bibitem [{\citenamefont {Berg}\ \emph {et~al.}(2009)\citenamefont {Berg}, \citenamefont {Fradkin}, \citenamefont {Kivelson},\ and\ \citenamefont {Tranquada}}]{Berg_2009}%
  \BibitemOpen
  \bibfield  {author} {\bibinfo {author} {\bibfnamefont {E.}~\bibnamefont {Berg}}, \bibinfo {author} {\bibfnamefont {E.}~\bibnamefont {Fradkin}}, \bibinfo {author} {\bibfnamefont {S.~A.}\ \bibnamefont {Kivelson}},\ and\ \bibinfo {author} {\bibfnamefont {J.~M.}\ \bibnamefont {Tranquada}},\ }\bibfield  {title} {\bibinfo {title} {Striped superconductors: how spin, charge and superconducting orders intertwine in the cuprates},\ }\href {https://doi.org/10.1088/1367-2630/11/11/115004} {\bibfield  {journal} {\bibinfo  {journal} {New Journal of Physics}\ }\textbf {\bibinfo {volume} {11}},\ \bibinfo {pages} {115004} (\bibinfo {year} {2009})}\BibitemShut {NoStop}%
\bibitem [{\citenamefont {Vojta}(2009)}]{Matthias_2009}%
  \BibitemOpen
  \bibfield  {author} {\bibinfo {author} {\bibfnamefont {M.}~\bibnamefont {Vojta}},\ }\bibfield  {title} {\bibinfo {title} {Lattice symmetry breaking in cuprate superconductors: stripes, nematics, and superconductivity},\ }\href {https://doi.org/10.1080/00018730903122242} {\bibfield  {journal} {\bibinfo  {journal} {Advances in Physics}\ }\textbf {\bibinfo {volume} {58}},\ \bibinfo {pages} {699} (\bibinfo {year} {2009})}\BibitemShut {NoStop}%
\bibitem [{\citenamefont {Zhang}\ \emph {et~al.}(2020)\citenamefont {Zhang}, \citenamefont {Lane}, \citenamefont {Furness}, \citenamefont {Barbiellini}, \citenamefont {Perdew}, \citenamefont {Markiewicz}, \citenamefont {Bansil},\ and\ \citenamefont {Sun}}]{Zhang68}%
  \BibitemOpen
  \bibfield  {author} {\bibinfo {author} {\bibfnamefont {Y.}~\bibnamefont {Zhang}}, \bibinfo {author} {\bibfnamefont {C.}~\bibnamefont {Lane}}, \bibinfo {author} {\bibfnamefont {J.~W.}\ \bibnamefont {Furness}}, \bibinfo {author} {\bibfnamefont {B.}~\bibnamefont {Barbiellini}}, \bibinfo {author} {\bibfnamefont {J.~P.}\ \bibnamefont {Perdew}}, \bibinfo {author} {\bibfnamefont {R.~S.}\ \bibnamefont {Markiewicz}}, \bibinfo {author} {\bibfnamefont {A.}~\bibnamefont {Bansil}},\ and\ \bibinfo {author} {\bibfnamefont {J.}~\bibnamefont {Sun}},\ }\bibfield  {title} {\bibinfo {title} {Competing stripe and magnetic phases in the cuprates from first principles},\ }\href {https://doi.org/10.1073/pnas.1910411116} {\bibfield  {journal} {\bibinfo  {journal} {Proceedings of the National Academy of Sciences}\ }\textbf {\bibinfo {volume} {117}},\ \bibinfo {pages} {68} (\bibinfo {year} {2020})}\BibitemShut {NoStop}%
\bibitem [{\citenamefont {Maldacena}(1999)}]{Maldacena1998}%
  \BibitemOpen
  \bibfield  {author} {\bibinfo {author} {\bibfnamefont {J.}~\bibnamefont {Maldacena}},\ }\bibfield  {title} {\bibinfo {title} {The large-n limit of superconformal field theories and supergravity},\ }\href {https://doi.org/10.1023/A:1026654312961} {\bibfield  {journal} {\bibinfo  {journal} {International Journal of Theoretical Physics}\ }\textbf {\bibinfo {volume} {38}},\ \bibinfo {pages} {1113} (\bibinfo {year} {1999})}\BibitemShut {NoStop}%
\bibitem [{\citenamefont {Witten}(1998)}]{Witten1998}%
  \BibitemOpen
  \bibfield  {author} {\bibinfo {author} {\bibfnamefont {E.}~\bibnamefont {Witten}},\ }\bibfield  {title} {\bibinfo {title} {{Anti de sitter space and holography}},\ }\href {https://doi.org/10.4310/ATMP.1998.v2.n2.a2} {\bibfield  {journal} {\bibinfo  {journal} {Advances in Theoretical and Mathematical Physics}\ }\textbf {\bibinfo {volume} {2}},\ \bibinfo {pages} {253} (\bibinfo {year} {1998})},\ \Eprint {https://arxiv.org/abs/hep-th/9802150} {arXiv:hep-th/9802150 [hep-th]} \BibitemShut {NoStop}%
\bibitem [{\citenamefont {Gubser}\ \emph {et~al.}(1998)\citenamefont {Gubser}, \citenamefont {Klebanov},\ and\ \citenamefont {Polyakov}}]{Gubser1998}%
  \BibitemOpen
  \bibfield  {author} {\bibinfo {author} {\bibfnamefont {S.}~\bibnamefont {Gubser}}, \bibinfo {author} {\bibfnamefont {I.}~\bibnamefont {Klebanov}},\ and\ \bibinfo {author} {\bibfnamefont {A.}~\bibnamefont {Polyakov}},\ }\bibfield  {title} {\bibinfo {title} {Gauge theory correlators from non-critical string theory},\ }\href {https://doi.org/https://doi.org/10.1016/S0370-2693(98)00377-3} {\bibfield  {journal} {\bibinfo  {journal} {Physics Letters B}\ }\textbf {\bibinfo {volume} {428}},\ \bibinfo {pages} {105} (\bibinfo {year} {1998})}\BibitemShut {NoStop}%
\bibitem [{\citenamefont {Faulkner}\ \emph {et~al.}(2011{\natexlab{a}})\citenamefont {Faulkner}, \citenamefont {Iqbal}, \citenamefont {Liu}, \citenamefont {McGreevy},\ and\ \citenamefont {Vegh}}]{Faulkner2011fixed}%
  \BibitemOpen
  \bibfield  {author} {\bibinfo {author} {\bibfnamefont {T.}~\bibnamefont {Faulkner}}, \bibinfo {author} {\bibfnamefont {N.}~\bibnamefont {Iqbal}}, \bibinfo {author} {\bibfnamefont {H.}~\bibnamefont {Liu}}, \bibinfo {author} {\bibfnamefont {J.}~\bibnamefont {McGreevy}},\ and\ \bibinfo {author} {\bibfnamefont {D.}~\bibnamefont {Vegh}},\ }\bibfield  {title} {\bibinfo {title} {Holographic non-fermi-liquid fixed points},\ }\href {https://doi.org/10.1098/rsta.2010.0354} {\bibfield  {journal} {\bibinfo  {journal} {Philosophical Transactions of the Royal Society A: Mathematical, Physical and Engineering Sciences}\ }\textbf {\bibinfo {volume} {369}},\ \bibinfo {pages} {1640} (\bibinfo {year} {2011}{\natexlab{a}})}\BibitemShut {NoStop}%
\bibitem [{\citenamefont {Hartnoll}\ \emph {et~al.}(2016)\citenamefont {Hartnoll}, \citenamefont {Lucas},\ and\ \citenamefont {Sachdev}}]{hartnoll2016holographic}%
  \BibitemOpen
  \bibfield  {author} {\bibinfo {author} {\bibfnamefont {S.~A.}\ \bibnamefont {Hartnoll}}, \bibinfo {author} {\bibfnamefont {A.}~\bibnamefont {Lucas}},\ and\ \bibinfo {author} {\bibfnamefont {S.}~\bibnamefont {Sachdev}},\ }\href {https://arxiv.org/abs/1612.07324} {\bibinfo {title} {Holographic quantum matter}} (\bibinfo {year} {2016})\BibitemShut {NoStop}%
\bibitem [{\citenamefont {Zaanen}\ \emph {et~al.}(2015)\citenamefont {Zaanen}, \citenamefont {Liu}, \citenamefont {Sun},\ and\ \citenamefont {Schalm}}]{Zaanen2015}%
  \BibitemOpen
  \bibfield  {author} {\bibinfo {author} {\bibfnamefont {J.}~\bibnamefont {Zaanen}}, \bibinfo {author} {\bibfnamefont {Y.}~\bibnamefont {Liu}}, \bibinfo {author} {\bibfnamefont {Y.-W.}\ \bibnamefont {Sun}},\ and\ \bibinfo {author} {\bibfnamefont {K.}~\bibnamefont {Schalm}},\ }\href {https://doi.org/10.1017/CBO9781139942492} {\emph {\bibinfo {title} {Holographic Duality in Condensed Matter Physics}}}\ (\bibinfo  {publisher} {Cambridge University Press},\ \bibinfo {year} {2015})\BibitemShut {NoStop}%
\bibitem [{\citenamefont {Ammon}\ and\ \citenamefont {Erdmenger}(2015)}]{ammon_erdmenger_2015}%
  \BibitemOpen
  \bibfield  {author} {\bibinfo {author} {\bibfnamefont {M.}~\bibnamefont {Ammon}}\ and\ \bibinfo {author} {\bibfnamefont {J.}~\bibnamefont {Erdmenger}},\ }\href {https://doi.org/10.1017/CBO9780511846373} {\emph {\bibinfo {title} {Gauge/Gravity Duality: Foundations and Applications}}}\ (\bibinfo  {publisher} {Cambridge University Press},\ \bibinfo {year} {2015})\BibitemShut {NoStop}%
\bibitem [{\citenamefont {Reber}\ \emph {et~al.}(2019)\citenamefont {Reber}, \citenamefont {Zhou}, \citenamefont {Plumb}, \citenamefont {Parham}, \citenamefont {Waugh}, \citenamefont {Cao}, \citenamefont {Sun}, \citenamefont {Li}, \citenamefont {Wang}, \citenamefont {Wen}, \citenamefont {Xu}, \citenamefont {Gu}, \citenamefont {Yoshida}, \citenamefont {Eisaki}, \citenamefont {Arnold},\ and\ \citenamefont {Dessau}}]{Reber2019}%
  \BibitemOpen
  \bibfield  {author} {\bibinfo {author} {\bibfnamefont {T.~J.}\ \bibnamefont {Reber}}, \bibinfo {author} {\bibfnamefont {X.}~\bibnamefont {Zhou}}, \bibinfo {author} {\bibfnamefont {N.~C.}\ \bibnamefont {Plumb}}, \bibinfo {author} {\bibfnamefont {S.}~\bibnamefont {Parham}}, \bibinfo {author} {\bibfnamefont {J.~A.}\ \bibnamefont {Waugh}}, \bibinfo {author} {\bibfnamefont {Y.}~\bibnamefont {Cao}}, \bibinfo {author} {\bibfnamefont {Z.}~\bibnamefont {Sun}}, \bibinfo {author} {\bibfnamefont {H.}~\bibnamefont {Li}}, \bibinfo {author} {\bibfnamefont {Q.}~\bibnamefont {Wang}}, \bibinfo {author} {\bibfnamefont {J.~S.}\ \bibnamefont {Wen}}, \bibinfo {author} {\bibfnamefont {Z.~J.}\ \bibnamefont {Xu}}, \bibinfo {author} {\bibfnamefont {G.}~\bibnamefont {Gu}}, \bibinfo {author} {\bibfnamefont {Y.}~\bibnamefont {Yoshida}}, \bibinfo {author} {\bibfnamefont {H.}~\bibnamefont {Eisaki}}, \bibinfo {author} {\bibfnamefont {G.~B.}\ \bibnamefont {Arnold}},\ and\ \bibinfo {author} {\bibfnamefont {D.~S.}\ \bibnamefont {Dessau}},\
  }\bibfield  {title} {\bibinfo {title} {A unified form of low-energy nodal electronic interactions in hole-doped cuprate superconductors},\ }\href {https://doi.org/10.1038/s41467-019-13497-4} {\bibfield  {journal} {\bibinfo  {journal} {Nature Communications}\ }\textbf {\bibinfo {volume} {10}},\ \bibinfo {pages} {5737} (\bibinfo {year} {2019})}\BibitemShut {NoStop}%
\bibitem [{\citenamefont {Varma}\ \emph {et~al.}(1989)\citenamefont {Varma}, \citenamefont {Littlewood}, \citenamefont {Schmitt-Rink}, \citenamefont {Abrahams},\ and\ \citenamefont {Ruckenstein}}]{PhysRevLett.63.1996}%
  \BibitemOpen
  \bibfield  {author} {\bibinfo {author} {\bibfnamefont {C.~M.}\ \bibnamefont {Varma}}, \bibinfo {author} {\bibfnamefont {P.~B.}\ \bibnamefont {Littlewood}}, \bibinfo {author} {\bibfnamefont {S.}~\bibnamefont {Schmitt-Rink}}, \bibinfo {author} {\bibfnamefont {E.}~\bibnamefont {Abrahams}},\ and\ \bibinfo {author} {\bibfnamefont {A.~E.}\ \bibnamefont {Ruckenstein}},\ }\bibfield  {title} {\bibinfo {title} {Phenomenology of the normal state of cu-o high-temperature superconductors},\ }\href {https://doi.org/10.1103/PhysRevLett.63.1996} {\bibfield  {journal} {\bibinfo  {journal} {Phys. Rev. Lett.}\ }\textbf {\bibinfo {volume} {63}},\ \bibinfo {pages} {1996} (\bibinfo {year} {1989})}\BibitemShut {NoStop}%
\bibitem [{\citenamefont {Iqbal}\ and\ \citenamefont {Liu}(2009)}]{Iqbal2009}%
  \BibitemOpen
  \bibfield  {author} {\bibinfo {author} {\bibfnamefont {N.}~\bibnamefont {Iqbal}}\ and\ \bibinfo {author} {\bibfnamefont {H.}~\bibnamefont {Liu}},\ }\bibfield  {title} {\bibinfo {title} {Real-time response in ads/cft with application to spinors},\ }\href {https://doi.org/https://doi.org/10.1002/prop.200900057} {\bibfield  {journal} {\bibinfo  {journal} {Fortschritte der Physik}\ }\textbf {\bibinfo {volume} {57}},\ \bibinfo {pages} {367} (\bibinfo {year} {2009})},\ \Eprint {https://arxiv.org/abs/https://onlinelibrary.wiley.com/doi/pdf/10.1002/prop.200900057} {https://onlinelibrary.wiley.com/doi/pdf/10.1002/prop.200900057} \BibitemShut {NoStop}%
\bibitem [{\citenamefont {Iqbal}\ \emph {et~al.}(2011)\citenamefont {Iqbal}, \citenamefont {Liu},\ and\ \citenamefont {Mezei}}]{Iqbal2011}%
  \BibitemOpen
  \bibfield  {author} {\bibinfo {author} {\bibfnamefont {N.}~\bibnamefont {Iqbal}}, \bibinfo {author} {\bibfnamefont {H.}~\bibnamefont {Liu}},\ and\ \bibinfo {author} {\bibfnamefont {M.}~\bibnamefont {Mezei}},\ }\bibfield  {title} {\bibinfo {title} {Lectures on holographic non-fermi liquids and quantum phase transitions},\ }in\ \href {https://doi.org/10.1142/9789814350525_0013} {\emph {\bibinfo {booktitle} {String Theory and Its Applications}}}\ (\bibinfo  {publisher} {{WORLD} {SCIENTIFIC}},\ \bibinfo {year} {2011})\BibitemShut {NoStop}%
\bibitem [{\citenamefont {Liu}\ \emph {et~al.}(2011)\citenamefont {Liu}, \citenamefont {McGreevy},\ and\ \citenamefont {Vegh}}]{Liu2011}%
  \BibitemOpen
  \bibfield  {author} {\bibinfo {author} {\bibfnamefont {H.}~\bibnamefont {Liu}}, \bibinfo {author} {\bibfnamefont {J.}~\bibnamefont {McGreevy}},\ and\ \bibinfo {author} {\bibfnamefont {D.}~\bibnamefont {Vegh}},\ }\bibfield  {title} {\bibinfo {title} {Non-fermi liquids from holography},\ }\href {https://doi.org/10.1103/PhysRevD.83.065029} {\bibfield  {journal} {\bibinfo  {journal} {Phys. Rev. D}\ }\textbf {\bibinfo {volume} {83}},\ \bibinfo {pages} {065029} (\bibinfo {year} {2011})}\BibitemShut {NoStop}%
\bibitem [{\citenamefont {Faulkner}\ \emph {et~al.}(2011{\natexlab{b}})\citenamefont {Faulkner}, \citenamefont {Liu}, \citenamefont {McGreevy},\ and\ \citenamefont {Vegh}}]{Faulkner2011}%
  \BibitemOpen
  \bibfield  {author} {\bibinfo {author} {\bibfnamefont {T.}~\bibnamefont {Faulkner}}, \bibinfo {author} {\bibfnamefont {H.}~\bibnamefont {Liu}}, \bibinfo {author} {\bibfnamefont {J.}~\bibnamefont {McGreevy}},\ and\ \bibinfo {author} {\bibfnamefont {D.}~\bibnamefont {Vegh}},\ }\bibfield  {title} {\bibinfo {title} {Emergent quantum criticality, fermi surfaces, and ${\mathrm{ads}}_{2}$},\ }\href {https://doi.org/10.1103/PhysRevD.83.125002} {\bibfield  {journal} {\bibinfo  {journal} {Phys. Rev. D}\ }\textbf {\bibinfo {volume} {83}},\ \bibinfo {pages} {125002} (\bibinfo {year} {2011}{\natexlab{b}})}\BibitemShut {NoStop}%
\bibitem [{\citenamefont {{\v{C}}ubrovi{\'{c}}}\ \emph {et~al.}(2009)\citenamefont {{\v{C}}ubrovi{\'{c}}}, \citenamefont {Zaanen},\ and\ \citenamefont {Schalm}}]{Cubrovic2009}%
  \BibitemOpen
  \bibfield  {author} {\bibinfo {author} {\bibfnamefont {M.}~\bibnamefont {{\v{C}}ubrovi{\'{c}}}}, \bibinfo {author} {\bibfnamefont {J.}~\bibnamefont {Zaanen}},\ and\ \bibinfo {author} {\bibfnamefont {K.}~\bibnamefont {Schalm}},\ }\bibfield  {title} {\bibinfo {title} {String theory, quantum phase transitions, and the emergent fermi liquid},\ }\href {https://doi.org/10.1126/science.1174962} {\bibfield  {journal} {\bibinfo  {journal} {Science}\ }\textbf {\bibinfo {volume} {325}},\ \bibinfo {pages} {439} (\bibinfo {year} {2009})}\BibitemShut {NoStop}%
\bibitem [{\citenamefont {Smit}\ \emph {et~al.}(2021)\citenamefont {Smit}, \citenamefont {Mauri}, \citenamefont {Bawden}, \citenamefont {Heringa}, \citenamefont {Gerritsen}, \citenamefont {van Heumen}, \citenamefont {Huang}, \citenamefont {Kondo}, \citenamefont {Takeuchi}, \citenamefont {Hussey}, \citenamefont {Kim}, \citenamefont {Cacho}, \citenamefont {Krikun}, \citenamefont {Schalm}, \citenamefont {Stoof},\ and\ \citenamefont {Golden}}]{Smit2021}%
  \BibitemOpen
  \bibfield  {author} {\bibinfo {author} {\bibfnamefont {S.}~\bibnamefont {Smit}}, \bibinfo {author} {\bibfnamefont {E.}~\bibnamefont {Mauri}}, \bibinfo {author} {\bibfnamefont {L.}~\bibnamefont {Bawden}}, \bibinfo {author} {\bibfnamefont {F.}~\bibnamefont {Heringa}}, \bibinfo {author} {\bibfnamefont {F.}~\bibnamefont {Gerritsen}}, \bibinfo {author} {\bibfnamefont {E.}~\bibnamefont {van Heumen}}, \bibinfo {author} {\bibfnamefont {Y.~K.}\ \bibnamefont {Huang}}, \bibinfo {author} {\bibfnamefont {T.}~\bibnamefont {Kondo}}, \bibinfo {author} {\bibfnamefont {T.}~\bibnamefont {Takeuchi}}, \bibinfo {author} {\bibfnamefont {N.~E.}\ \bibnamefont {Hussey}}, \bibinfo {author} {\bibfnamefont {T.~K.}\ \bibnamefont {Kim}}, \bibinfo {author} {\bibfnamefont {C.}~\bibnamefont {Cacho}}, \bibinfo {author} {\bibfnamefont {A.}~\bibnamefont {Krikun}}, \bibinfo {author} {\bibfnamefont {K.}~\bibnamefont {Schalm}}, \bibinfo {author} {\bibfnamefont {H.~T.~C.}\ \bibnamefont {Stoof}},\ and\ \bibinfo {author} {\bibfnamefont {M.~S.}\
  \bibnamefont {Golden}},\ }\href@noop {} {\bibinfo {title} {Momentum-dependent scaling exponents of nodal self-energies measured in strange metal cuprates and modelled using semi-holography}} (\bibinfo {year} {2021}),\ \Eprint {https://arxiv.org/abs/arXiv:2112.06576} {arXiv:2112.06576} \BibitemShut {NoStop}%
\bibitem [{\citenamefont {Negele}\ and\ \citenamefont {Orland}(1998)}]{Negele1998Quantum}%
  \BibitemOpen
  \bibfield  {author} {\bibinfo {author} {\bibfnamefont {J.~W.}\ \bibnamefont {Negele}}\ and\ \bibinfo {author} {\bibfnamefont {H.}~\bibnamefont {Orland}},\ }\href {http://www.worldcat.org/isbn/0738200522} {\emph {\bibinfo {title} {Quantum Many-particle Systems}}}\ (\bibinfo  {publisher} {Westview Press},\ \bibinfo {year} {1998})\BibitemShut {NoStop}%
\bibitem [{\citenamefont {Jin}\ \emph {et~al.}(2019)\citenamefont {Jin}, \citenamefont {Tsyplyatyev}, \citenamefont {Moreno}, \citenamefont {Anthore}, \citenamefont {Tan}, \citenamefont {Griffiths}, \citenamefont {Farrer}, \citenamefont {Ritchie}, \citenamefont {Glazman}, \citenamefont {Schofield},\ and\ \citenamefont {Ford}}]{Jin2019}%
  \BibitemOpen
  \bibfield  {author} {\bibinfo {author} {\bibfnamefont {Y.}~\bibnamefont {Jin}}, \bibinfo {author} {\bibfnamefont {O.}~\bibnamefont {Tsyplyatyev}}, \bibinfo {author} {\bibfnamefont {M.}~\bibnamefont {Moreno}}, \bibinfo {author} {\bibfnamefont {A.}~\bibnamefont {Anthore}}, \bibinfo {author} {\bibfnamefont {W.~K.}\ \bibnamefont {Tan}}, \bibinfo {author} {\bibfnamefont {J.~P.}\ \bibnamefont {Griffiths}}, \bibinfo {author} {\bibfnamefont {I.}~\bibnamefont {Farrer}}, \bibinfo {author} {\bibfnamefont {D.~A.}\ \bibnamefont {Ritchie}}, \bibinfo {author} {\bibfnamefont {L.~I.}\ \bibnamefont {Glazman}}, \bibinfo {author} {\bibfnamefont {A.~J.}\ \bibnamefont {Schofield}},\ and\ \bibinfo {author} {\bibfnamefont {C.~J.~B.}\ \bibnamefont {Ford}},\ }\bibfield  {title} {\bibinfo {title} {Momentum-dependent power law measured in an interacting quantum wire beyond the luttinger limit},\ }\href {https://doi.org/10.1038/s41467-019-10613-2} {\bibfield  {journal} {\bibinfo  {journal} {Nature Communications}\ }\textbf {\bibinfo
  {volume} {10}},\ \bibinfo {pages} {2821} (\bibinfo {year} {2019})}\BibitemShut {NoStop}%
\bibitem [{\citenamefont {G{\"{u}}rsoy}\ \emph {et~al.}(2012)\citenamefont {G{\"{u}}rsoy}, \citenamefont {Plauschinn}, \citenamefont {Stoof},\ and\ \citenamefont {Vandoren}}]{Gursoy2012}%
  \BibitemOpen
  \bibfield  {author} {\bibinfo {author} {\bibfnamefont {U.}~\bibnamefont {G{\"{u}}rsoy}}, \bibinfo {author} {\bibfnamefont {E.}~\bibnamefont {Plauschinn}}, \bibinfo {author} {\bibfnamefont {H.}~\bibnamefont {Stoof}},\ and\ \bibinfo {author} {\bibfnamefont {S.}~\bibnamefont {Vandoren}},\ }\bibfield  {title} {\bibinfo {title} {{Holography and ARPES sum-rules}},\ }\href {https://doi.org/10.1007/JHEP05(2012)018} {\bibfield  {journal} {\bibinfo  {journal} {Journal of High Energy Physics}\ }\textbf {\bibinfo {volume} {2012}},\ \bibinfo {pages} {18} (\bibinfo {year} {2012})},\ \Eprint {https://arxiv.org/abs/1112.5074} {1112.5074} \BibitemShut {NoStop}%
\bibitem [{\citenamefont {Gubser}\ and\ \citenamefont {Ren}(2012)}]{Gubser2012}%
  \BibitemOpen
  \bibfield  {author} {\bibinfo {author} {\bibfnamefont {S.~S.}\ \bibnamefont {Gubser}}\ and\ \bibinfo {author} {\bibfnamefont {J.}~\bibnamefont {Ren}},\ }\bibfield  {title} {\bibinfo {title} {Analytic fermionic green's functions from holography},\ }\href {https://doi.org/10.1103/PhysRevD.86.046004} {\bibfield  {journal} {\bibinfo  {journal} {Phys. Rev. D}\ }\textbf {\bibinfo {volume} {86}},\ \bibinfo {pages} {046004} (\bibinfo {year} {2012})}\BibitemShut {NoStop}%
\bibitem [{\citenamefont {Gubser}\ and\ \citenamefont {Rocha}(2010)}]{Gubser_2010}%
  \BibitemOpen
  \bibfield  {author} {\bibinfo {author} {\bibfnamefont {S.~S.}\ \bibnamefont {Gubser}}\ and\ \bibinfo {author} {\bibfnamefont {F.~D.}\ \bibnamefont {Rocha}},\ }\bibfield  {title} {\bibinfo {title} {Peculiar properties of a charged dilatonic black hole in ${\mathrm{ads}}_{5}$},\ }\href {https://doi.org/10.1103/PhysRevD.81.046001} {\bibfield  {journal} {\bibinfo  {journal} {Phys. Rev. D}\ }\textbf {\bibinfo {volume} {81}},\ \bibinfo {pages} {046001} (\bibinfo {year} {2010})}\BibitemShut {NoStop}%
\bibitem [{Note1()}]{Note1}%
  \BibitemOpen
  \bibinfo {note} {Without disorder, our model is translationally invariant and it, hence, has an infinite conductivity at all $T$. Given that a strange metal is defined by its linear-in-$T$ resistivity, it would be more appropriate to refer to the model as that of a \protect \textit {non-Fermi liquid}. However, since we do not consider transport here but only the fermionic self-energy that we compare to the spectral function measured by ARPES of the \protect \textit {strange metal} phase of a cuprate, we still sometimes refer to the model as one for a strange metal with a little abuse of nomenclature.}\BibitemShut {Stop}%
\bibitem [{\citenamefont {Balm}\ \emph {et~al.}(2023)\citenamefont {Balm}, \citenamefont {Chagnet}, \citenamefont {Arend}, \citenamefont {Aretz}, \citenamefont {Grosvenor}, \citenamefont {Janse}, \citenamefont {Moors}, \citenamefont {Post}, \citenamefont {Ohanesjan}, \citenamefont {Rodriguez-Fernandez}, \citenamefont {Schalm},\ and\ \citenamefont {Zaanen}}]{Balm2023}%
  \BibitemOpen
  \bibfield  {author} {\bibinfo {author} {\bibfnamefont {F.}~\bibnamefont {Balm}}, \bibinfo {author} {\bibfnamefont {N.}~\bibnamefont {Chagnet}}, \bibinfo {author} {\bibfnamefont {S.}~\bibnamefont {Arend}}, \bibinfo {author} {\bibfnamefont {J.}~\bibnamefont {Aretz}}, \bibinfo {author} {\bibfnamefont {K.}~\bibnamefont {Grosvenor}}, \bibinfo {author} {\bibfnamefont {M.}~\bibnamefont {Janse}}, \bibinfo {author} {\bibfnamefont {O.}~\bibnamefont {Moors}}, \bibinfo {author} {\bibfnamefont {J.}~\bibnamefont {Post}}, \bibinfo {author} {\bibfnamefont {V.}~\bibnamefont {Ohanesjan}}, \bibinfo {author} {\bibfnamefont {D.}~\bibnamefont {Rodriguez-Fernandez}}, \bibinfo {author} {\bibfnamefont {K.}~\bibnamefont {Schalm}},\ and\ \bibinfo {author} {\bibfnamefont {J.}~\bibnamefont {Zaanen}},\ }\bibfield  {title} {\bibinfo {title} {$t$-linear resistivity, optical conductivity, and planckian transport for a holographic local quantum critical metal in a periodic potential},\ }\href {https://doi.org/10.1103/PhysRevB.108.125145}
  {\bibfield  {journal} {\bibinfo  {journal} {Phys. Rev. B}\ }\textbf {\bibinfo {volume} {108}},\ \bibinfo {pages} {125145} (\bibinfo {year} {2023})}\BibitemShut {NoStop}%
\bibitem [{Note2()}]{Note2}%
  \BibitemOpen
  \bibinfo {note} {As you can see, the dimension of the operator in standard and alternative quantization is related by $m \rightarrow -m$, if we indeed always define the source as the leading order term in the expansion in Eq.\ \protect \eqref {eq:psi_near_boundary}, then changing the sign of $m$ exchanges the roles of the coefficients as the source and operator response, effectively going from standard to alternative quantization.}\BibitemShut {Stop}%
\bibitem [{\citenamefont {Jacobs}\ \emph {et~al.}(2014)\citenamefont {Jacobs}, \citenamefont {Vandoren},\ and\ \citenamefont {Stoof}}]{Jacobs2014}%
  \BibitemOpen
  \bibfield  {author} {\bibinfo {author} {\bibfnamefont {V.~P.}\ \bibnamefont {Jacobs}}, \bibinfo {author} {\bibfnamefont {S.~J.}\ \bibnamefont {Vandoren}},\ and\ \bibinfo {author} {\bibfnamefont {H.~T.}\ \bibnamefont {Stoof}},\ }\bibfield  {title} {\bibinfo {title} {{Holographic interaction effects on transport in Dirac semimetals}},\ }\bibfield  {journal} {\bibinfo  {journal} {Physical Review B - Condensed Matter and Materials Physics}\ }\textbf {\bibinfo {volume} {90}},\ \href {https://doi.org/10.1103/PhysRevB.90.045108} {10.1103/PhysRevB.90.045108} (\bibinfo {year} {2014}),\ \Eprint {https://arxiv.org/abs/1403.3608} {arXiv:1403.3608} \BibitemShut {NoStop}%
\bibitem [{\citenamefont {Jacobs}\ \emph {et~al.}(2015)\citenamefont {Jacobs}, \citenamefont {Grubinskas},\ and\ \citenamefont {Stoof}}]{Jacobs2015}%
  \BibitemOpen
  \bibfield  {author} {\bibinfo {author} {\bibfnamefont {V.~P.~J.}\ \bibnamefont {Jacobs}}, \bibinfo {author} {\bibfnamefont {S.}~\bibnamefont {Grubinskas}},\ and\ \bibinfo {author} {\bibfnamefont {H.~T.~C.}\ \bibnamefont {Stoof}},\ }\bibfield  {title} {\bibinfo {title} {Towards a field-theory interpretation of bottom-up holography},\ }\bibfield  {journal} {\bibinfo  {journal} {Journal of High Energy Physics}\ }\textbf {\bibinfo {volume} {2015}},\ \href {https://doi.org/10.1007/jhep04(2015)033} {10.1007/jhep04(2015)033} (\bibinfo {year} {2015})\BibitemShut {NoStop}%
\bibitem [{\citenamefont {Son}\ and\ \citenamefont {Starinets}(2002)}]{Son_2002}%
  \BibitemOpen
  \bibfield  {author} {\bibinfo {author} {\bibfnamefont {D.~T.}\ \bibnamefont {Son}}\ and\ \bibinfo {author} {\bibfnamefont {A.~O.}\ \bibnamefont {Starinets}},\ }\bibfield  {title} {\bibinfo {title} {Minkowski-space correlators in ads/cft correspondence: recipe and applications},\ }\href {https://doi.org/10.1088/1126-6708/2002/09/042} {\bibfield  {journal} {\bibinfo  {journal} {Journal of High Energy Physics}\ }\textbf {\bibinfo {volume} {2002}},\ \bibinfo {pages} {042} (\bibinfo {year} {2002})}\BibitemShut {NoStop}%
\bibitem [{\citenamefont {Faulkner}\ and\ \citenamefont {Polchinski}(2011)}]{Faulkner:2010tq}%
  \BibitemOpen
  \bibfield  {author} {\bibinfo {author} {\bibfnamefont {T.}~\bibnamefont {Faulkner}}\ and\ \bibinfo {author} {\bibfnamefont {J.}~\bibnamefont {Polchinski}},\ }\bibfield  {title} {\bibinfo {title} {{Semi-Holographic Fermi Liquids}},\ }\href {https://doi.org/10.1007/JHEP06(2011)012} {\bibfield  {journal} {\bibinfo  {journal} {Journal of High Energy Physics}\ }\textbf {\bibinfo {volume} {06}},\ \bibinfo {pages} {012} (\bibinfo {year} {2011})},\ \Eprint {https://arxiv.org/abs/1001.5049} {1001.5049} \BibitemShut {NoStop}%
\bibitem [{\citenamefont {Kordyuk}\ \emph {et~al.}(2005)\citenamefont {Kordyuk}, \citenamefont {Borisenko}, \citenamefont {Koitzsch}, \citenamefont {Fink}, \citenamefont {Knupfer},\ and\ \citenamefont {Berger}}]{Kordyuk2005}%
  \BibitemOpen
  \bibfield  {author} {\bibinfo {author} {\bibfnamefont {A.~A.}\ \bibnamefont {Kordyuk}}, \bibinfo {author} {\bibfnamefont {S.~V.}\ \bibnamefont {Borisenko}}, \bibinfo {author} {\bibfnamefont {A.}~\bibnamefont {Koitzsch}}, \bibinfo {author} {\bibfnamefont {J.}~\bibnamefont {Fink}}, \bibinfo {author} {\bibfnamefont {M.}~\bibnamefont {Knupfer}},\ and\ \bibinfo {author} {\bibfnamefont {H.}~\bibnamefont {Berger}},\ }\bibfield  {title} {\bibinfo {title} {Bare electron dispersion from experiment: Self-consistent self-energy analysis of photoemission data},\ }\href {https://doi.org/10.1103/PhysRevB.71.214513} {\bibfield  {journal} {\bibinfo  {journal} {Phys. Rev. B}\ }\textbf {\bibinfo {volume} {71}},\ \bibinfo {pages} {214513} (\bibinfo {year} {2005})}\BibitemShut {NoStop}%
\bibitem [{Note3()}]{Note3}%
  \BibitemOpen
  \bibinfo {note} {For the phonon, we considered the Fermi velocity doping independent while we allowed the coupling $G_\protect \text {ph}$ and the factor $\Omega $ to change with doping.}\BibitemShut {Stop}%
\end{thebibliography}%

\end{document}